\documentclass[usenatbib,useAMS]{mn2e}
\usepackage{epsfig,color}
\usepackage{amsmath}
\usepackage{mathptmx}
\usepackage{amssymb}
\usepackage{units}
\usepackage{arydshln}
\usepackage[tight]{subfigure}

\newcommand{\LEAFS}{{\sc{Leafs}}}

\newcommand{\gcc}{\ \mathrm{g\ cm^{-3}}}
\newcommand{\cms}{\ \mathrm{cm\ s^{-1}}}
\newcommand{\kms}{\ \mathrm{km\ s^{-1}}}
\newcommand{\cm}{\ \mathrm{cm}}
\newcommand{\s}{\ \mathrm{s}}
\newcommand{\ions}[2]{#1\,{\sc #2}}

\newcommand{\nuc}[2]{\ensuremath{\mathrm{^{#1}#2}}}
\newcommand{\ye}{\ensuremath{Y_\mathrm{e}}}
\newcommand{\msun}{\ensuremath{\mathrm{M}_\odot}}
\newcommand{\simgt}{\,\hbox{\lower0.6ex\hbox{$\sim$}\llap{\raise0.6ex\hbox{$>$}}}\,}
\title[3D delayed detonation models with nucleosynthesis for SNe~Ia]{Three-dimensional delayed-detonation models with
  nucleosynthesis for Type Ia supernovae}
\author[Seitenzahl et al. 2012]{Ivo~R.~Seitenzahl$^{1,2}$
  \thanks{email:irs@mpa-garching.mpg.de},
  Franco~Ciaraldi-Schoolmann$^{2}$,
  Friedrich~K.~R\"opke$^{1}$, \newauthor
  Michael~Fink$^{1}$, 
  Wolfgang~Hillebrandt$^{2}$, 
  Markus~Kromer$^{2}$, 
  R\"udiger~Pakmor$^{3}$,\newauthor
  Ashley~J.~Ruiter$^{2}$,
  Stuart~A.~Sim$^{4}$, 
  Stefan Taubenberger$^{2}$\\ 
$^{1}$Institut f\"ur Theoretische Physik und Astrophysik, Universit\"at
  W\"urzburg, Emil-Fischer-Stra{\ss}e 31, 97074 W\"urzburg\\
$^{2}$Max-Planck-Institut f\"ur Astrophysik,
  Karl-Schwarzschild-Stra{\ss}e 1, 85748 Garching, Germany\\
$^{3}$Heidelberger Institut f\"{u}r Theoretische Studien,
  Schloss-Wolfsbrunnenweg 35, 69118 Heidelberg, Germany\\
$^{4}$Research School of Astronomy and Astrophysics, Mount
  Stromlo Observatory, Cotter Road, Weston Creek, ACT 2611,
  Australia\\}
\date{\today}
\pubyear{}
\begin{document}
\maketitle
\begin{abstract}
  We present results for a suite of fourteen three-dimensional, high
  resolution hydrodynamical simulations of delayed-detonation models
  of Type Ia supernova (SN~Ia) explosions.  This model suite comprises
  the first set of three-dimensional SN~Ia simulations with detailed
  isotopic yield information.  As such, it may serve as a database for
  Chandrasekhar-mass delayed-detonation model nucleosynthetic yields
  and for deriving synthetic observables such as spectra and light
  curves.  We employ a physically motivated, stochastic model based on
  turbulent velocity fluctuations and fuel density to calculate in
  situ the deflagration to detonation transition (DDT) probabilities.
  To obtain different strengths of the deflagration phase and thereby
  different degrees of pre-expansion, we have chosen a sequence of
  initial models with 1, 3, 5, 10, 20, 40, 100, 150, 200, 300, and
  1600 (two different realizations) ignition kernels in a hydrostatic
  white dwarf with central density of $2.9 \times 10^9 \gcc$, plus in
  addition one high central density ($5.5 \times 10^9 \gcc$) and one
  low central density ($1.0 \times 10^9 \gcc$) rendition of the 100
  ignition kernel configuration.  For each simulation we determined
  detailed nucleosynthetic yields by post-processing $10^6$ tracer
  particles with a 384 nuclide reaction network.  All delayed
  detonation models result in explosions unbinding the white dwarf,
  producing a range of \nuc{56}{Ni} masses from 0.32 to 1.11~\msun.
  As a general trend, the models predict that the stable neutron-rich
  iron group isotopes are not found at the lowest velocities, but
  rather at intermediate velocities (${\sim}3,000 - 10,000 \kms$) in a
  shell surrounding a \nuc{56}{Ni}-rich core. The models further
  predict relatively low velocity oxygen and carbon, with typical
  minimum velocities around $4,000$ and $10,000 \kms$, respectively.
\end{abstract}

\begin{keywords}{nuclear reactions, nucleosynthesis, abundances ---
    supernovae: general --- white dwarfs}
\end{keywords}
\section{Introduction}
\label{sec:int}
Type Ia supernovae (SNe Ia) play essential roles in the basic
frameworks of many branches of astrophysics: In star formation and
galaxy dynamics by heating cold interstellar gas
\citep[e.g.][]{scannapieco2008a}, in high energy astrophysics as
sources of Galactic positrons \citep[e.g.][]{chan1993a,prantzos2011a},
in galactic chemical evolution by enriching the interstellar gas with
$\alpha$, Fe-peak, and possibly p-process elements
\citep[e.g.][]{timmes1995a,kobayashi2009a,travaglio2011a} and last but
not least in cosmology as distance indicators
\citep[e.g.][]{riess1998a,schmidt1998a,perlmutter1999a}.  In spite of
their ubiquitous presence in astrophysics, no progenitor systems have
been observed and unambiguous identification of their nature remains
elusive. Essentially by means of exclusion, white dwarf (WD) stars in
interacting binary systems are the only viable proposed progenitor
systems \citep[cf.][]{bloom2012a}.

In the last few years we have witnessed revived interest of the double
detonation He-accretion channel
\citep[e.g.][]{livne1990a,livne1990b,woosley1994a,hoeflich1996a,
  nugent1997a,fink2007a,fink2010a,kromer2010a,woosley2011a} and
increasing support from theorists for the double degenerate channel
\citep{gilfanov2010a,pakmor2010a,pakmor2011a,pakmor2012a}, culminating
in mounting evidence that SN 2011fe was possibly due to a merger of
two white dwarfs \citep[e.g.][]{bloom2012a, chomiuk2012a,roepke2012a}.
In spite of these recent developments, Chandrasekhar-mass models,
which had been the favoured explosion scenario by many in the last two
decades, still retain a list of strong arguments in their favor.  For
example, blue shifted Na-absorption features along the lines of sight
towards SNe Ia are interpreted as a clear signature of the single
degenerate channel \citep{patat2007a,sternberg2011a}.  Further support
to the single degenerate scenario is given by the fact that two
recurrent nova systems are known where the accreting WD is near the
Chandrasekhar-limit (RS Oph and U Sco), which tells us that potential
progenitor systems do exist in nature. In fact, the supernova PTF 11kx
is best described by a symbiotic nova progenitor system
\citep{dilday2012a}.

The issue is further complicated by the fact that the evolution of
zero-age binary systems towards potential SN Ia progenitor systems is
still not well understood and remains a very active area of research
today
\citep[e.g.][]{ruiter2009a,ruiter2011a,mennekens2010a,wang2010a}.
Binary population simulations, in which a large number
(${\simgt}10^{6}$) of binaries can be rapidly evolved from the
zero-age main sequence for a Hubble time, are the only way in which
one can obtain reliable estimates of relative birthrates for different
progenitor scenarios.  However, the physics of binary star evolution
is very complex and some evolutionary phases are poorly understood
\citep[e.g.,][]{vandersluys2010a}.  For example, the evolution of
single degenerate Chandrasekhar mass progenitors strongly relies on
the assumptions made about WD mass accretion rates and retention
efficiencies, for which there are differing camps of thought
\citep[cf.][]{prialnik1995a,han2004a}.  Despite the different
assumptions made in various binary population codes, many of the codes
do predict that single degenerate (Chandrasekhar mass) scenario
progenitors are still promising candidates for at least some fraction
of Type Ia supernovae \citep[see][table 1]{nelemans2012a}.

Near Chandrasekhar-mass explosion models in the single degenerate
channel have long been considered as favorites to explain SNe Ia.  The
realization that a detonation burning through a near
Chandrasekhar-mass white dwarf in hydrostatic equilibrium
\citep{arnett1969a} produces mainly material that has been processed
to nuclear statistical equilibrium (NSE) and not enough intermediate
mass elements (IMEs), such as silicon or sulfur, has lead to the
introduction of ``delayed detonation'' models
\citep{khokhlov1989a}. The key features of delayed detonation models
are the following:
\begin{itemize}
\item First nuclear burning is ignited in a deflagration flame
  producing mainly iron group elements (IGEs) in the initial burning
  phase at high density.
\item The energy released in this sub-sonically propagating mode of
  nuclear burning leads to an expansion of the star, moving unburned
  nuclear fuel to lower density.
\item After some time delay a supersonically moving mode of nuclear
  burning -- a detonation -- emerges.
\item The supersonically moving detonation front quickly overruns much
  of the remaining fuel, a significant fraction of which only burns to
  IMEs owing to the reduced burning densities resulting from the
  pre-expansion.
\end{itemize}

Several three-dimensional models of delayed detonation
Chandrasekhar-mass explosions have been published in the variants of
deflagration to detonation transition (DDT)
\citep{gamezo2005a,roepke2007b,bravo2008a,seitenzahl2011a,roepke2012a},
gravitationally confined detonation (GCD)
\citep{jordan2008a,jordan2012a}, and pulsational reverse detonation
(PRD) models \citep{bravo2009a,bravo2009b}.  Generally, information on
the chemical composition of the ejecta for these models is either not
given or limited to \nuc{56}{Ni} and a coarse description of elemental
yields or major nucleosynthesis groups, such as unburned carbon or
oxygen, IMEs, or IGEs. The exception are \citet{bravo2009b}, who show
a table of 24 isotopes for two PRD explosion models.  While full
isotopic information of the ejecta for three-dimensional \emph{pure
  deflagration} explosions exists in the literature
\citep[e.g.][]{travaglio2004a,roepke2006b} such detailed information
for delayed detonation explosion models is currently only available
for a few two-dimensional \citep{meakin2009a,maeda2010a} or
one-dimensional explosion models
\citep[e.g.][]{iwamoto1999a,brachwitz2000a}. Here, for the first time,
we present detailed nucleosynthetic yields for a suite of high
resolution, three dimensional hydrodynamical explosion simulations,
producing a large range of \nuc{56}{Ni} masses between ${\sim}0.32$
and $1.11$ \msun.

In Section \ref{sec:sims} we introduce our initial stellar models and
the ignition setups, briefly describe our thermonuclear supernova
hydrodynamics code \LEAFS, elaborate on how we model the deflagration
to detonation transition, and discuss the morphologies of the
explosion models.  In Section \ref{sec:nucleo} we summarize how we
obtain the full compositional information by post-processing of tracer
particles and we present the yields for all models.  In Section \ref
{sec:discussion} we discuss the relevance of our contribution and
conclude with an outlook.

\section{Hydrodynamic Simulations}
\label{sec:sims}

\subsection{Initial models}
\label{sec:inimod}

\begin{figure*}
  \begin{center}
    \subfigure[Central ignition region deep inside the WD (for model
    N100)]
    {\includegraphics[width=0.90\columnwidth]{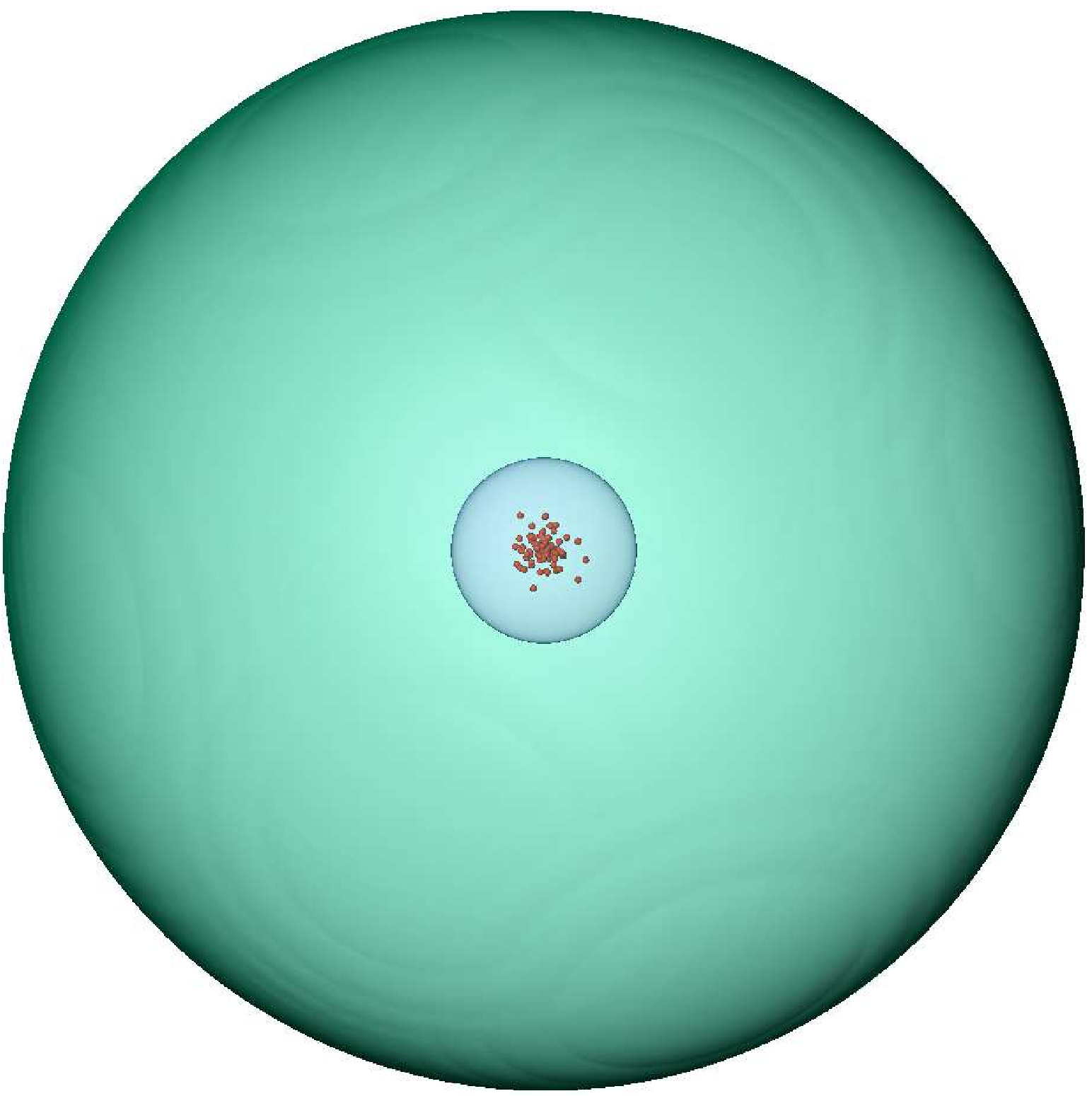}} \\
    \subfigure[N1] {\includegraphics[width=0.45\columnwidth]{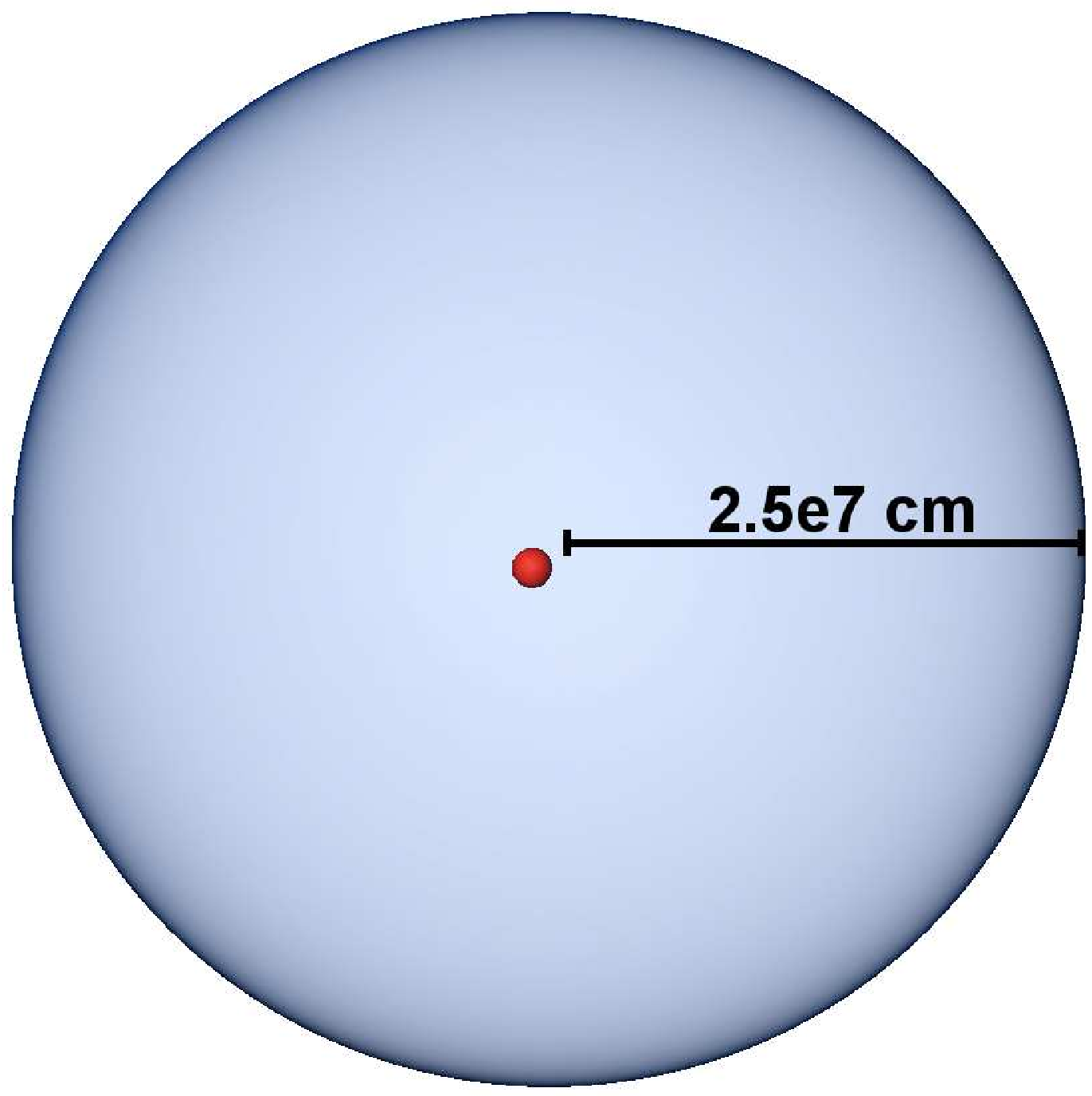}}
    \subfigure[N3] {\includegraphics[width=0.45\columnwidth]{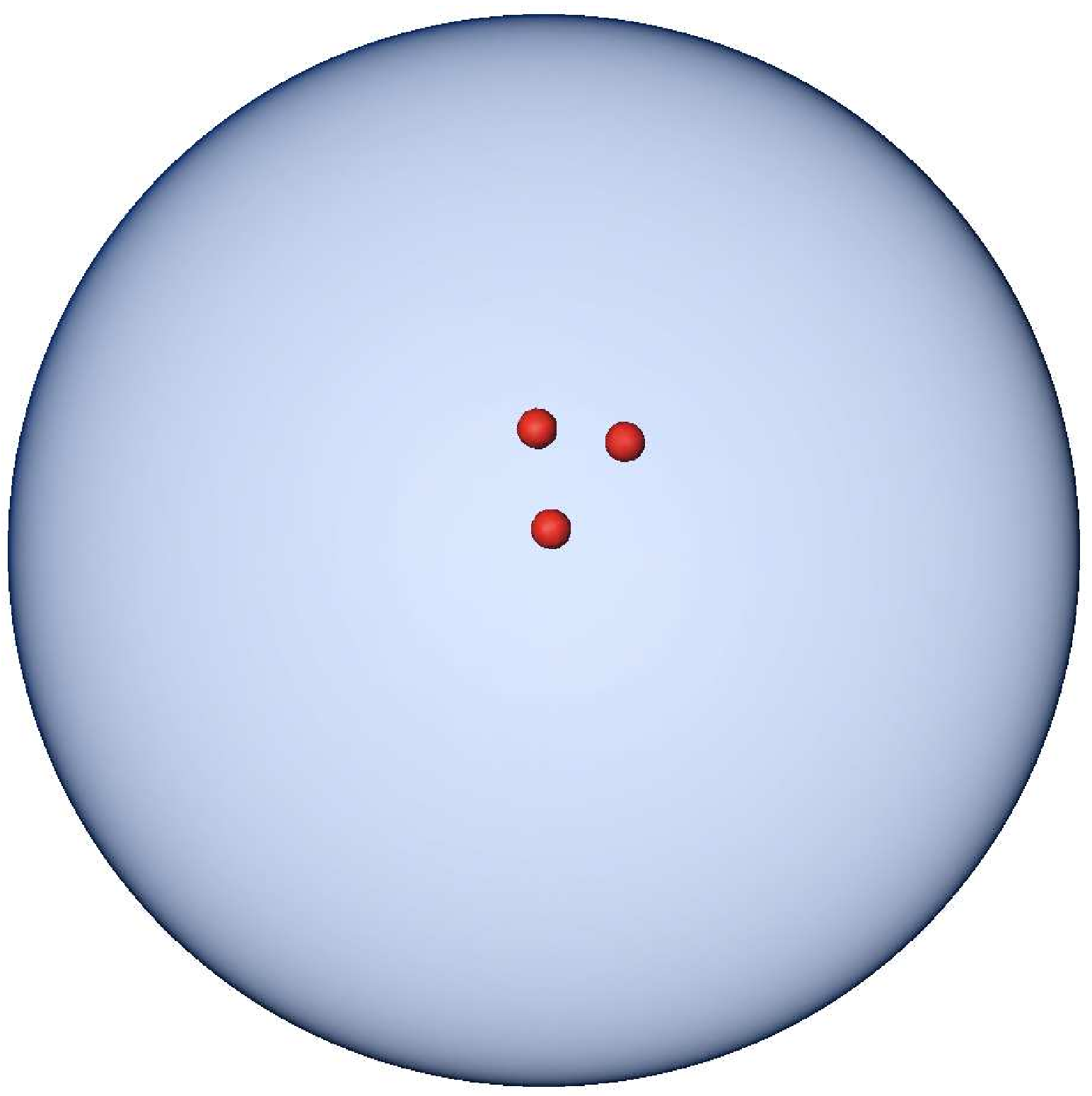}}
    \subfigure[N5] {\includegraphics[width=0.45\columnwidth]{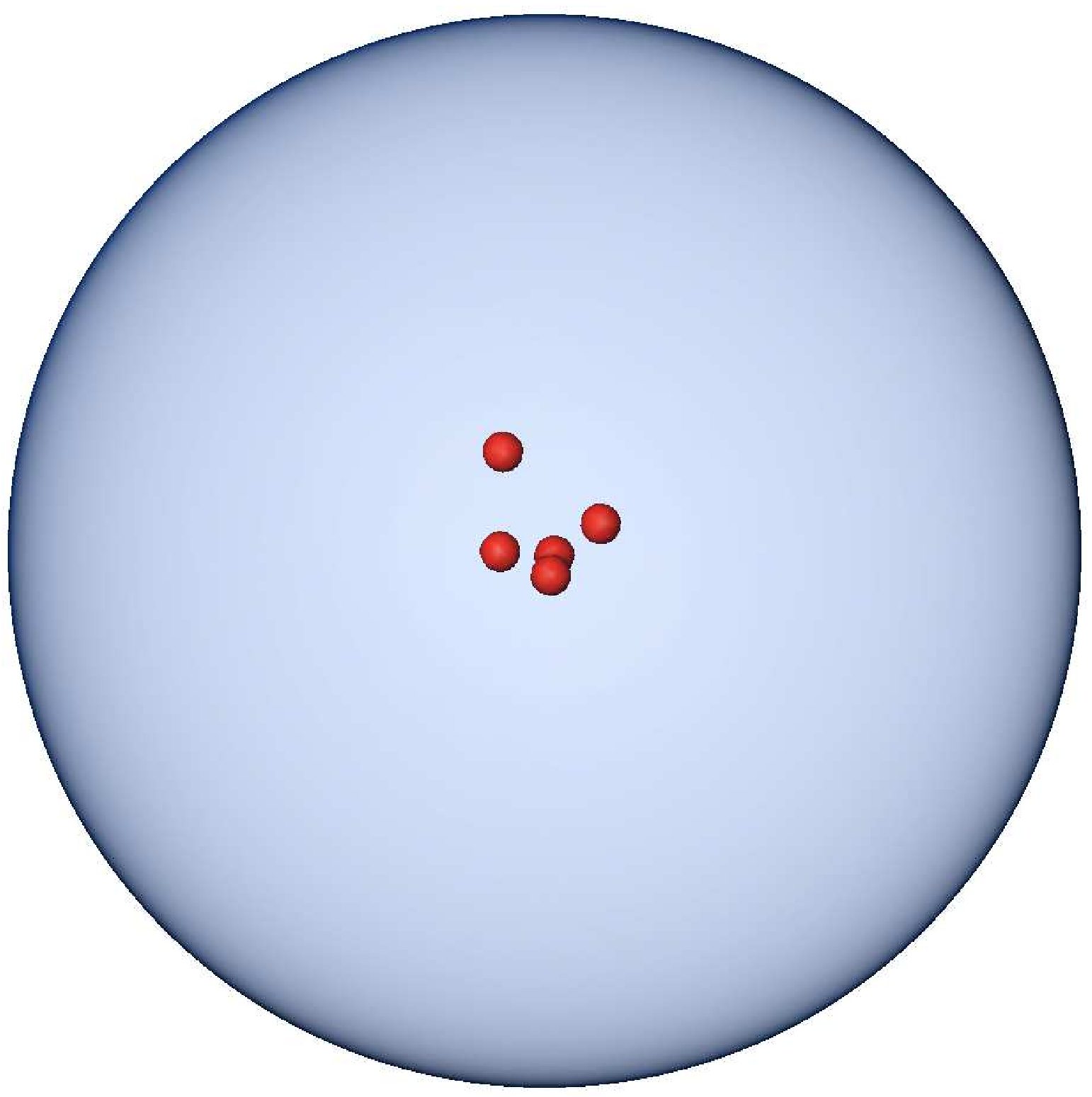}}
    \subfigure[N10] {\includegraphics[width=0.45\columnwidth]{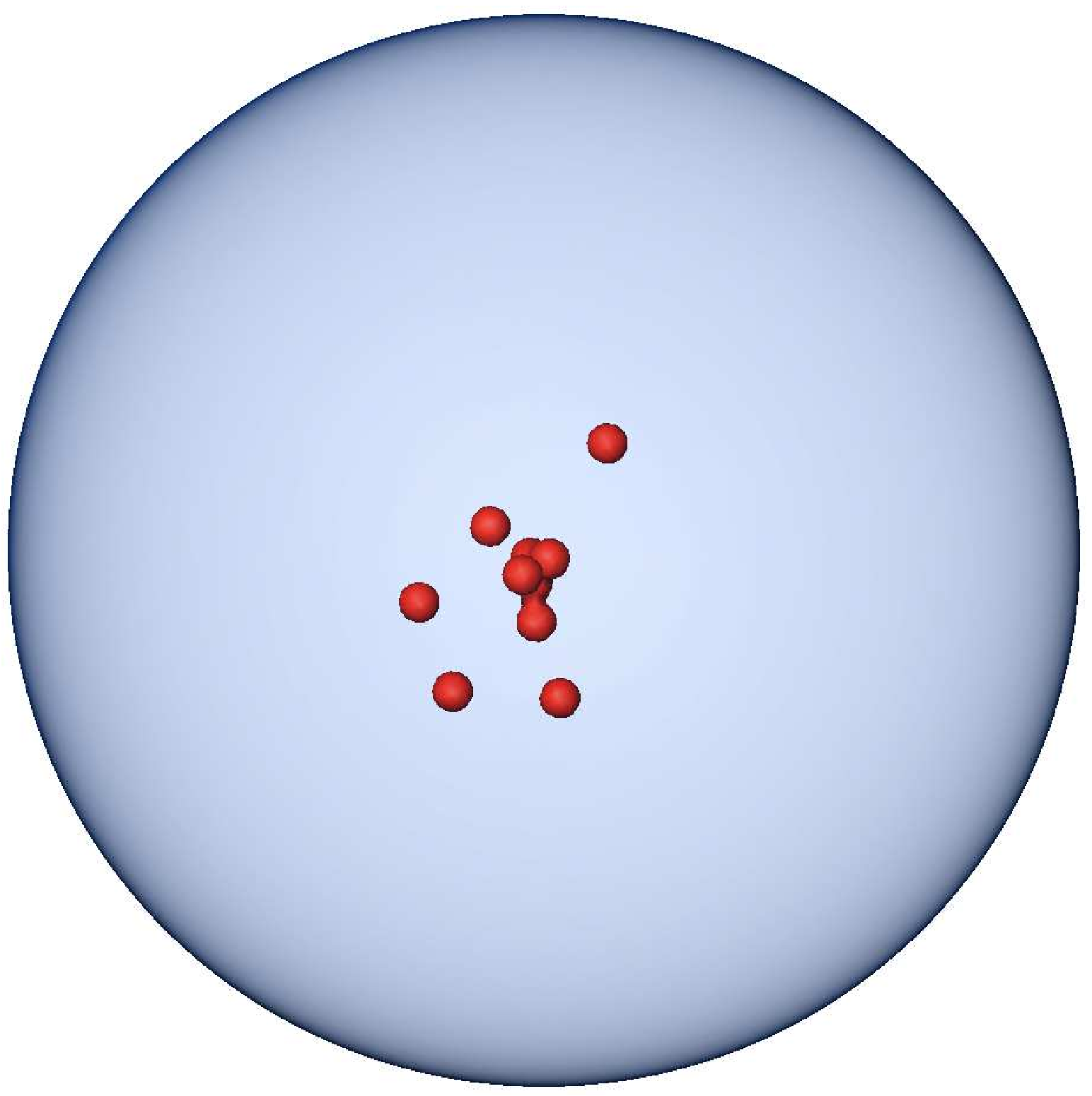}}
    \subfigure[N20] {\includegraphics[width=0.45\columnwidth]{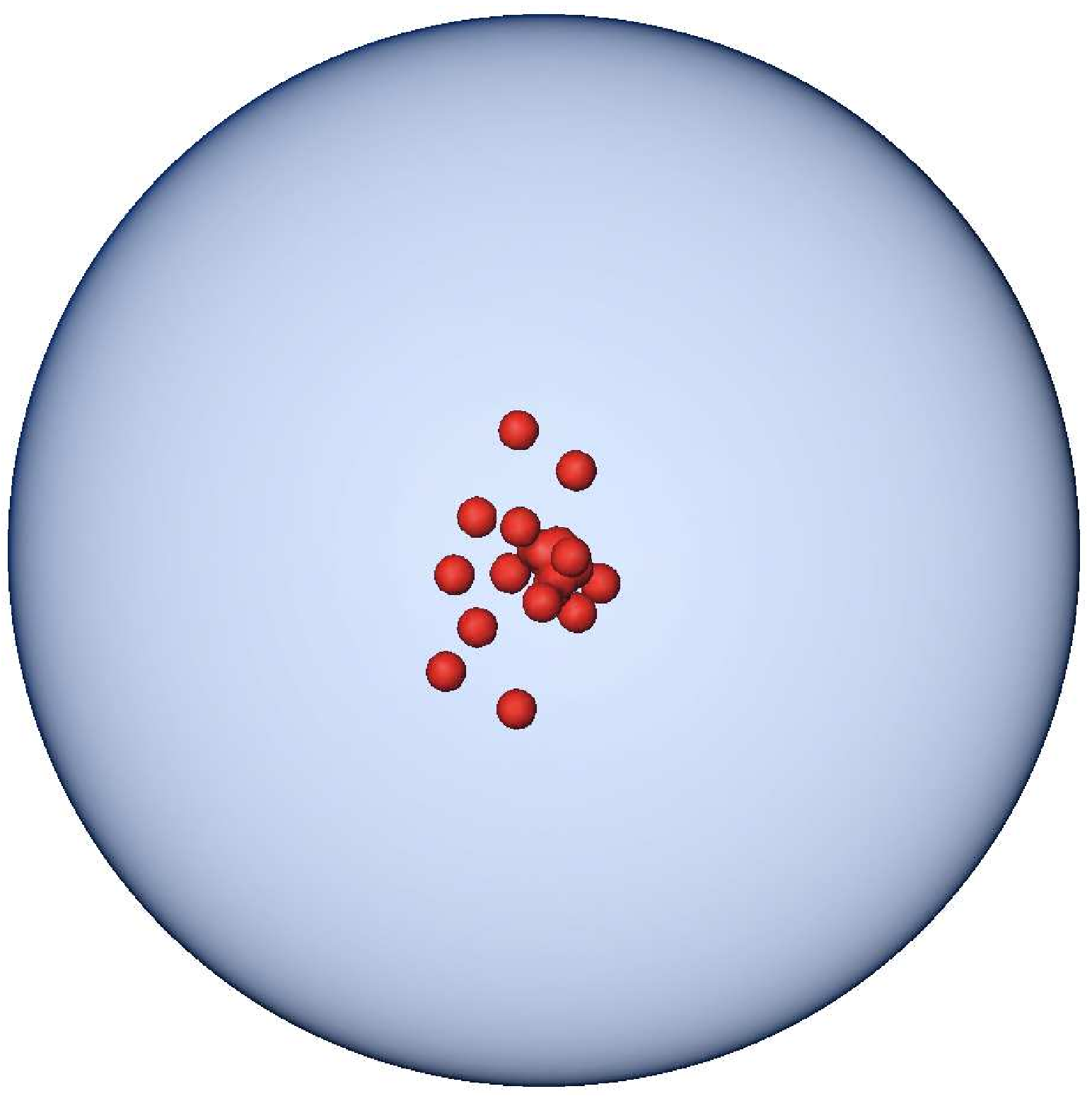}}
    \subfigure[N40] {\includegraphics[width=0.45\columnwidth]{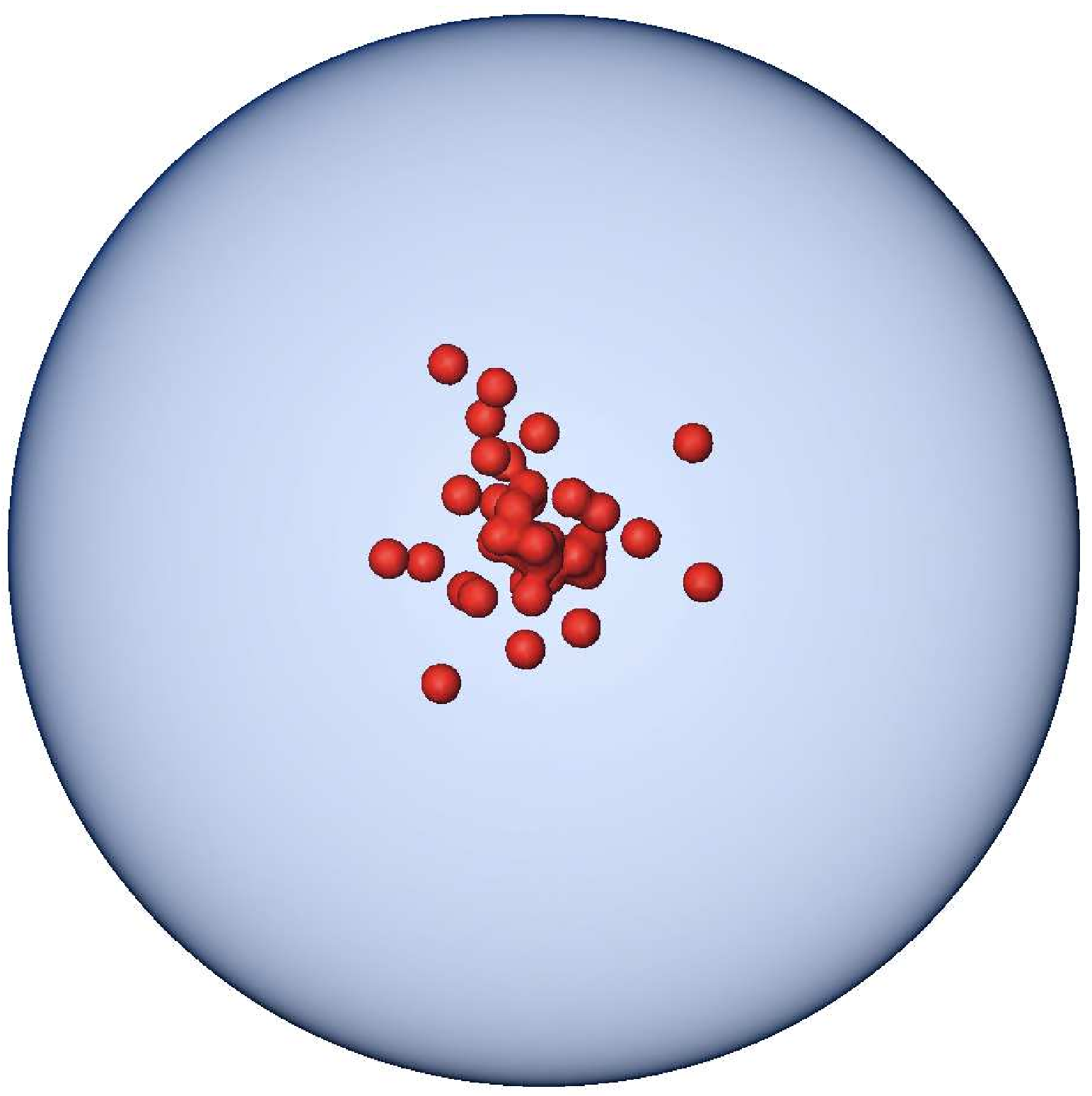}}
    \subfigure[N100(L,H)]
    {\includegraphics[width=0.45\columnwidth]{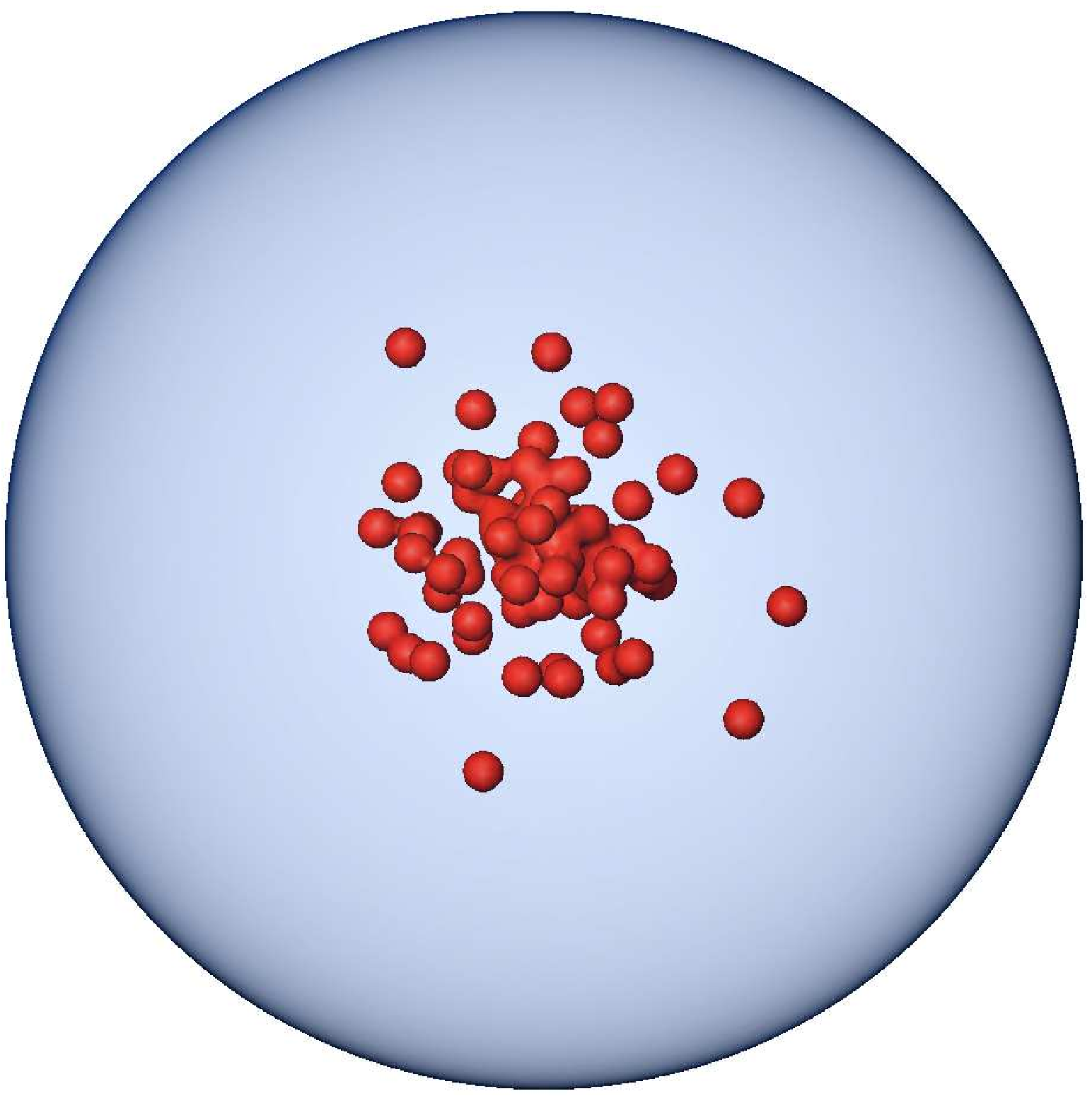}} \subfigure[N150]
    {\includegraphics[width=0.45\columnwidth]{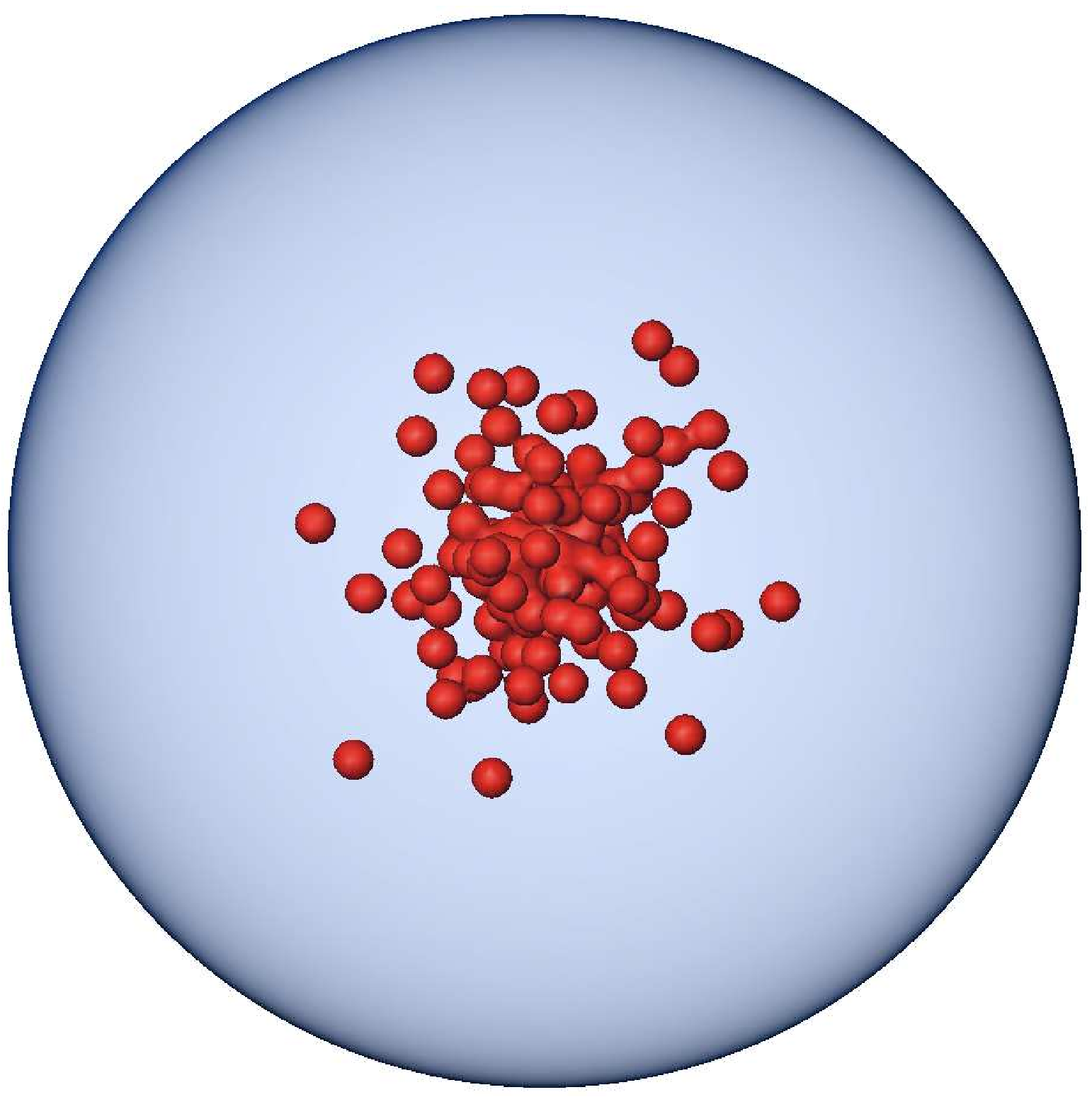}} \subfigure[N200]
    {\includegraphics[width=0.45\columnwidth]{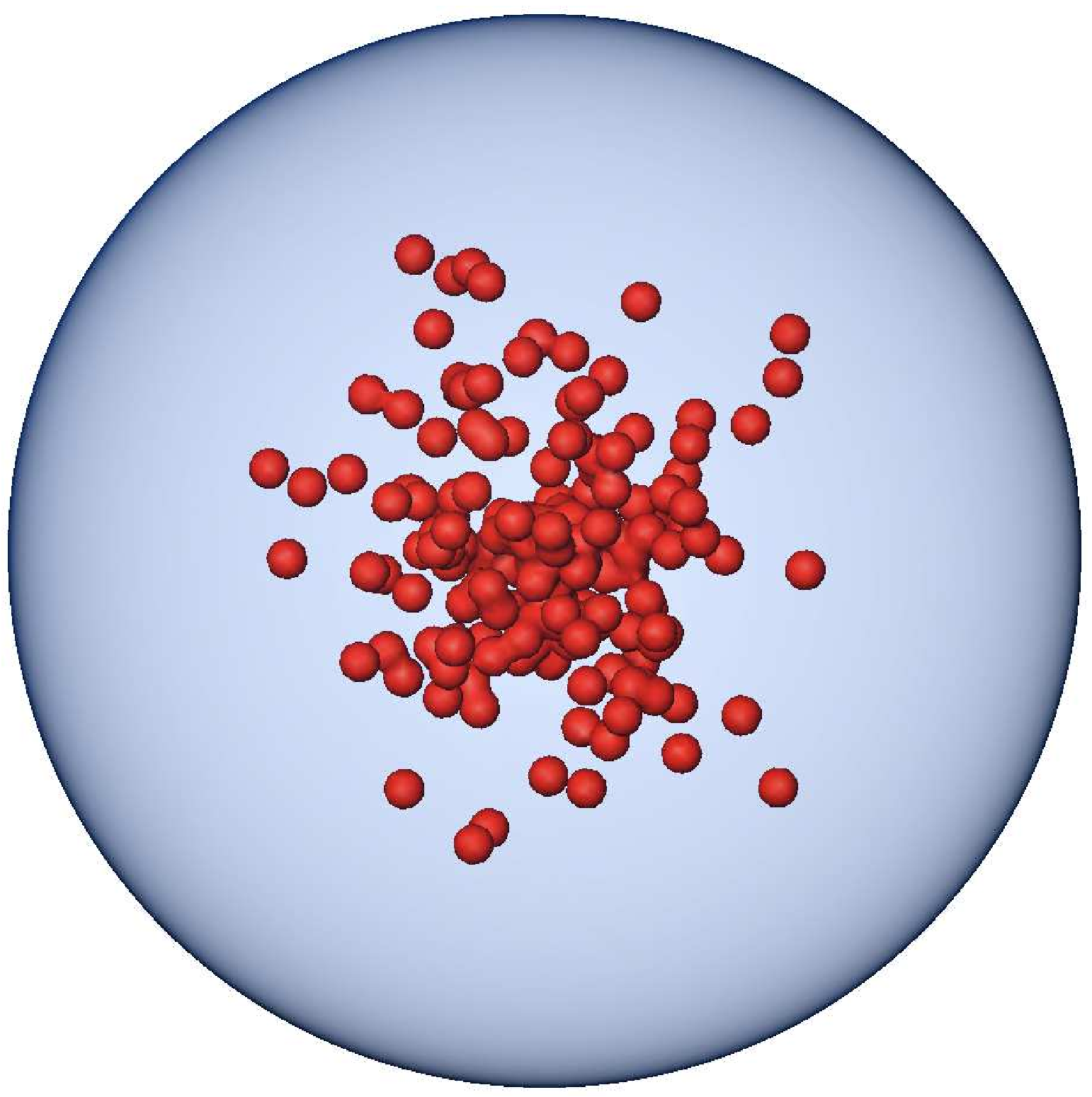}}
    \subfigure[N300C]
    {\includegraphics[width=0.45\columnwidth]{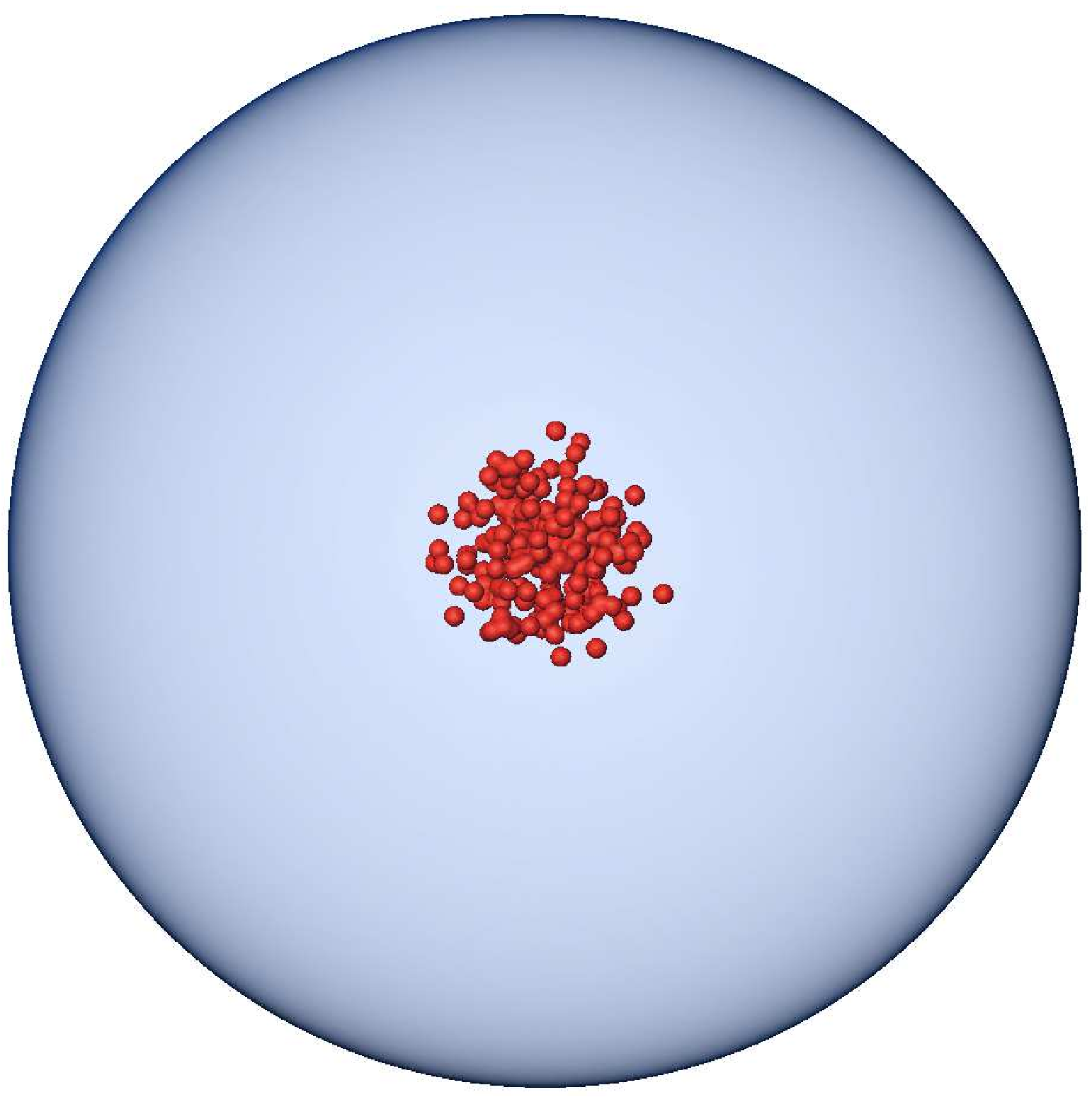}}
    \subfigure[N1600]
    {\includegraphics[width=0.45\columnwidth]{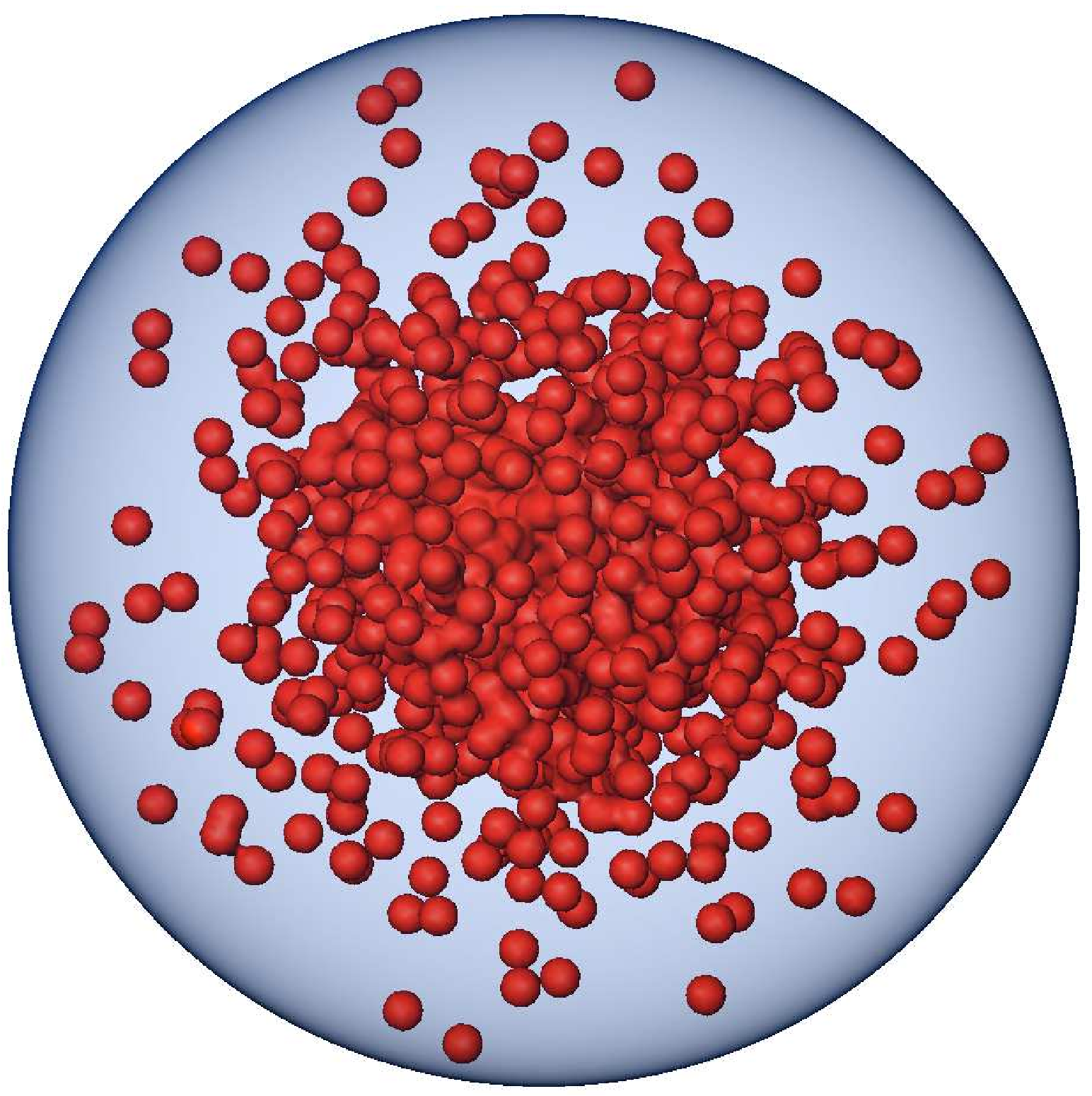}}
    \subfigure[N1600C]
    {\includegraphics[width=0.45\columnwidth]{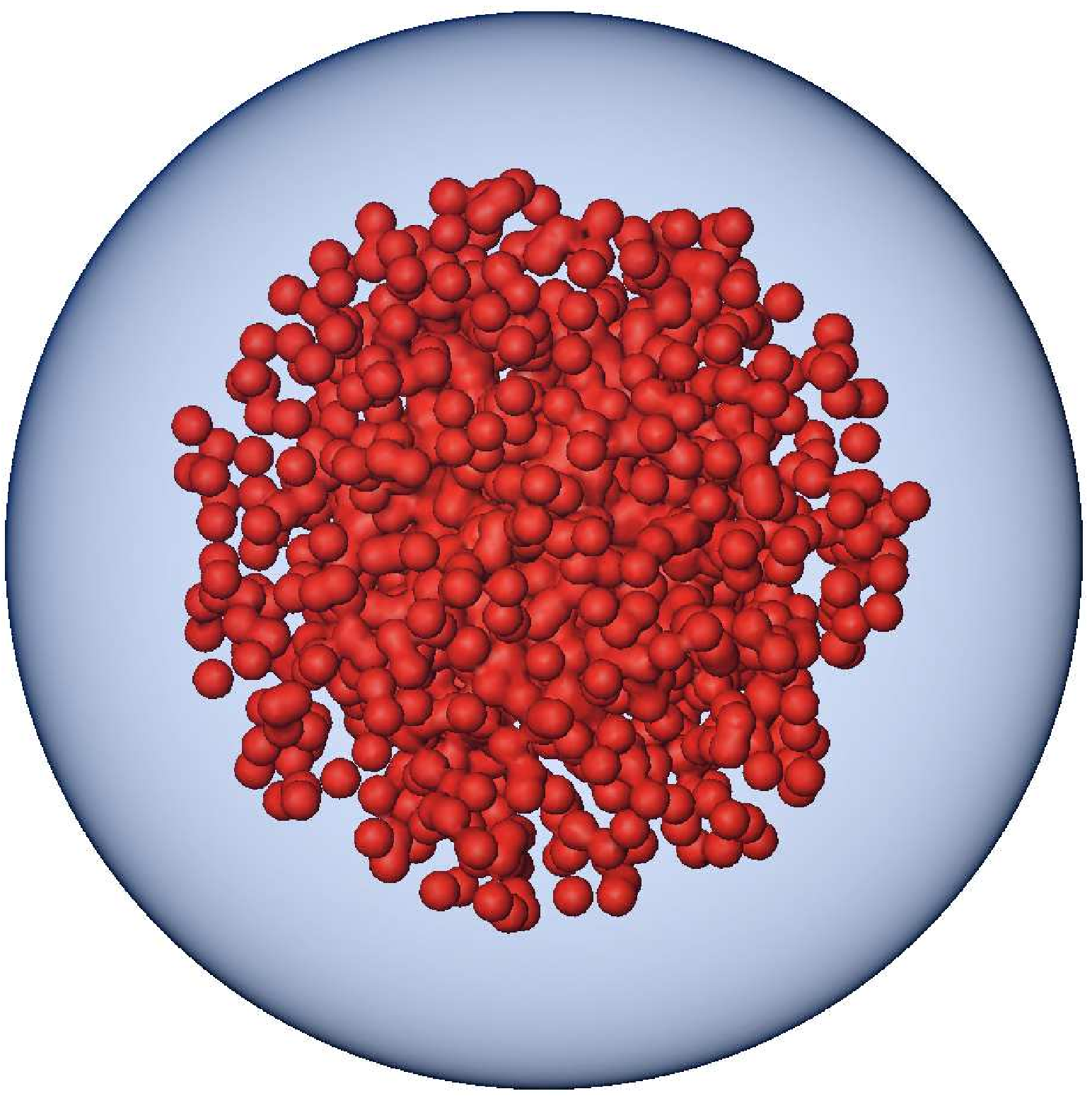}}
    \caption{Ignition geometries of the deflagration for all
      models. We show the arranged ignition kernels (red spheres) and
      a transparent blue contour where the distance to the center is
      $2.5\times 10^7$~cm. For the models with $\rho_\mathrm{c} =
      2.9\times 10^9\gcc$ this radius corresponds to a density of
      $\rho_\mathrm{fuel} = 2.2\times 10^9\gcc$.  Models N300C and
      N1600C have a very compact and dense arrangement of the ignition
      kernels resulting in a setup of high spherical symmetry.}
    \label{fig:IGN}
  \end{center}
\end{figure*}

\begin{figure*}
  \begin{center}
    \subfigure[N3; $t=0.80\s$]
    {\includegraphics[width=0.8\columnwidth]{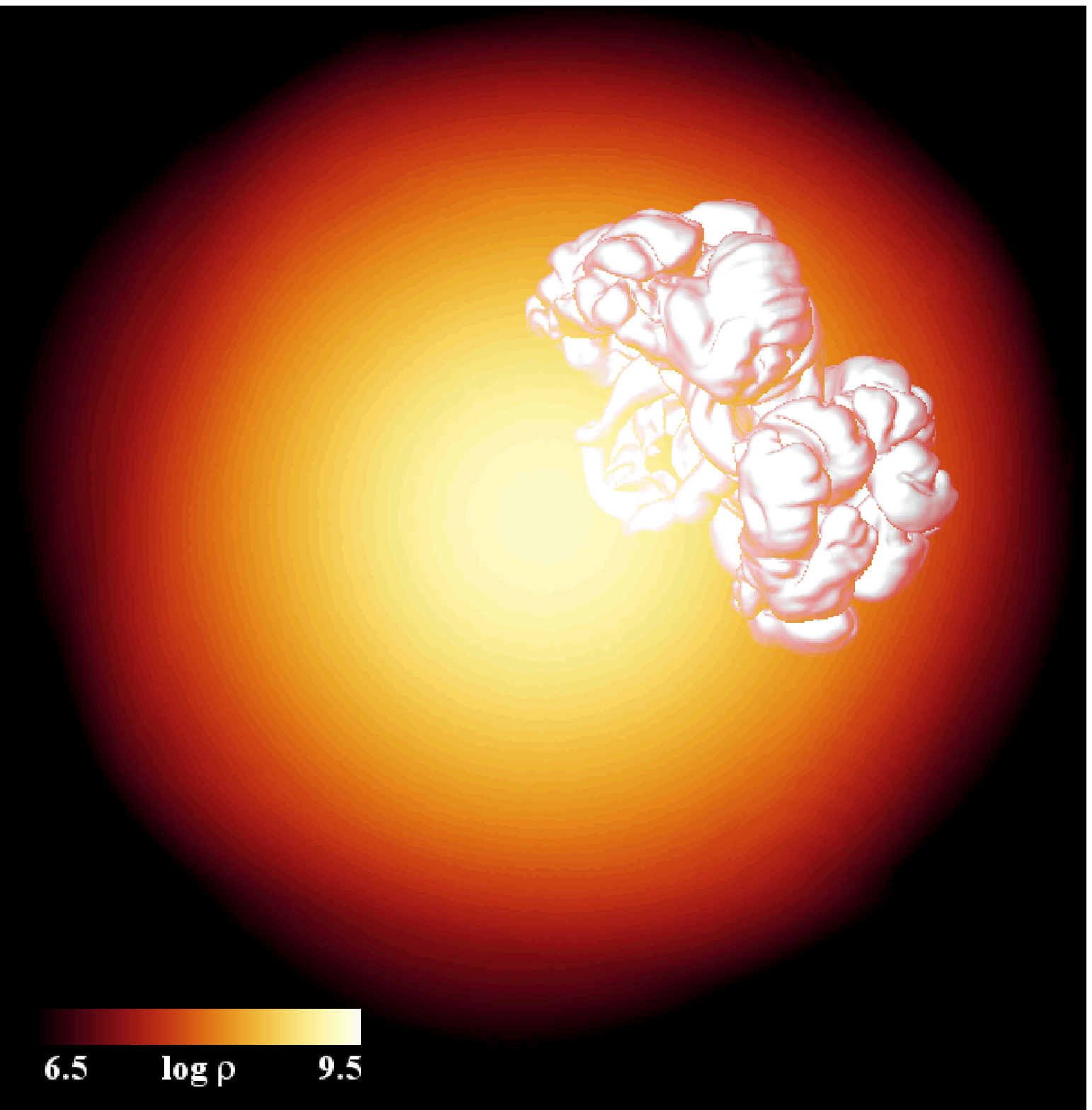}}
    \subfigure[N100; $t=0.70\s$]
    {\includegraphics[width=0.8\columnwidth]{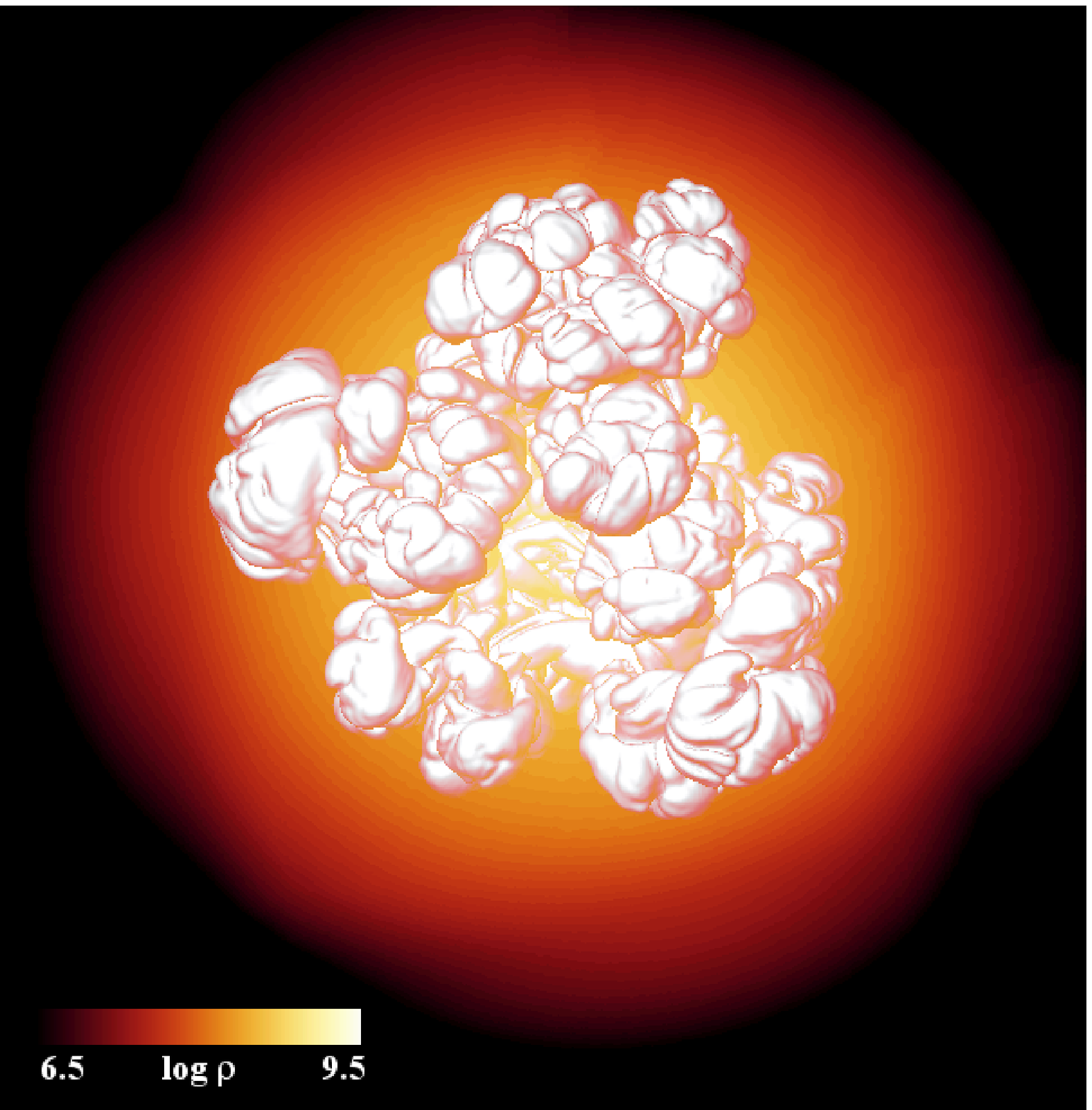}}
    \subfigure[N3; $t=1.05\s$]
    {\includegraphics[width=0.8\columnwidth]{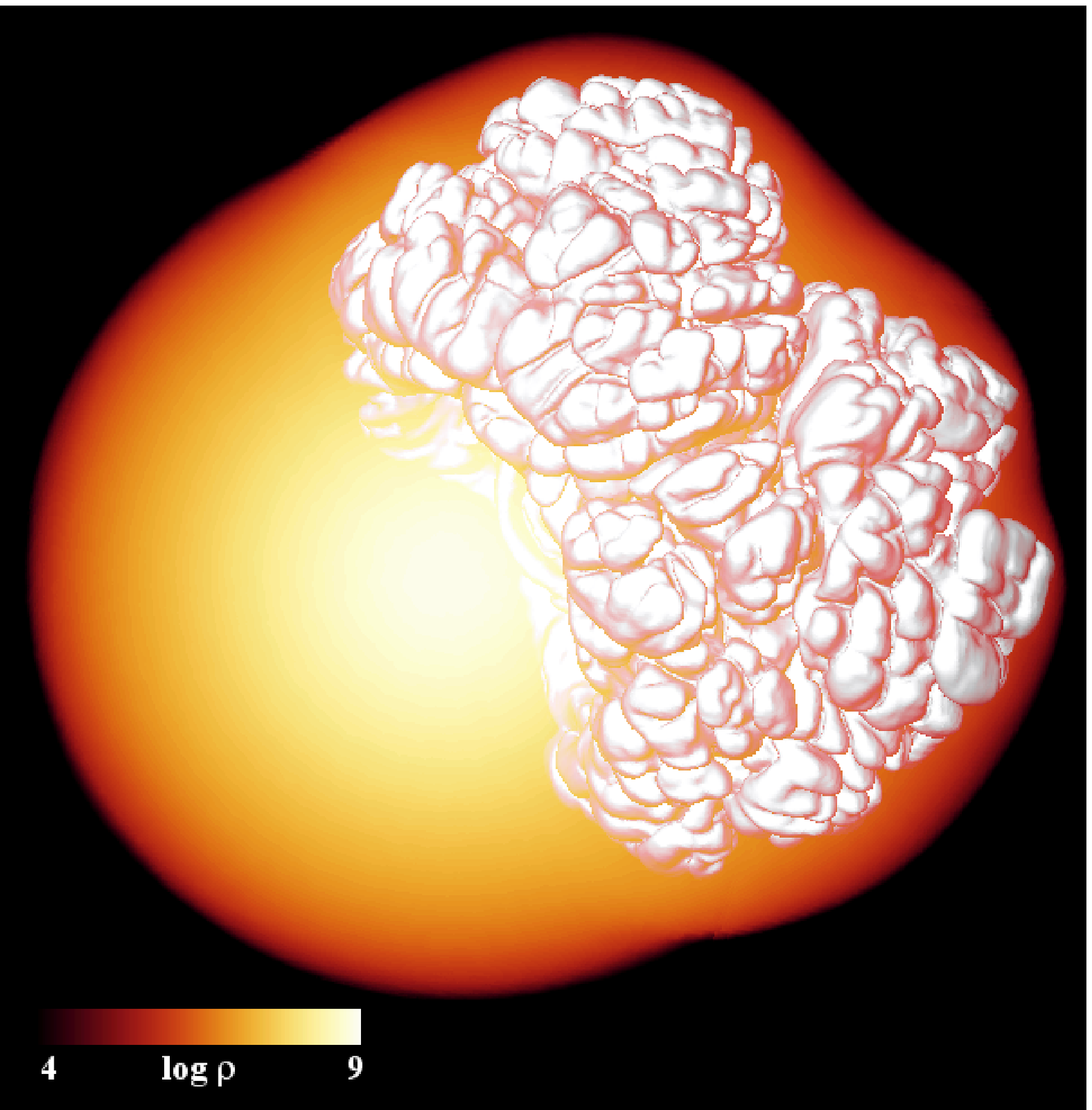}}
    \subfigure[N100; $t=0.93\s$]
    {\includegraphics[width=0.8\columnwidth]{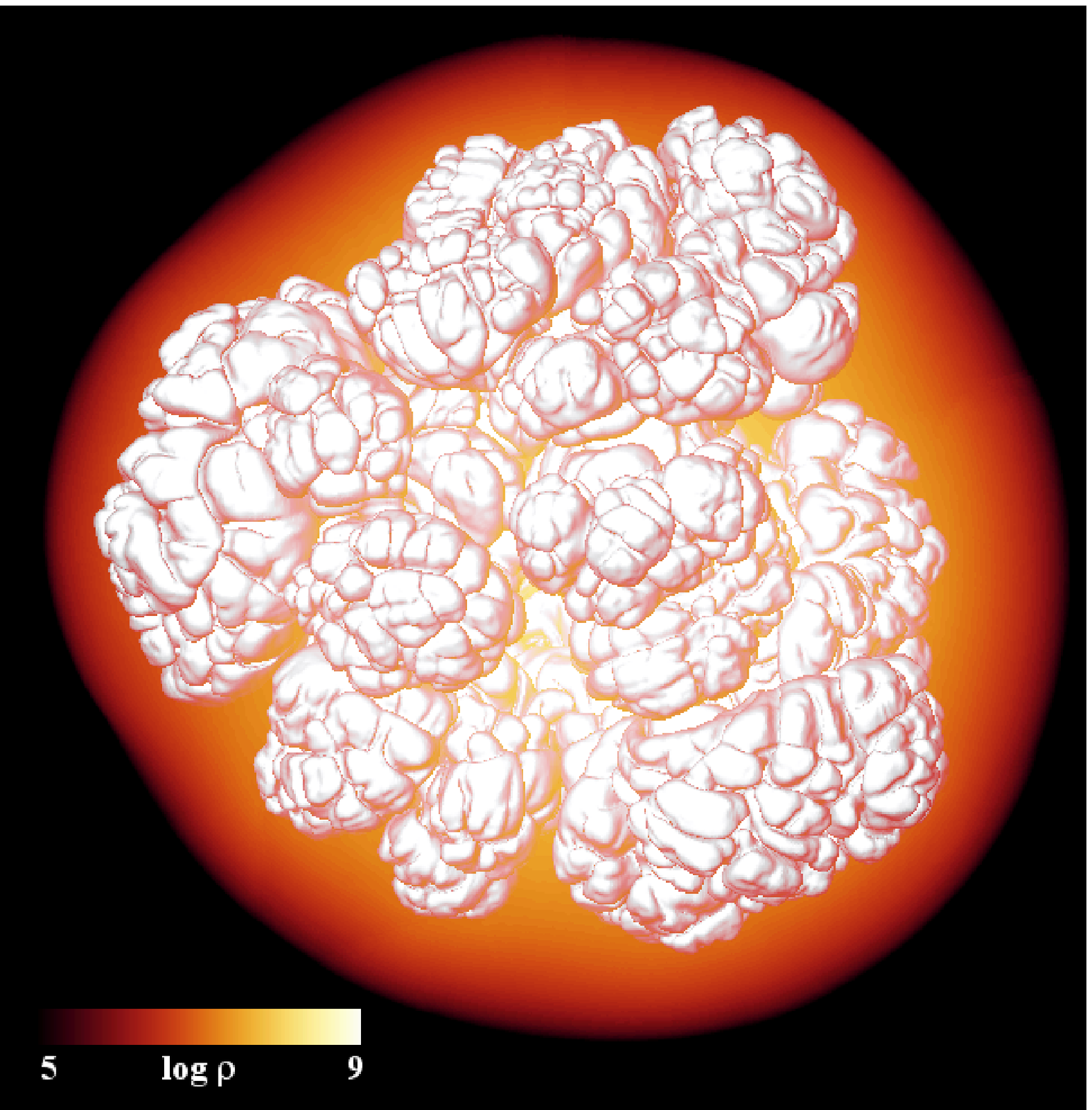}}
    \subfigure[N3; $t=1.15\s$]
    {\includegraphics[width=0.8\columnwidth]{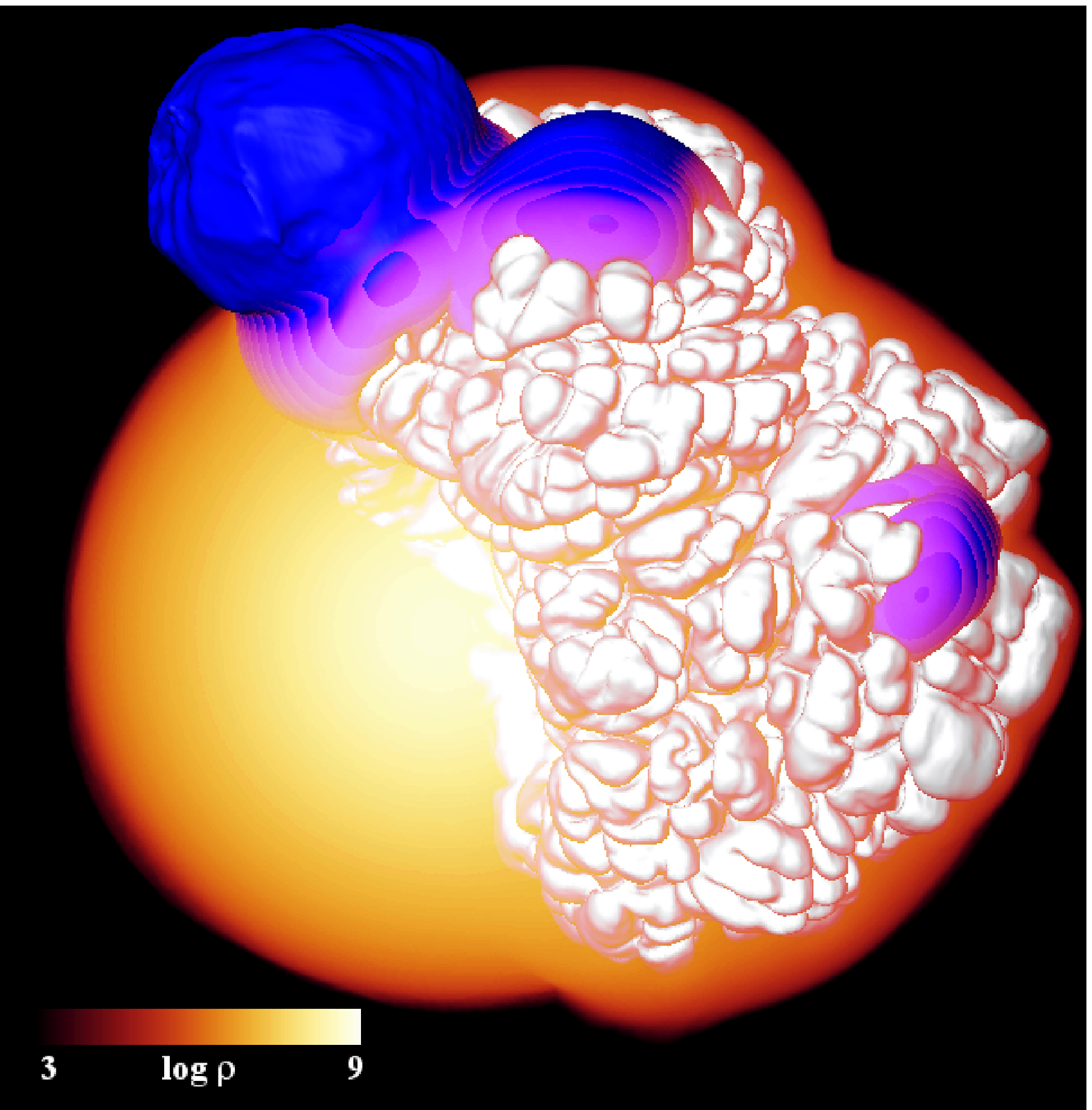}}
    \subfigure[N100; $t=1.00\s$]
    {\includegraphics[width=0.8\columnwidth]{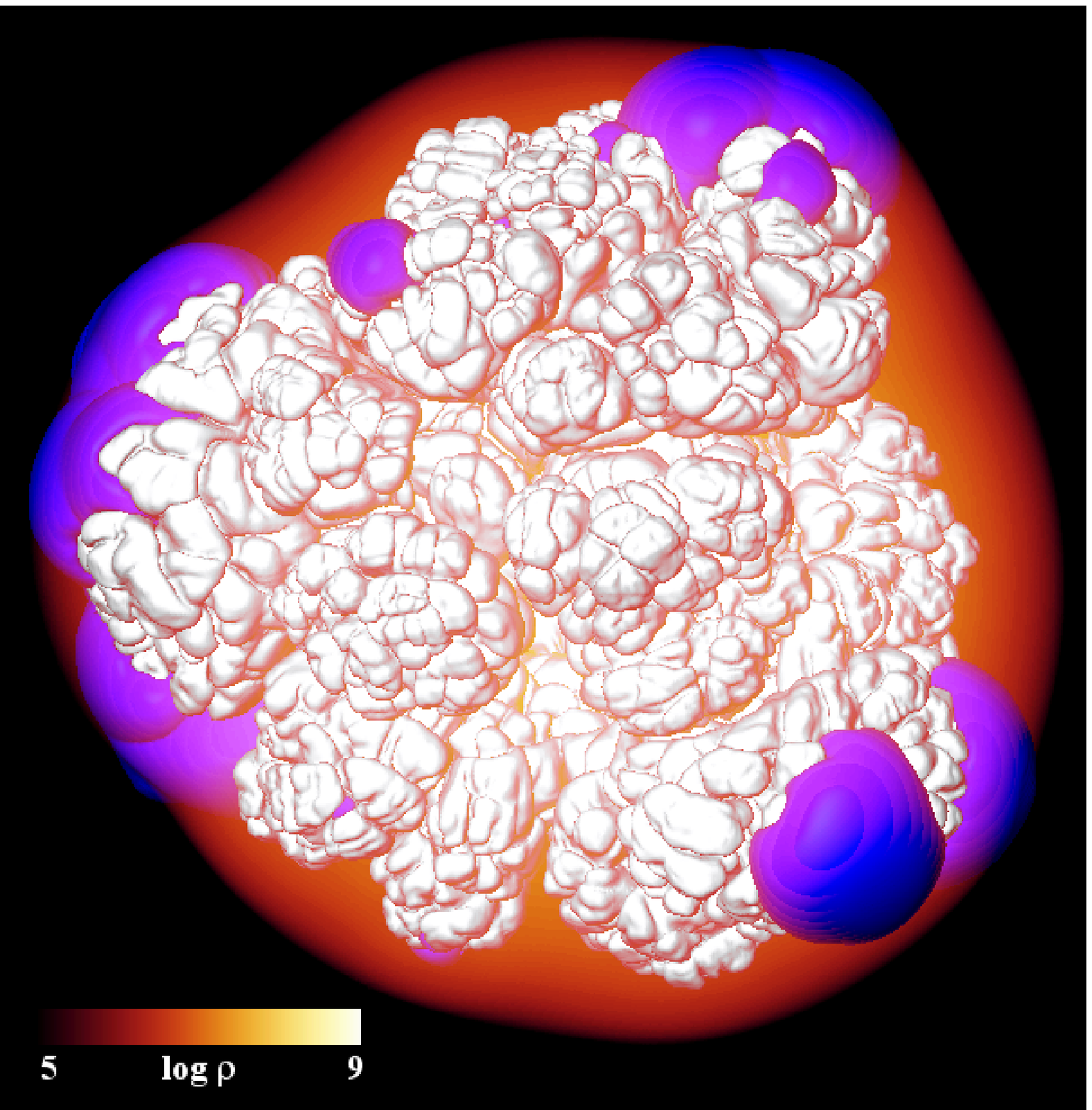}}
    \caption{Snapshots of the hydrodynamic evolution of the N3 model
      (left column) and the N100 model (right column).  The top row
      shows the rising plumes of the deflagration level set (white)
      during the Rayleigh-Taylor unstable stage of the deflagration
      phase embedded in a volume rendering of the density (in $\gcc$).
      The middle row shows the density and deflagration level set
      structure at the time the first DDT occurs.  The bottom row
      shows the subsequent spreading of the detonation level set
      (blue) from the DDT initiation sites.}
    \label{fig:lsets}
  \end{center}
\end{figure*}

For initial stellar models we have chosen isothermal non-rotating WDs
in hydrostatic equilibrium.  The models have a central density of
$\rho_{\mathrm{c}}=2.9\times10^9 \gcc$ ($M = 1.400\ \msun$, $R = 1.96
\times 10^8 \cm$), but we also include a low central density
$\rho_\mathrm{c}=1.0 \times 10^9 \gcc$ ($M = 1.361\ \msun$, $R = 1.96
\times 10^8 \cm$) and a high central density $\rho_\mathrm{c}=5.5
\times 10^9 \gcc$ ($M = 1.416\ \msun$, $R = 1.96 \times 10^8 \cm$)
version of the N100 model \citep{roepke2012a}.  These low and high
central density models are labeled N100L and N100H, respectively.  

{For all models we set up a cold ($T=5\times10^5\,\mathrm{K}$),
hydrostatic WD for a homogeneous composition of 
\nuc{12}{C} and \nuc{16}{O} in equal parts by mass.
To account for an assumed solar  metallicity of the zero-age main-sequence progenitor, we set the
electron fraction to $\ye = 0.49886$, which corresponds to a \nuc{22}{Ne} content of
 2.5 per cent by mass (see Sec~\ref{sec:nucleo}).}

The ignition of the deflagration is critical for the outcome of the
explosion. This stage, however, is difficult to model. A century of
convective carbon burning precedes the thermonuclear runaway. The flow
field in this ``simmering phase'' is expected to be highly turbulent
and thus hard to address numerically \citep[but see
e.g.,][]{hoeflich2002a,kuhlen2006a,zingale2009a, nonaka2012a} and not
finally settled. Not explicitly modeling the ignition phase, we
therefore take a pragmatic approach and treat the ignition geometry as
a free parameter. This has two aspects: (i) the ignition shape has to
seed unstable modes in a realistic way and (ii) it should capture the
correct position of the ignition region relative to the WD's center.
Regarding (i), we chose to ignite the deflagration in a number of
spherical sparks near the center of the WD. {Although
  recent simulations suggest that a high number of nearly simultaneous
  ignitions appears unlikely \citep{nonaka2012a}, it has been argued
  before that such a multi-spot ignition is a probable outcome of the
  convective carbon simmering phase leading up to the thermonuclear
  runaway
  \citep[e.g.][]{garcia1995a,woosley2004a,iapichino2006a}}. Such a
configuration has numerical advantages because it gives rise to a
broad spectrum of perturbations that seed the Rayleigh-Taylor
instability, while smoother ignition geometries leave this to
numerical discretization errors which depend on resolution
\citep{roepke2007a}. Regarding (ii), in cases with many ignition
sparks we set off burning on different sides of the center.  This
corresponds to a nearly central ignition of the deflagration.  In
contrast, the recent simulations of \citet{zingale2009a} and
\citet{nonaka2012a} suggest off-center ignitions due to a dominantly
bipolar flow through the center.  {However, even slight
  rotation would disrupt the bipolar flow and consequently asymmetric
  ignition becomes less likely \citep{kuhlen2006a}.  Even in the
  non-rotating case, for realistic Reynolds numbers (${\sim}10^{11}$
  times larger than what is achieved in numerical simulations), the
  dipole pattern would likely give way to a highly turbulent and
  chaotic flow, which again would suggest a more symmetric ignition.}

For this work we investigate a set of fourteen different explosion
models, corresponding to different ignition setups that are summarized
in Table~\ref{tab:setup}. These setups are generated from a
Monte-Carlo based algorithm. The main parameter of the ignition
scenario is the number of spherical ignition kernels $N_\mathrm{k}$
that are aligned in the central area of the WD following a Gaussian
distribution in radius.  To avoid extreme outliers, we require that
the ignition kernels have to be contained within a sphere of radius
$R=2.5 \sigma$, where $\sigma$ is the variance of the Gaussian
distribution.  We mention that for ignition models with low
$N_\mathrm{k}$ the placement of the kernels cannot be considered as a
real Gaussian distribution anymore. In particular model N1 with
$N_\mathrm{k} = 1$ constitutes a simple single off-center
configuration where $R = \sigma$ holds.  The ignition models N300C and
N1600C (representing ``compact'' ignition configurations) are based on
a configuration that has been used in \citet{roepke2007c}. By the
generation of these models $R=2.5 \sigma$ holds, but after all
ignition kernels are placed, the ones with a larger distance than
$\sigma$ are removed and replaced until for the whole configuration $R
= \sigma$ holds (as in the case for $N_\mathrm{k} = 1$). As a result
we obtain a very dense and compact arrangement of the ignition
kernels, where the whole configuration exhibits a highly spherical
symmetry. We use a uniform radius for all ignition kernels set to
$r_\mathrm{k} = 10^6$ cm for all models except for the most centrally
concentrated and compact ignition model N300C, where $r_\mathrm{k} =
5\times 10^5$ cm was used. The length $d_\mathrm{k}$ defines a minimum
separation distance between the centers of the ignition kernels. Note
that for large $N_\mathrm{k}$, $r_\mathrm{k} > d_\mathrm{k}$, hence
the ignition kernels may partially overlap, in particular in the
vicinity of the center of the white dwarf.  For a visualization of
ignition setups see Fig.~\ref{fig:IGN}.  We have chosen the ignition
setups to obtain a suite of models covering a large range of strengths
of the deflagration phase and associated pre-expansion of the WD prior
to the onset of the detonation. This way, we obtain events that cover
the same range of \nuc{56}{Ni} masses that is derived for normal
SNe~Ia in \citet{stritzinger2006b}.

\begin{table}
\begin{center}
\caption{Parameters of the ignition setup of the deflagration. \label{tab:setup}}
\begin{tabular}{lrcccc}
Model & 
$N_\mathrm{k}$ & 
$\sigma$ & 
$r_\mathrm{k}$ & 
$d_\mathrm{k}$ &
$\rho_\mathrm{c}$ \\
& 
& 
$[10^7 \cm] $ & 
$[10^6 \cm]$ & 
$[10^6 \cm] $ &
$[10^9 \gcc]$ \\ \hline
\textbf{N1}    & 1    & 0.36 & 1.00 &  -   &2.9\\ 
\textbf{N3}    & 3    & 0.50 & 1.00 & 3.00 &2.9\\ 
\textbf{N5}    & 5    & 0.60 & 1.00 & 1.00 &2.9\\ 
\textbf{N10}   & 10   & 0.60 & 1.00 & 1.00 &2.9\\ 
\textbf{N20}   & 20   & 0.60 & 1.00 & 0.60 &2.9\\ 
\textbf{N40}   & 40   & 0.60 & 1.00 & 1.00 &2.9\\ 
\textbf{N100L} & 100  & 0.60 & 1.00 & 0.30 &1.0\\ 
\textbf{N100}  & 100  & 0.60 & 1.00 & 0.30 &2.9\\ 
\textbf{N100H} & 100  & 0.60 & 1.00 & 0.30 &5.5\\ 
\textbf{N150}  & 150  & 0.60 & 1.00 & 0.35 &2.9\\ 
\textbf{N200}  & 200  & 0.75 & 1.00 & 0.30 &2.9\\ 
\textbf{N300C} & 300  & 0.50 & 0.50 & 0.05 &2.9\\ 
\textbf{N1600} & 1600 & 1.00 & 1.00 & 0.05 &2.9\\ 
\textbf{N1600C}& 1600 & 1.80 & 1.00 & 0.05 &2.9\\ \hline
\end{tabular}
\end{center}
\end{table}

\subsection{Computational method}
\label{sec:method}

All hydrodynamic explosion simulations presented here were performed
with our thermonuclear supernova code \LEAFS, a three-dimensional
finite-volume discretization of the reactive Euler equations. The
hydrodynamics solver is based on the PROMETHEUS implementation
\citep{fryxell1989a} of the ``piecewise parabolic method'' (PPM) by
\citet{colella1984a}.

Deflagration and detonation fronts are modeled as separate
discontinuities between carbon-oxygen fuel and nuclear ash and their
propagation is tracked with a level-set scheme \citep{reinecke1999a,
  osher1988a, smiljanovski1997a}. All material crossed by these fronts
is converted to nuclear ash with a composition and energy release
depending on fuel density. Although modeled with the same method, the
propagation velocity, the ash composition and the nuclear energy
release is different for deflagrations and detonations at a given fuel
density. For detonations, the corresponding data is taken from the
tables of \citet{fink2010a}, and another table for deflagrations was
determined in the same way as described there. In our hydrodynamic
simulations, we model the composition of the material with only five
species: carbon, oxygen, a representative proxy for an
intermediate-mass isotope, and a mixture of nickel and alpha particles
representing nuclear statistical equilibrium configurations. The
latter are adjusted according to the prevailing thermodynamic
conditions. We keep track of neutronization (and its effect on the
equation of state) by following the evolution of $Y_\mathrm{e}$ as a
separate and independent passive scalar.

The speed of the detonations is modeled as in \citet{fink2010a}: at
high densities ($\rho \gtrsim 10^7 \gcc$, pathological case), speeds
are taken from \citet{gamezo1999a}; at low densities, Chapman--Jouguet
like speeds were calculated for the incomplete burning yields in our
detonation tables.  After a very short phase of laminar burning
following ignition, the propagation of deflagrations is dominated by
buoyancy and shear-induced instabilities and interactions with a
complex turbulent flow field. The unresolved acceleration of the flame
due to turbulence is accounted for by a sub-grid scale model
\citep{schmidt2006b,schmidt2006c}.

To follow the explosion ejecta to $100\, \mathrm{s}$, where homologous
expansion becomes a good approximation, we employ a moving mesh
technique \citep{roepke2005c,roepke2006a} with an outer coarse
inhomogeneous grid tracking the WD's expansion and an inner finer
homogeneous one encompassing the deflagration region. We include
self-gravity with a monopole gravity solver. The total grid resolution
is $512 \times 512 \times 512$ cells for all simulations presented
here.

\subsection{Criterion for deflagration to detonation transition (DDT)}
\label{sec:ddt}
Although details of the transition process from a subsonic
deflagration to a supersonic detonation are not well understood, it is
generally believed that strong turbulent velocity fluctuations $v'$ at
the deflagration flame are required for a DDT to occur
\citep{niemeyer1997b, lisewski2000b, woosley2007a, woosley2009a}.
Under this condition, mixed regions of hot burned and cold unburned
material emerge that exceed a certain length scale of the order of
$\ell_\mathrm{crit}\approx 10^6$ cm \citep{dursi2006a, niemeyer1997b,
  khokhlov1997a, seitenzahl2009b}. Based on the Zel'dovich gradient
mechanism \citep{zeldovich1970a} a spontaneous ignition in such
regions may lead to a sufficiently strong shock for the formation of a
detonation front \citep{blinnikov1986a, khokhlov1991a, khokhlov1991c,
  khokhlov1997a, seitenzahl2009c}. \citet{lisewski2000b} pointed out
that $v'$ has to exceed $10^8 \cms$, but more recent studies of
\citet{woosley2009a} suggested that smaller values of $v'\approx 5
\times 10^7 \cms$ may be sufficient to trigger the DDT. In
three-dimensional simulations of pure deflagrations
\citet{roepke2007d} found that the probability $P (v' \geq 10^8 \cms)$
of finding $v'$ of at least $10^8 \cms$ may be high enough for a DDT
to occur.

The length scale $\ell_\mathrm{crit}$ is not resolved in our
large-scale simulations.  Therefore a subgrid scale model is employed
that evaluates the probability $P (v' \geq 10^8 \cms)(t)$ at the time
$t$ at specific places at the deflagration flame that obey the
following DDT constraints.  To restrict our analysis to the immediate
area of the flame, we define $X_\mathrm{fuel} = 0.4$--$0.6$ as the
allowed fuel fraction (the carbon/oxygen composition) in the grid
cells, which ensures that the level set propagates on average through
the center of these cells. We further define $\rho_\mathrm{fuel} =
(0.6$--$0.7)\times 10^7 \gcc$ as the allowed fuel densities in these
cells, where this interval constitutes a subrange of the suggested
fuel densities of \citet{woosley2007a} where DDTs are expected to
occur. The number $N_\mathrm{flame}^{*}(t)$ and length $\Delta(t)$ of
the grid cells that meet the above-mentioned constraints define a
flame surface area $A_{\mathrm{flame}}^{*}(t) =
N_\mathrm{flame}^{*}(t) \times (\Delta (t))^D$, where $D$ is the
fractal dimension.  Following \citet{kerstein1988a},
\citet{sreenivasan1991a} and \citet{woosley2007a} we use $D =
2.36$. We define now $A_\mathrm{det}(t) = P (v' \geq 10^8 \cms)(t)
\times A_{\mathrm{flame}}^{*}(t)$ as the part of the flame that is
capable of performing a DDT.  If $A_\mathrm{det}(t)$ exceeds a certain
threshold $A_\mathrm{crit}$ (we use $A_{\mathrm{crit}} =
\ell_\mathrm{crit}^D$, again with $D = 2.36$), it is checked whether
this condition holds at least for half an eddy turnover time
$\tau_{\mathrm{eddy}_{1/2}} = 0.5 \times \ell_\mathrm{crit}/v' = 5
\times 10^{-3}$ s. This is to ensure that fuel and ash become
sufficiently mixed on the scale $\ell_\mathrm{crit}$.  If finally
$A_{\mathrm{det}}(t) \geq A_{\mathrm{crit}}$ holds for
$\tau_{\mathrm{eddy}_{1/2}}$, detonations are initialized in the grid
cells that contain the highest velocity fluctuations, where the number
of ignitions is given by the current ratio $A_\mathrm{det}(t)$ to
$A_\mathrm{crit}$. This DDT criterion is consistent with findings of
microscopical studies \citep{woosley2007a, woosley2009a} and is
largely resolution-independent (for further details see
Ciaraldi-Schoolmann \& R\"opke, in preparation).

\subsection{Hydrodynamic evolution and explosion morphologies}
\label{sec:morph}
First we point out that all fourteen simulations result in a
completely gravitationally unbound remnant and are in this regard to
be considered as ``successful'' explosions. The basic underlying
evolution of all models is rather similar to other extant
three-dimensional delayed detonation simulations.

The burning front propagates initially as a laminar conductive flame
\citep{timmes1992a}. The nuclear burning releases energy by converting
\nuc{12}{C} and \nuc{16}{O} to a mix of more tightly bound
nuclei. This energy release is sufficient to at least partially lift
the degeneracy of the burned material, which means that the ash is not
only hotter but also less dense than the surrounding background
material.  Buoyancy forces then lead to a rise of the deflagration ash
material against gravity.  Once the deflagration bubble exceeds a
critical size it becomes subject to the Rayleigh-Taylor (RT)
instability \citep[e.g.][]{khokhlov1995a,townsley2007a}, which
wrinkles the flame front and causes a rapid acceleration of the growth
of its surface area. As these RT-unstable plumes of hot deflagration
ash continue to rise towards the stellar surface, the low Reynolds
numbers coupled with the rather high relative speeds lead to
Kelvin-Helmholtz induced shear on the sides of the rising plumes. This
shear-induced turbulence in turn results in the high velocity
fluctuations that lead to the fulfillment of our DDT criterion (see
Sec.~\ref{sec:ddt}). The nascent detonation fronts propagate
supersonically from their initiation sites and quickly burn any
accessible (i.e. topologically connected, cf. \citealt{maier2006a})
fuel that remains above the respective density threshold to IMEs and
IGEs.  As an aside, we mention that the detonation level set may be
advected into low density ($\rho<5 \times 10^5 \gcc$)
regions. However, this does not result in any burning there since we
use the fuel density dependent energy release tables of
\citet{fink2010a}. Once the detonation has burned all remaining and
accessible fuel, the supernova enters the stage of ballistic
expansion. In Fig.~\ref{fig:lsets} we show the morphology for two
representative models (N3 and N100) at three different stages of the
evolution.

\begin{figure*}
  \begin{center}
    \subfigure[N3; X(\nuc{56}{Ni})]
    {\includegraphics[height=3.2cm]{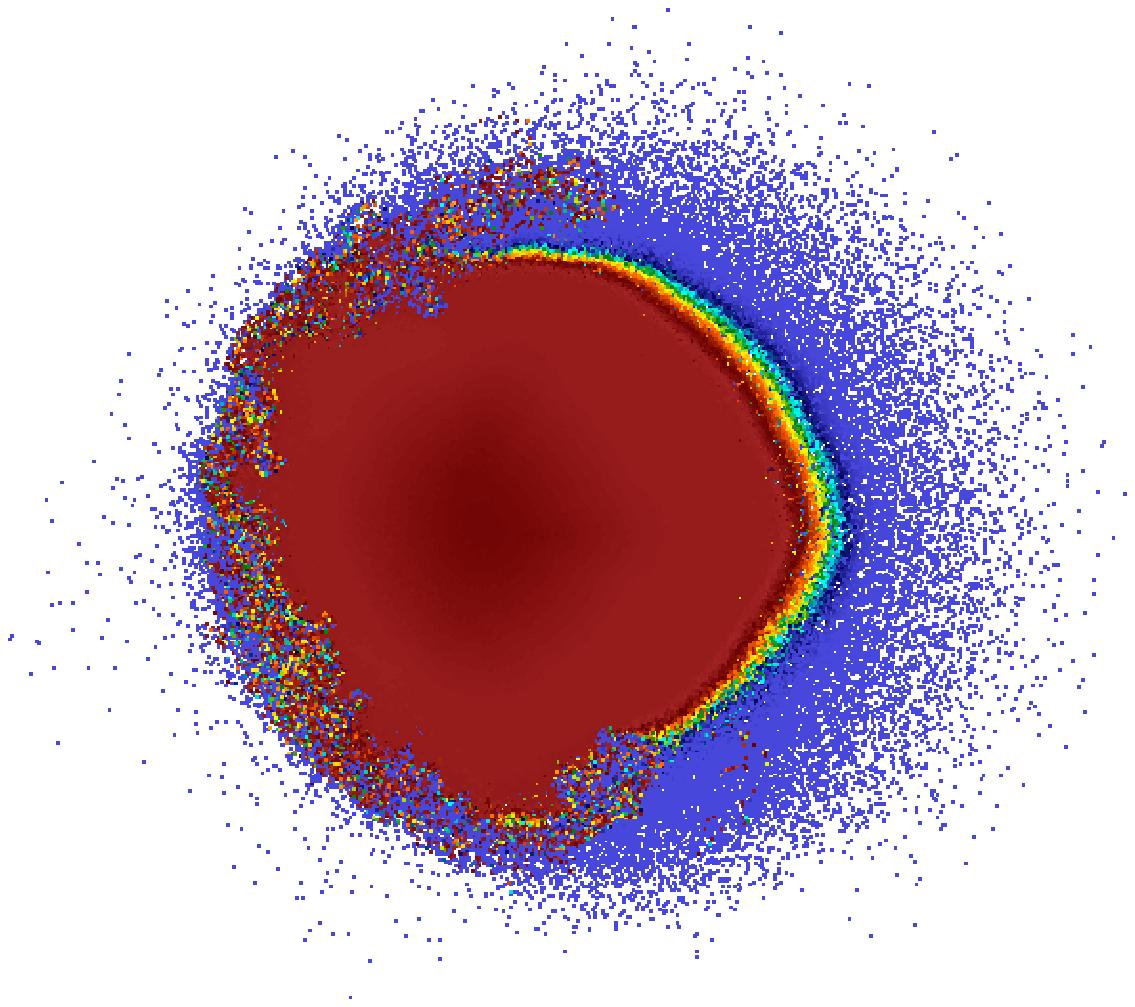}}
    \subfigure[X(\nuc{54}{Fe}+\nuc{56}{Fe}+\nuc{58}{Ni})]
    {\includegraphics[height=3.2cm]{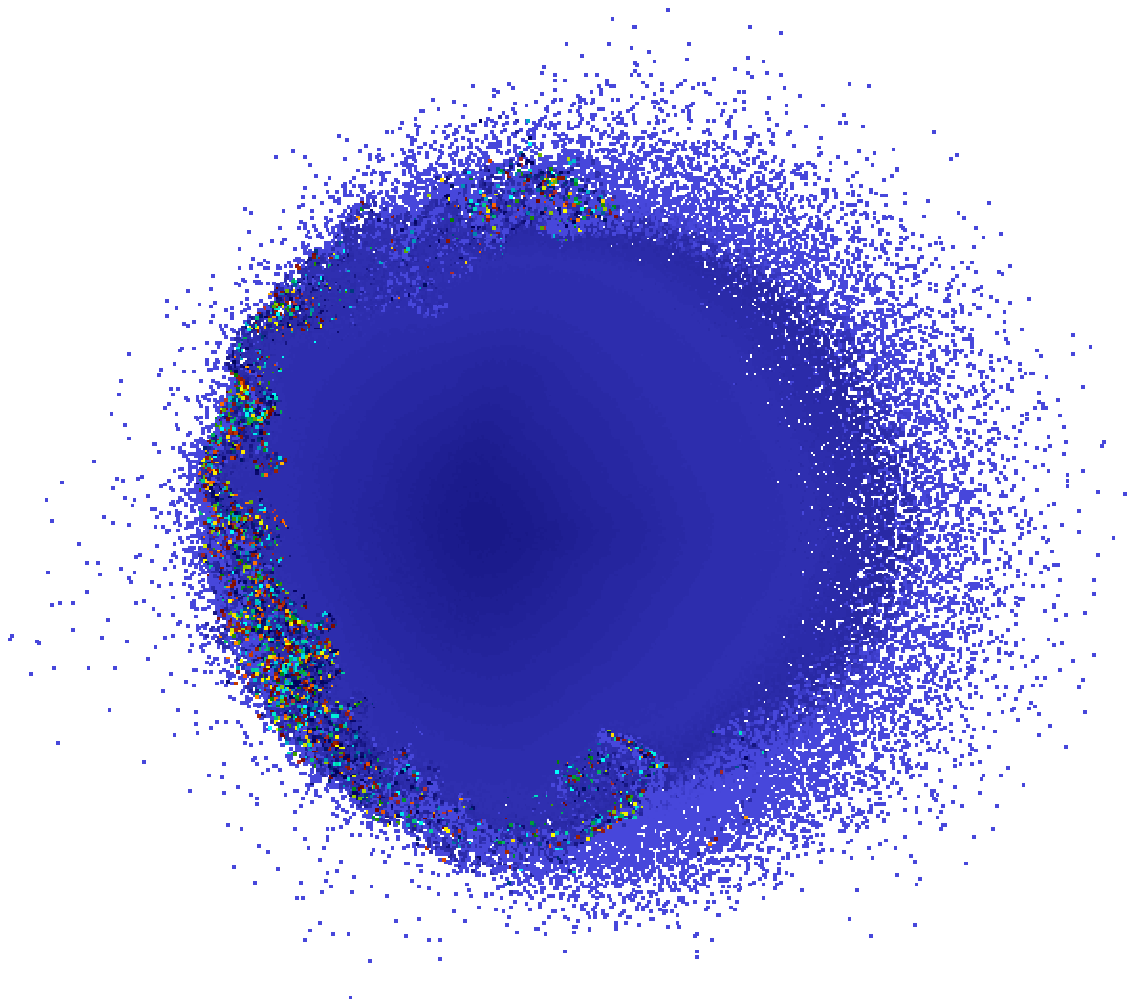}}
    \subfigure[X(\nuc{28}{Si})]
    {\includegraphics[height=3.2cm]{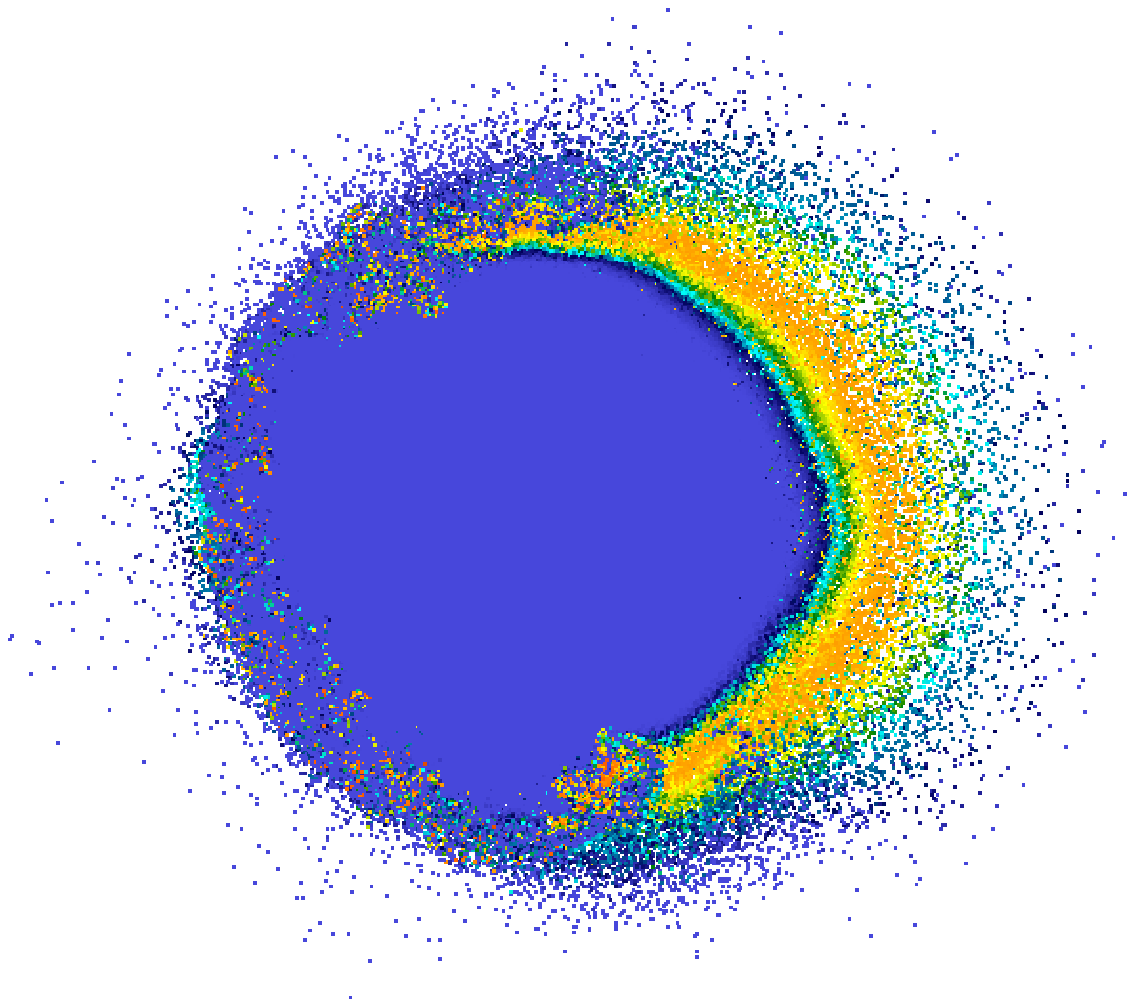}}
    \subfigure[X(\nuc{16}{O})]
    {\includegraphics[height=3.2cm]{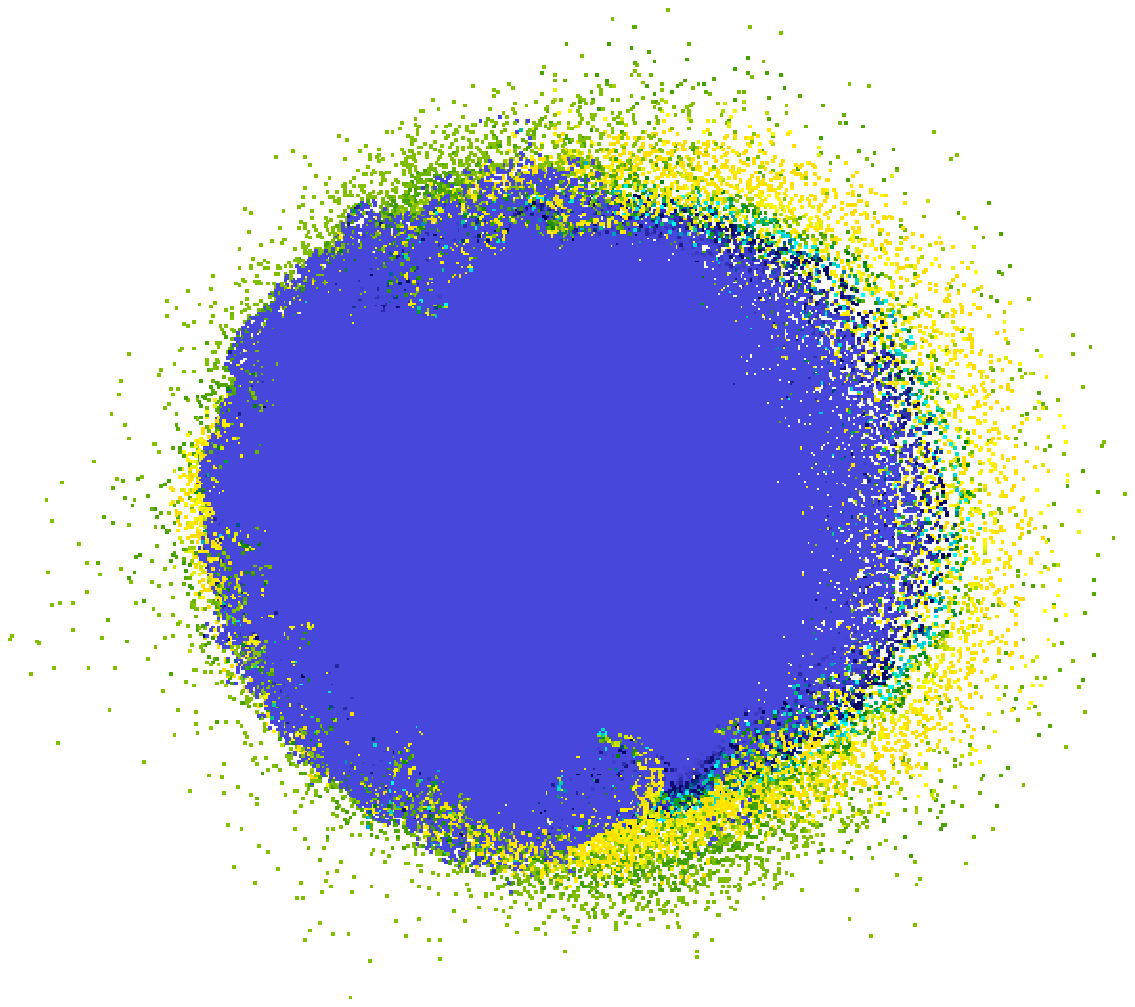}}
    \subfigure[X(\nuc{12}{C})]
    {\includegraphics[height=3.2cm]{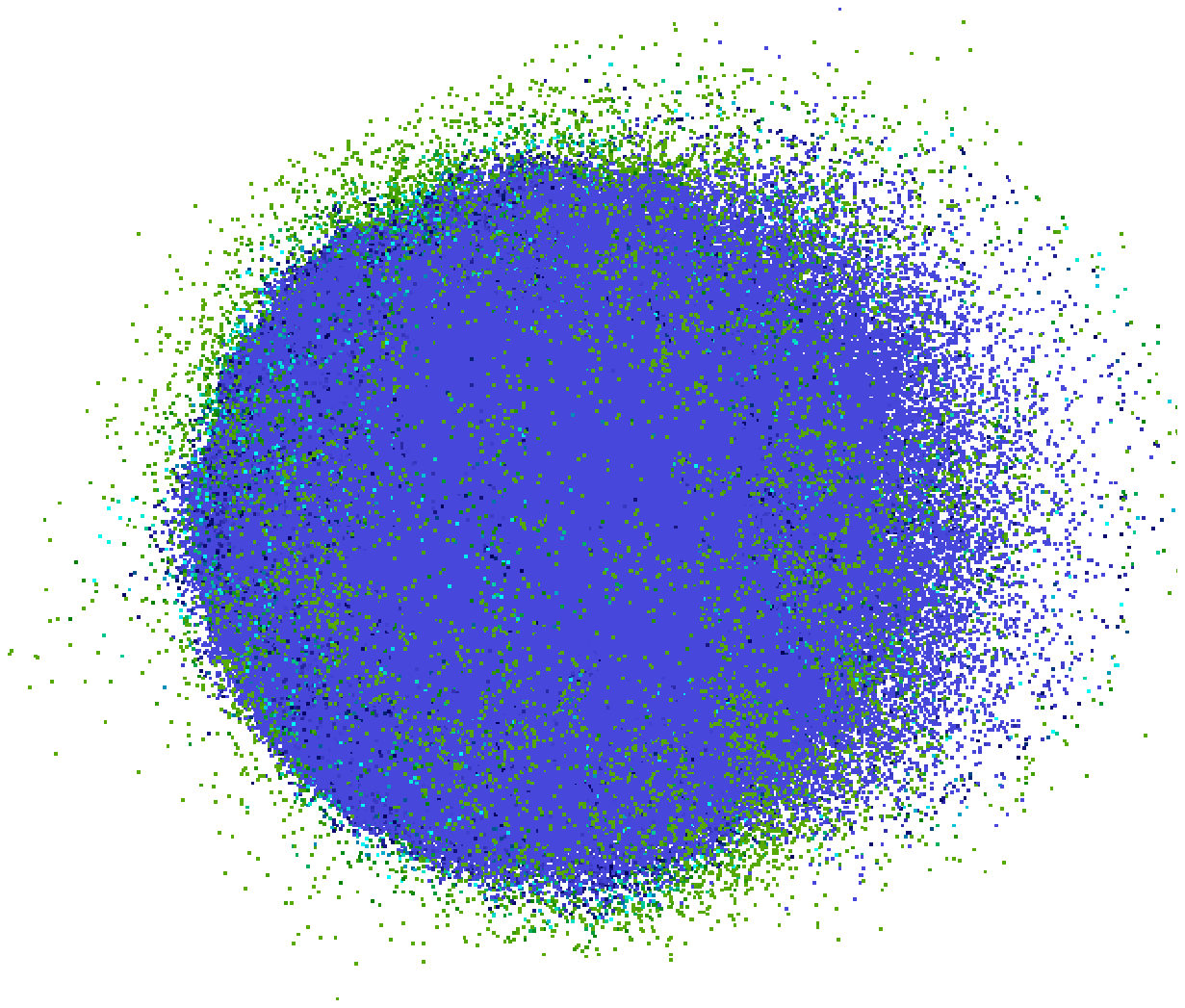}} \subfigure[N40;
    X(\nuc{56}{Ni})] {\includegraphics[height=3.2cm]{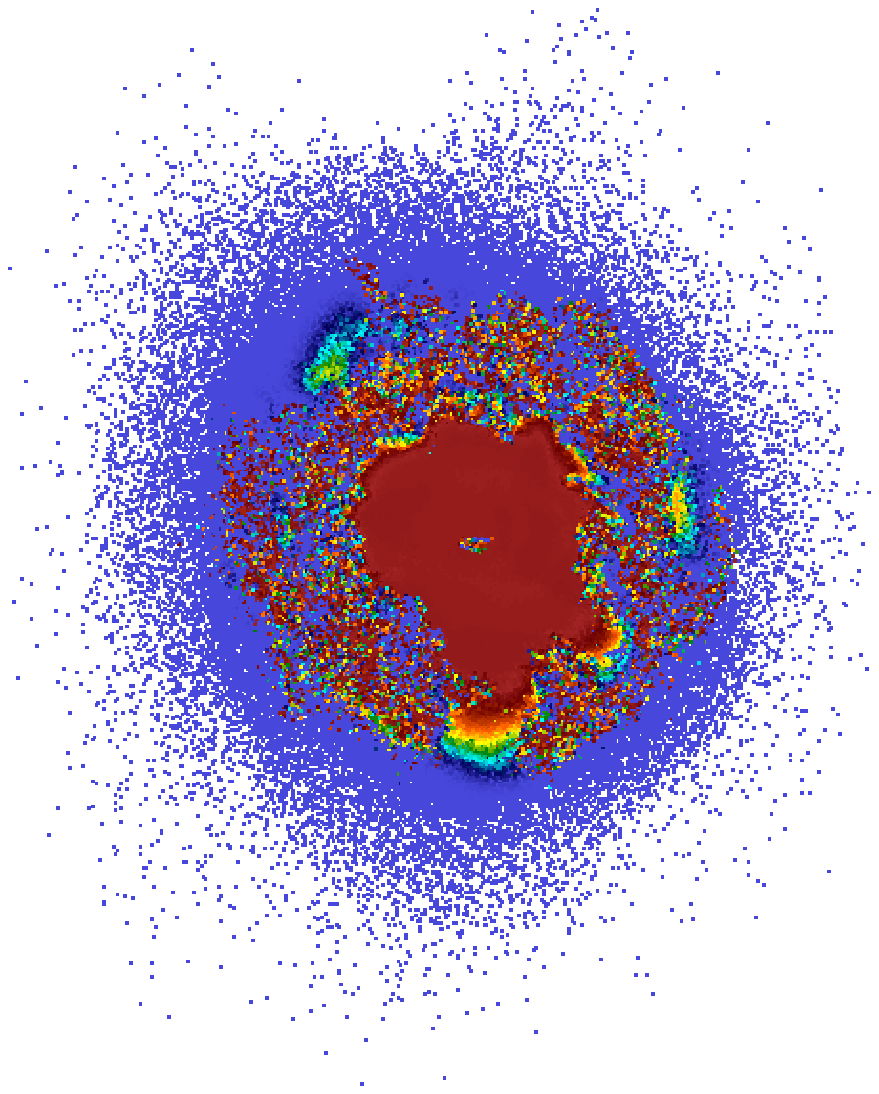}}
    \subfigure[X(\nuc{54}{Fe}+\nuc{56}{Fe}+\nuc{58}{Ni})]
    {\includegraphics[height=3.2cm]{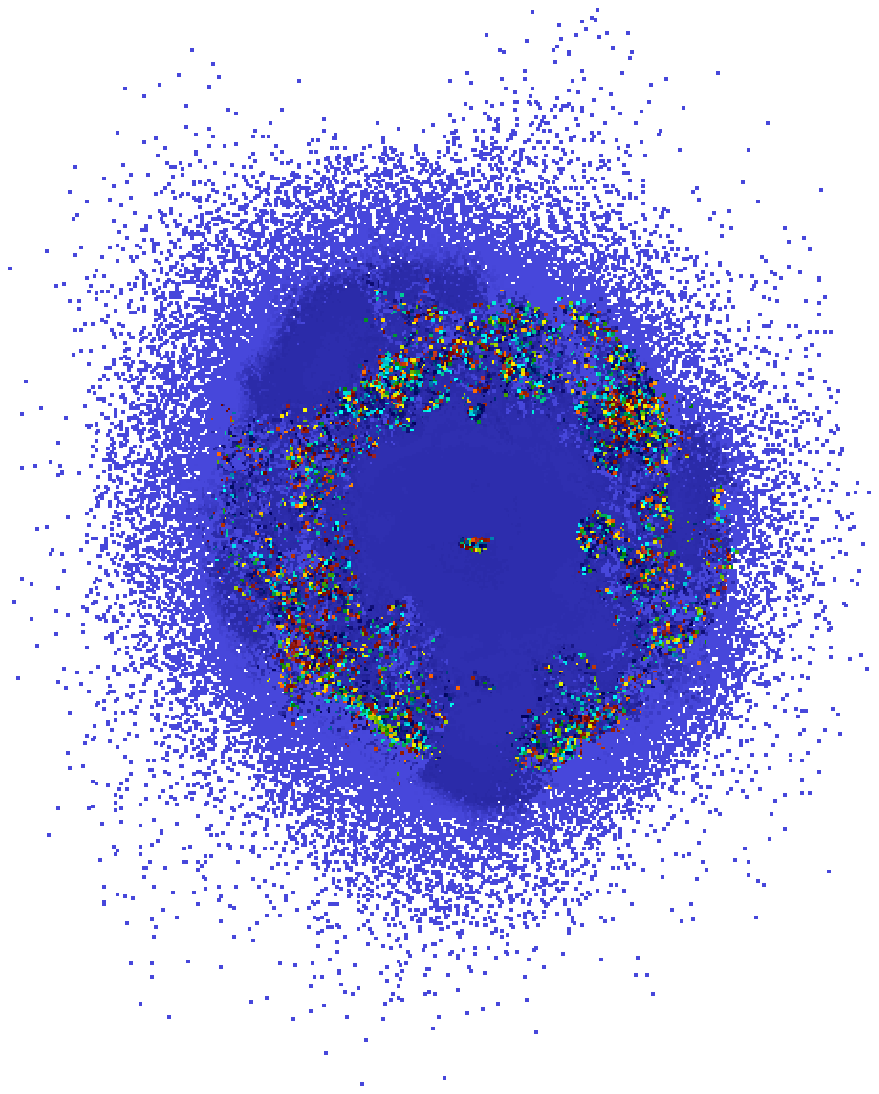}}
    \subfigure[X(\nuc{28}{Si})]
    {\includegraphics[height=3.2cm]{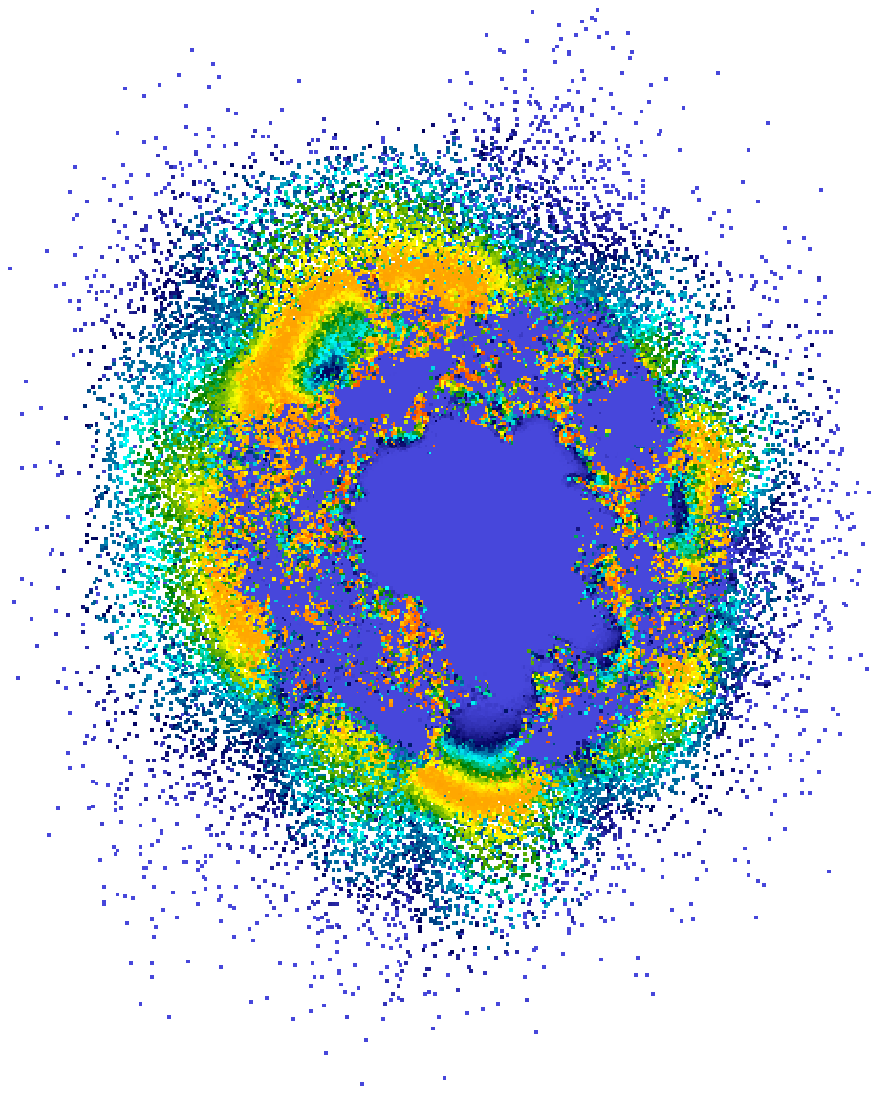}}
    \subfigure[X(\nuc{16}{O})]
    {\includegraphics[height=3.2cm]{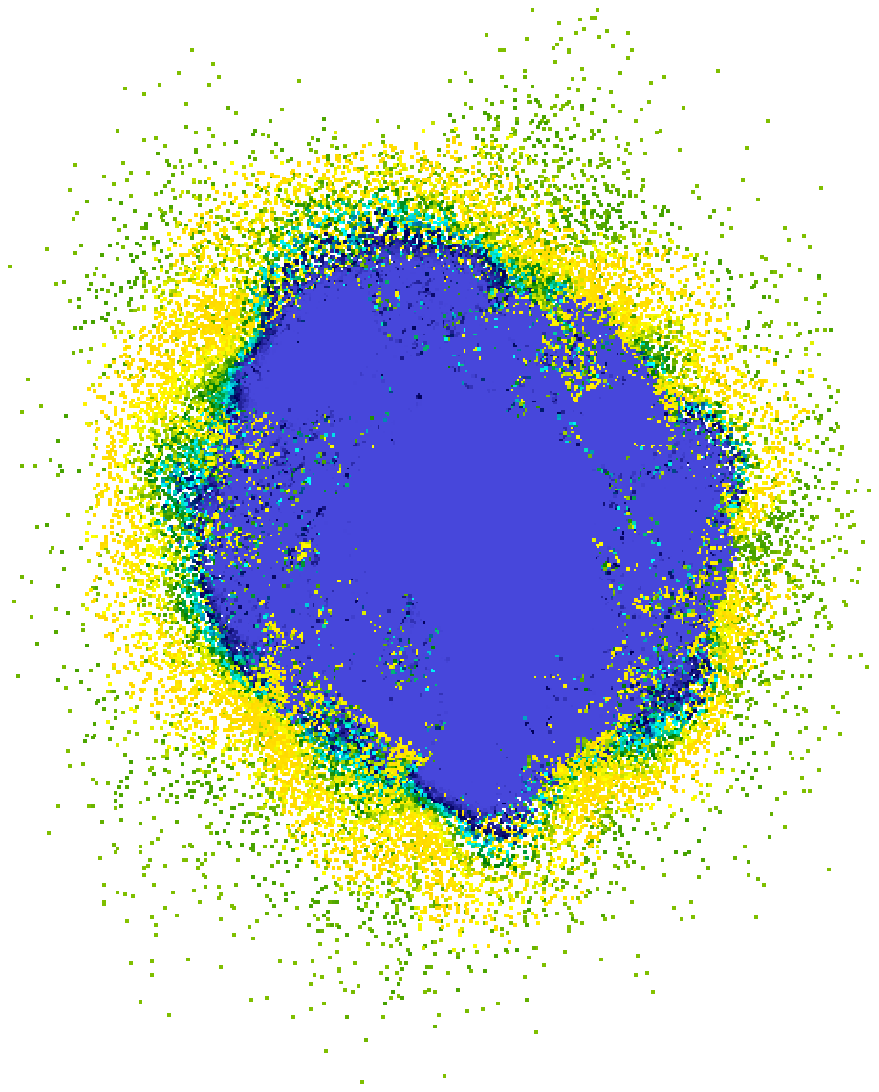}}
    \subfigure[X(\nuc{12}{C})]
    {\includegraphics[height=3.2cm]{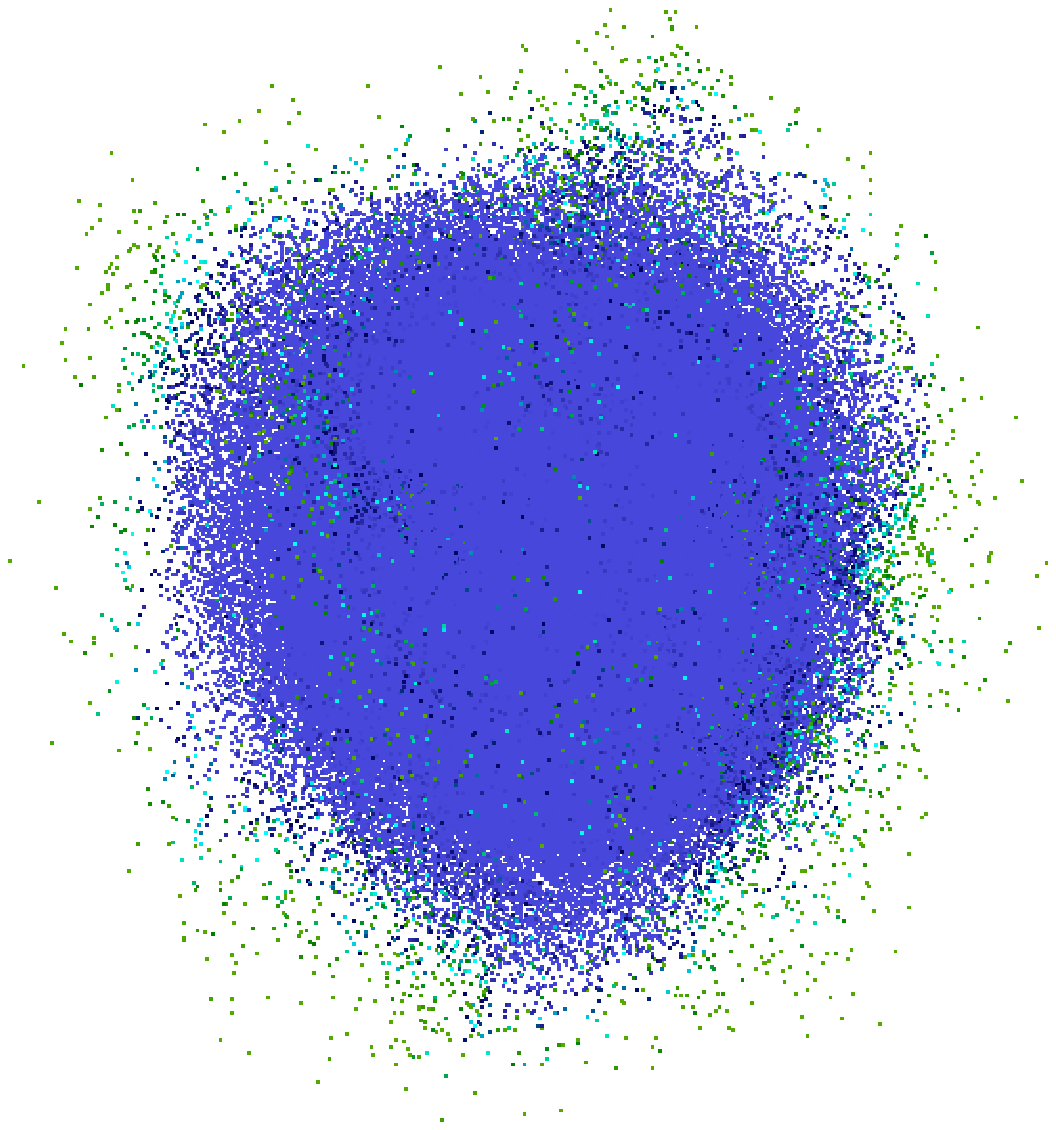}} \subfigure[N300C;
    X(\nuc{56}{Ni})] {\includegraphics[height=3.2cm]{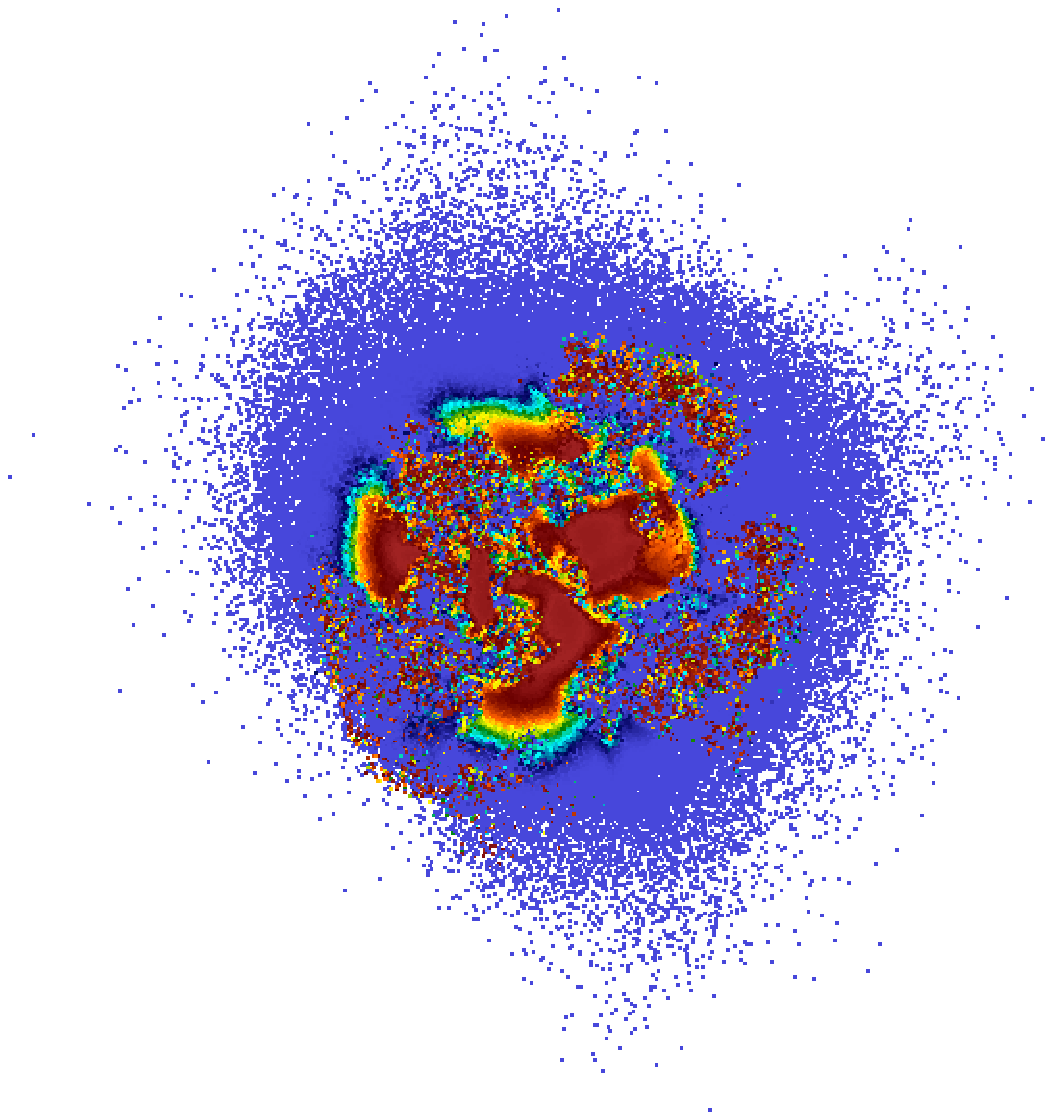}}
    \subfigure[X(\nuc{54}{Fe}+\nuc{56}{Fe}+\nuc{58}{Ni})]
    {\includegraphics[height=3.2cm]{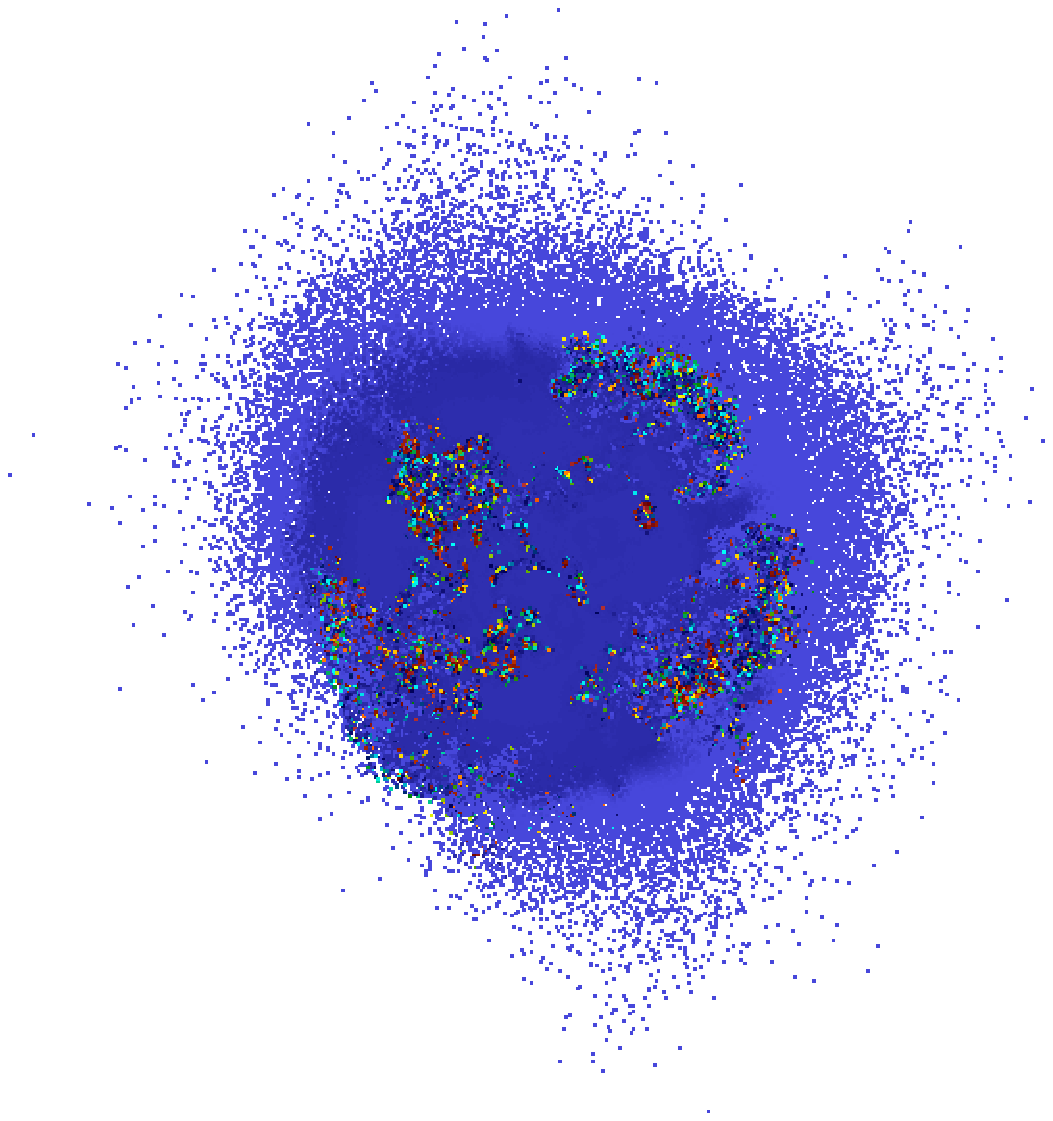}}
    \subfigure[X(\nuc{28}{Si})]
    {\includegraphics[height=3.2cm]{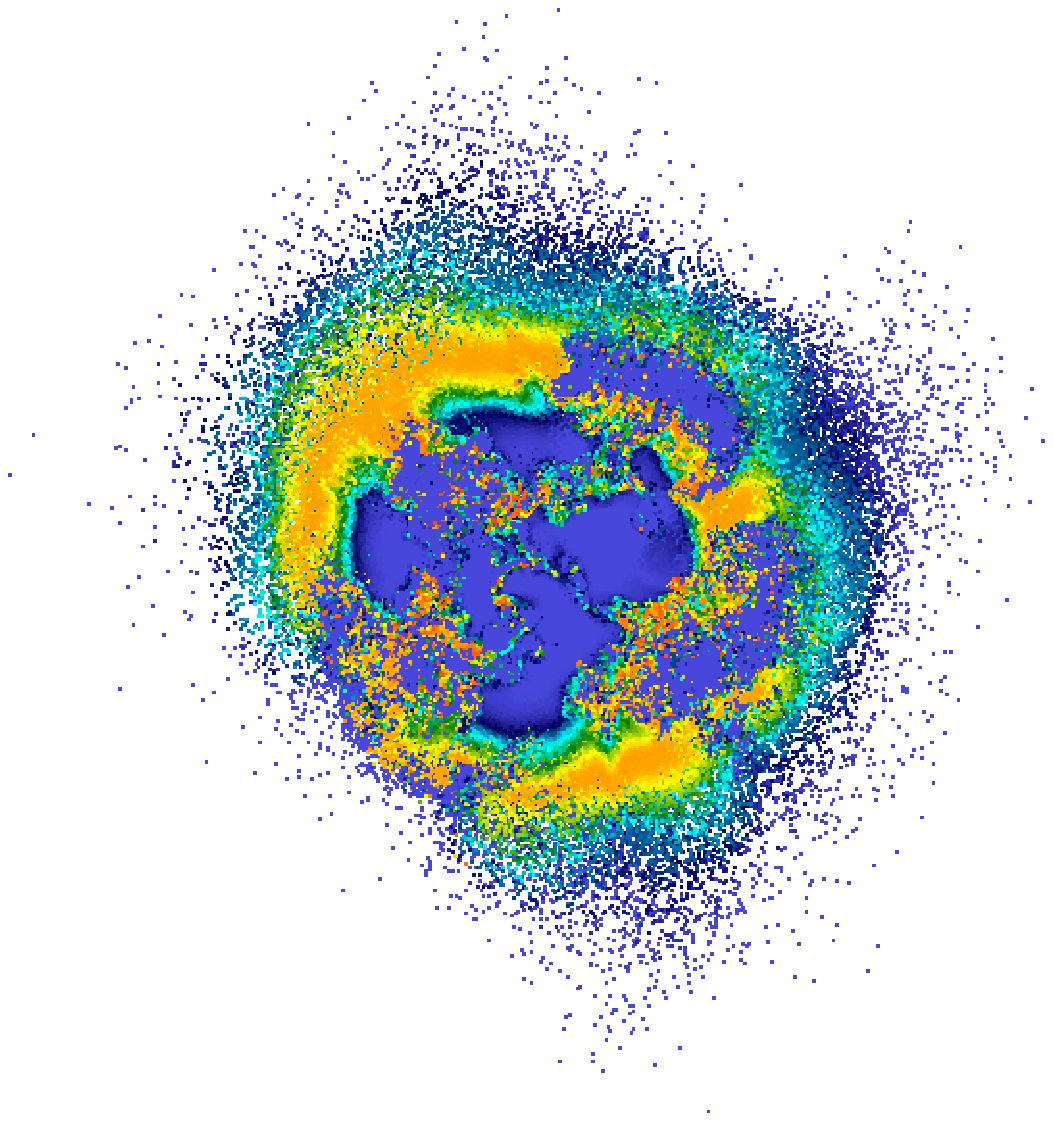}}
    \subfigure[X(\nuc{16}{O})]
    {\includegraphics[height=3.2cm]{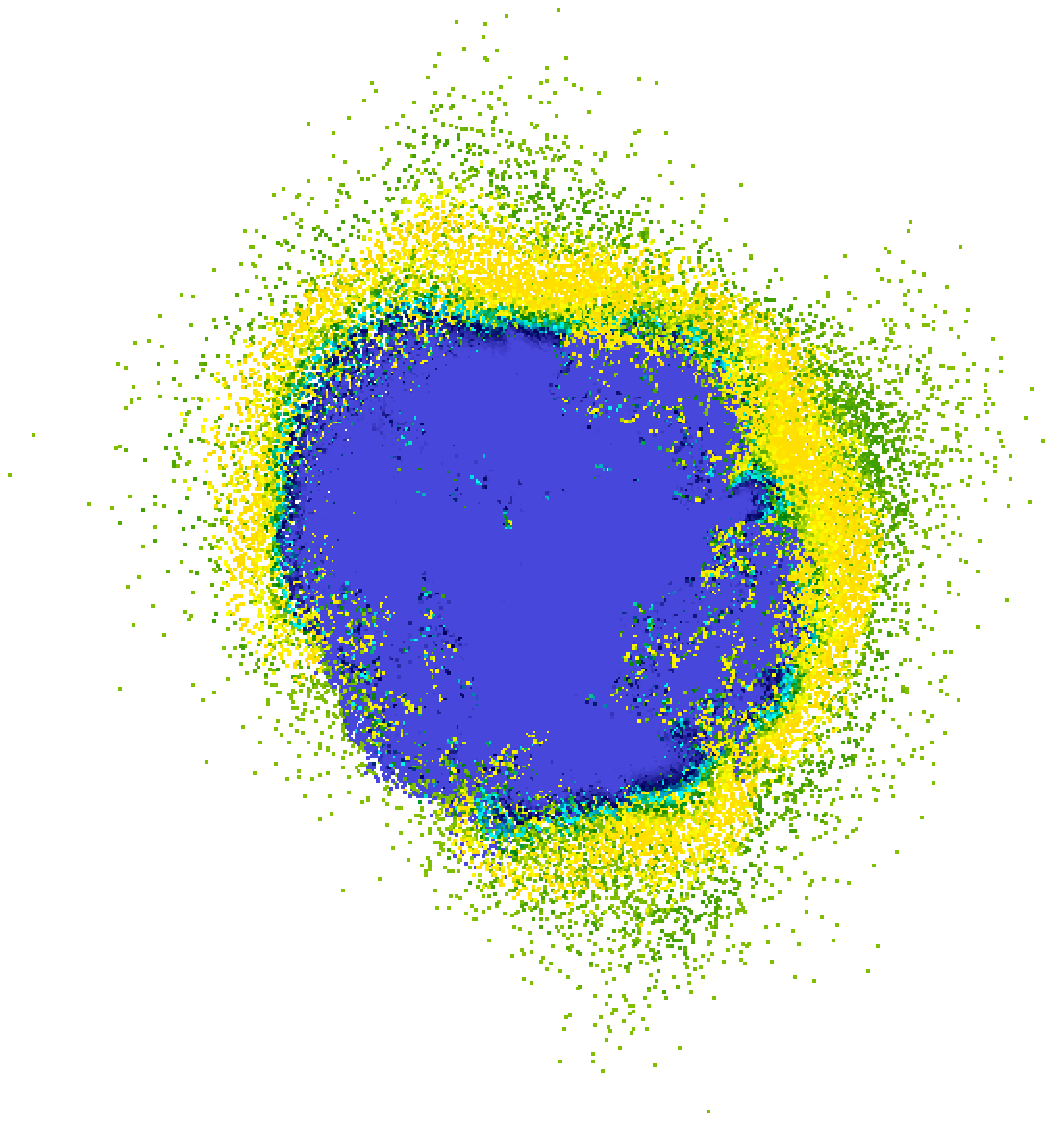}}
    \subfigure[X(\nuc{12}{C})]
    {\includegraphics[height=3.2cm]{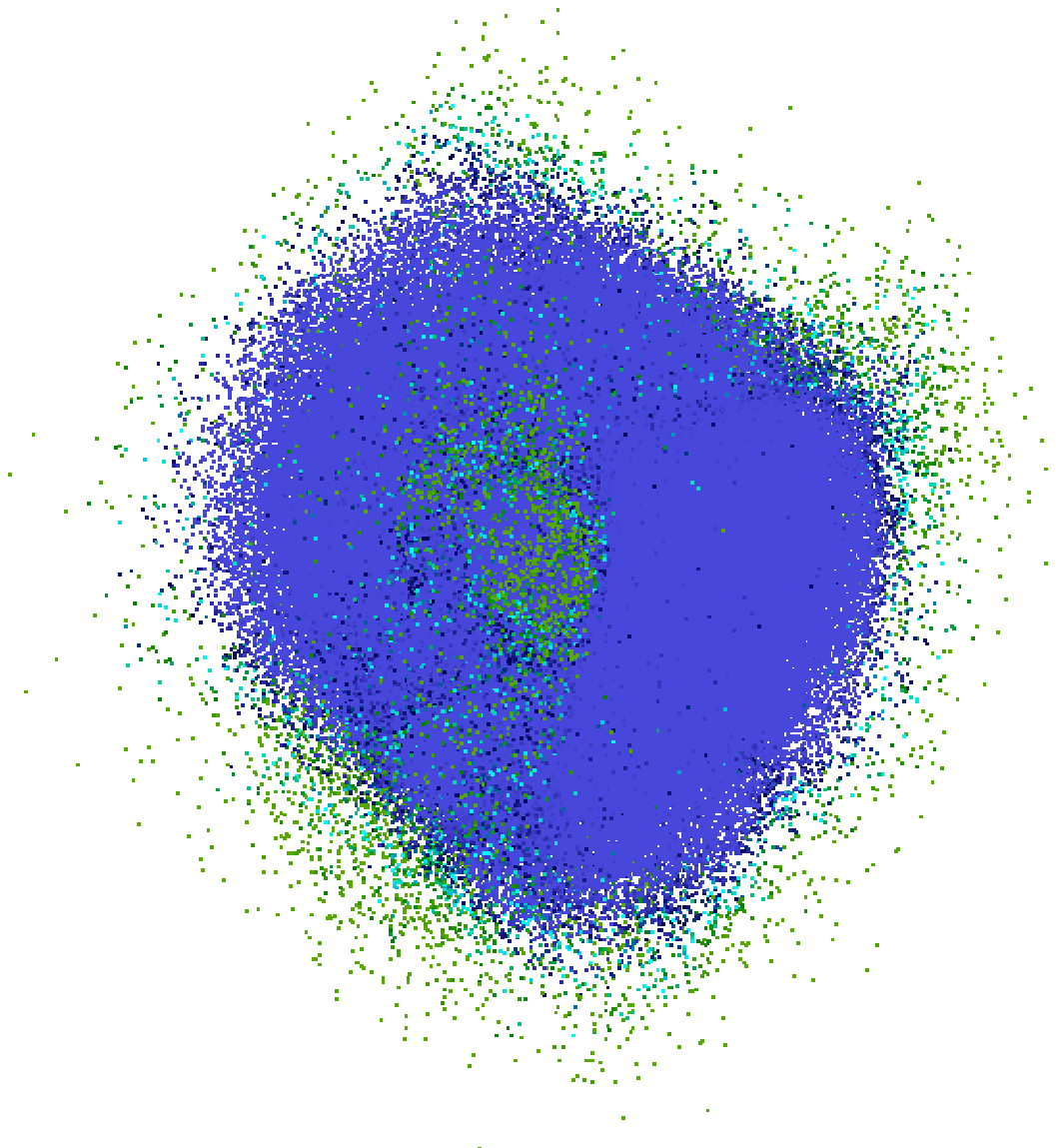}}
    \subfigure[N1600; X(\nuc{56}{Ni})]
    {\includegraphics[height=3.2cm]{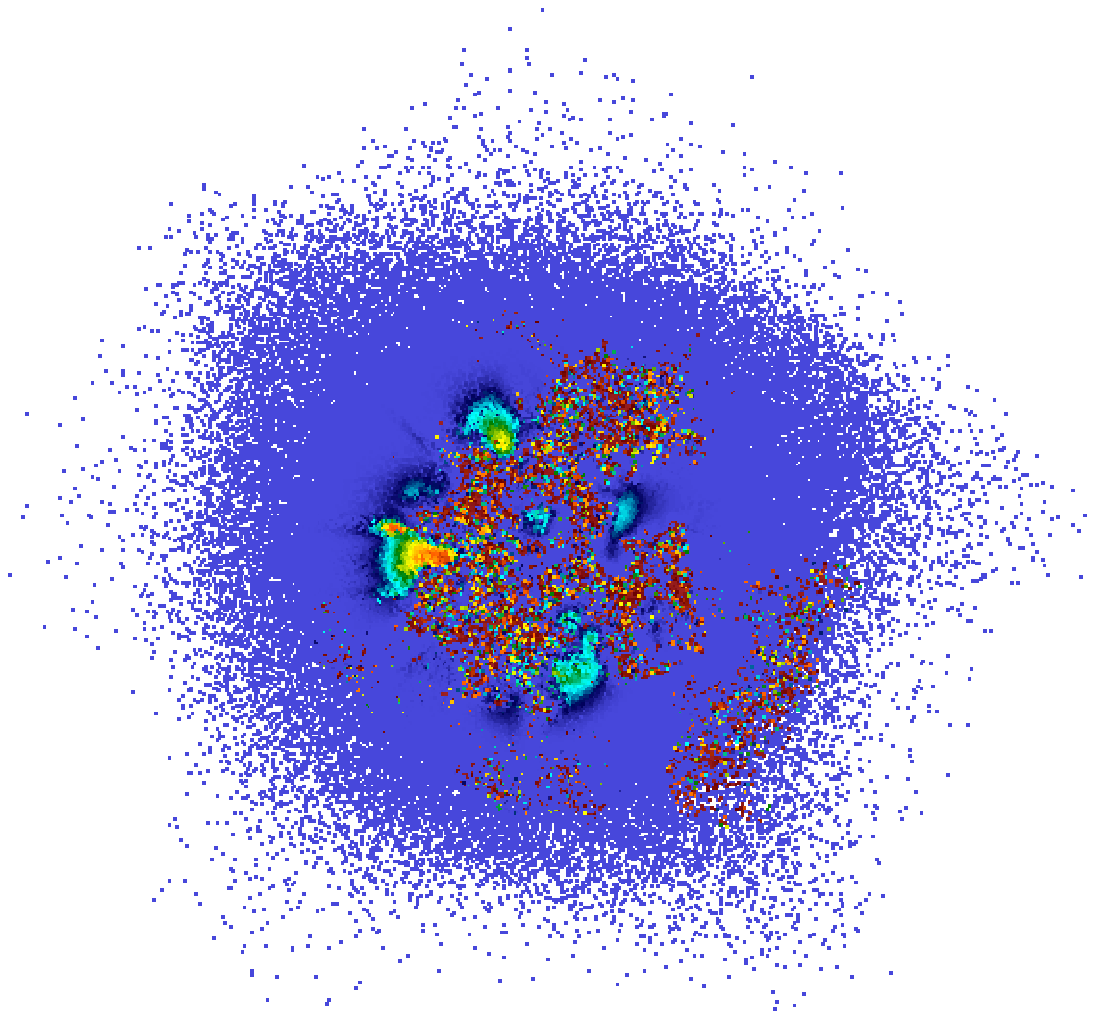}}
    \subfigure[X(\nuc{54}{Fe}+\nuc{56}{Fe}+\nuc{58}{Ni})]
    {\includegraphics[height=3.2cm]{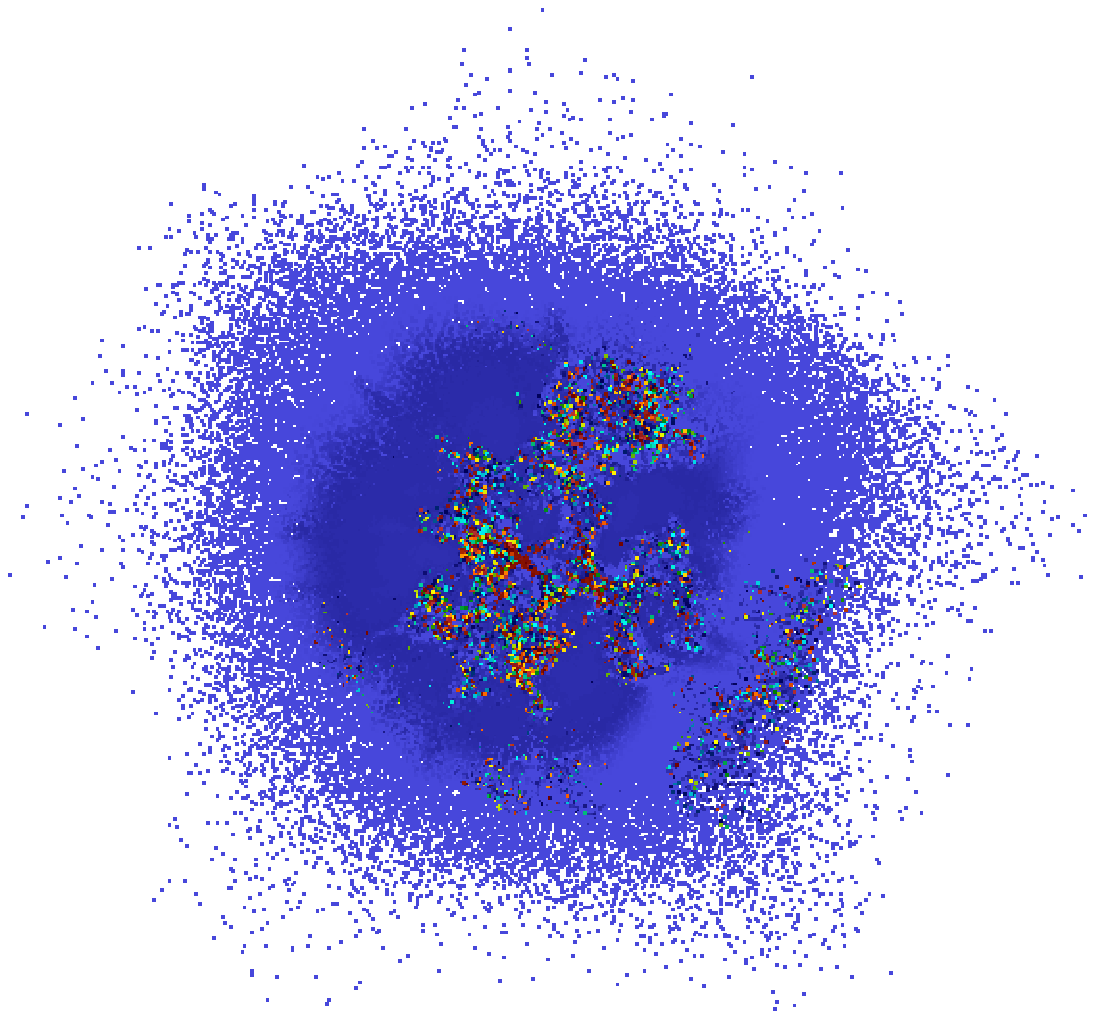}}
    \subfigure[X(\nuc{28}{Si})]
    {\includegraphics[height=3.2cm]{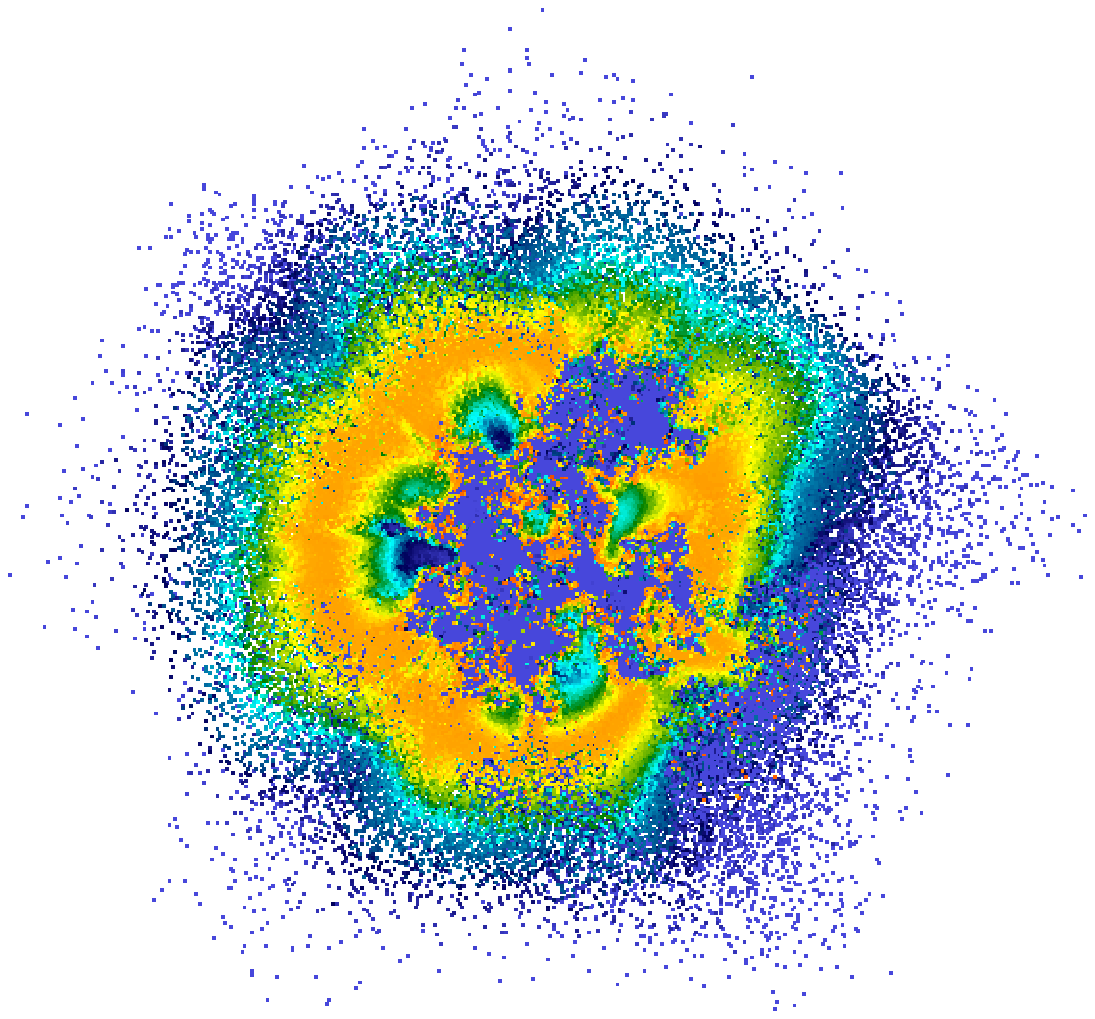}}
    \subfigure[X(\nuc{16}{O})]
    {\includegraphics[height=3.2cm]{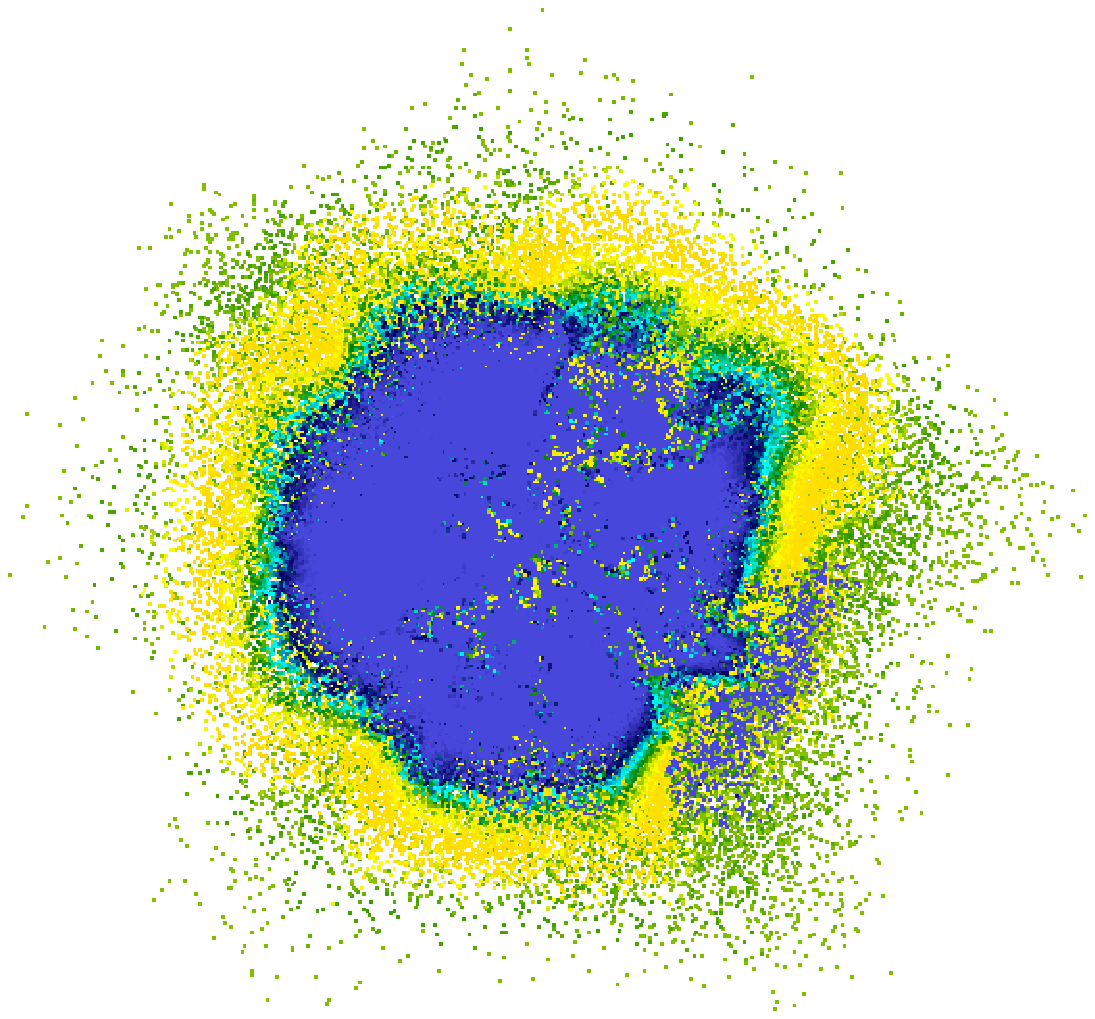}}
    \subfigure[X(\nuc{12}{C})]
    {\includegraphics[height=3.2cm]{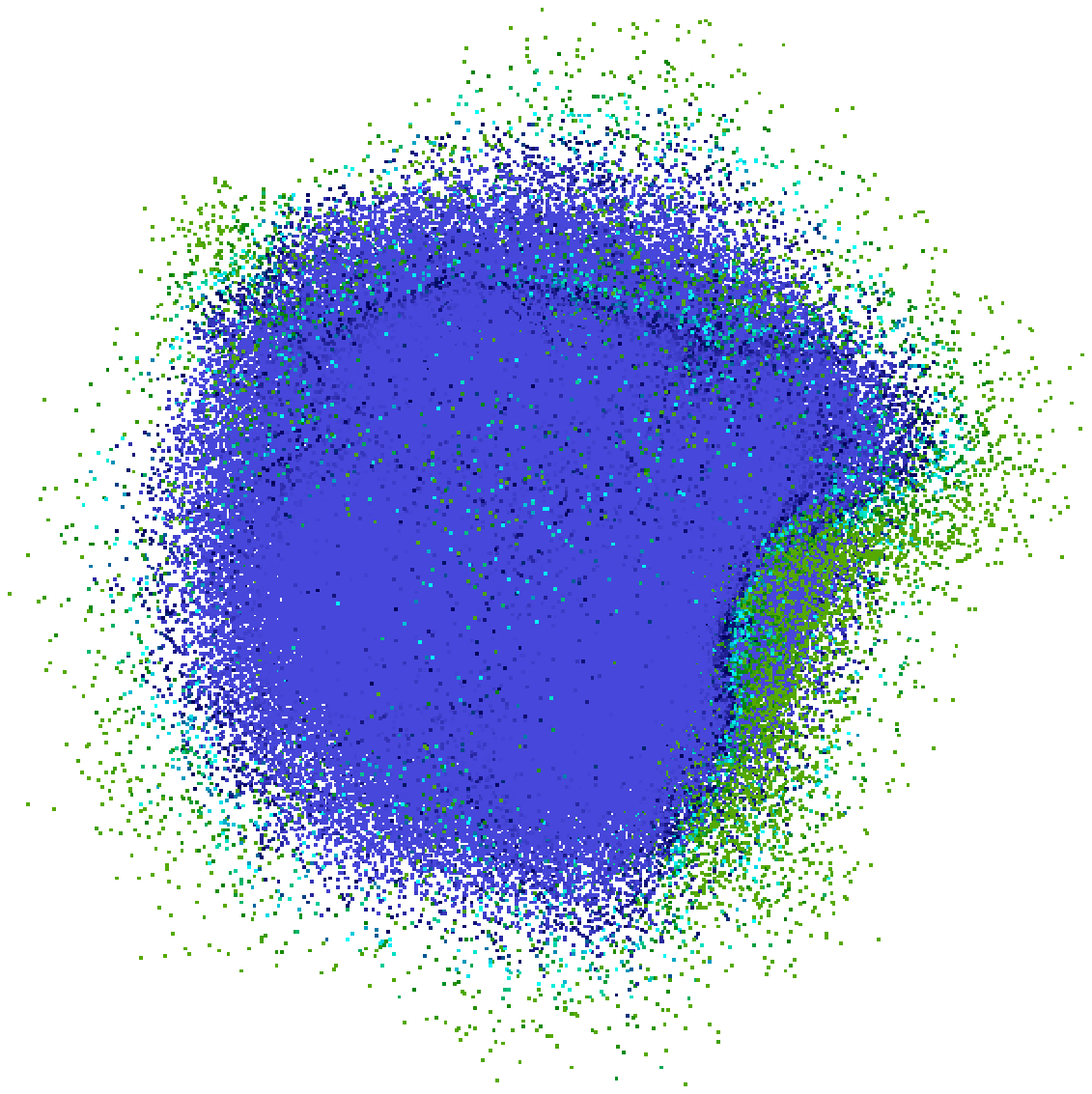}}
    {\includegraphics[width=0.7\columnwidth]{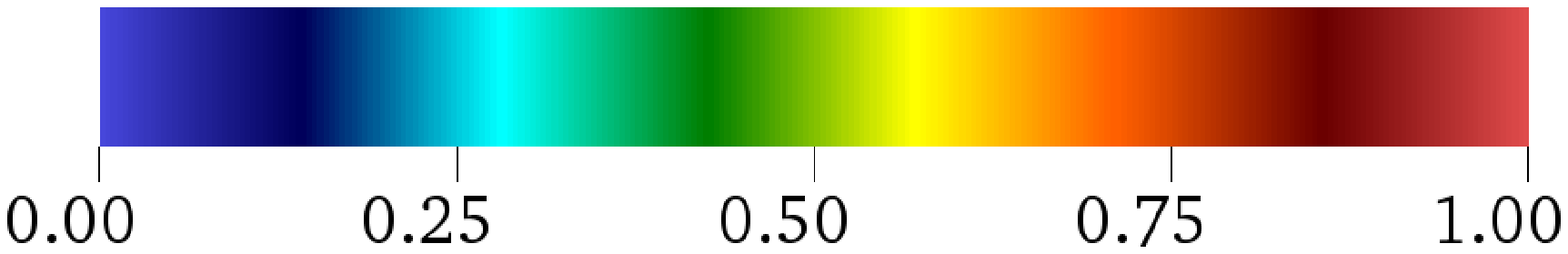}}
    \caption{Tracer particle positions for the N3, N40, N300C, and
      N1600 models (from top to bottom). The tracer particles are
      colored by the mass fractions of \nuc{56}{Ni} (first column),
      \nuc{54}{Fe}+\nuc{56}{Fe}+\nuc{58}{Ni} (representing stable iron
      group nuclei, second column), \nuc{28}{Si} (third column),
      \nuc{16}{O} (forth column) and \nuc{12}{C} (fifth column) at
      $t=100\s$. For \nuc{56}{Ni}, \nuc{28}{Si}, \nuc{16}{O} and the
      stable iron group nuclei, the tracer particle cloud is cut in
      the plane of the page and only the bottom hemisphere is
      shown. Plotting the data this way allows the viewer to also see
      the abundance distributions in the deep core while still
      retaining the three dimensional nature of the tracer particle
      locations. Only for \nuc{12}{C} the full spherical cloud is
      shown since \nuc{12}{C} largely resides near the surface.  }
    \label{fig:tracer}
  \end{center}
\end{figure*}

\begin{figure*}
  \begin{center}
    \subfigure[N100L; X(\nuc{56}{Ni})]
    {\includegraphics[height=3.2cm]{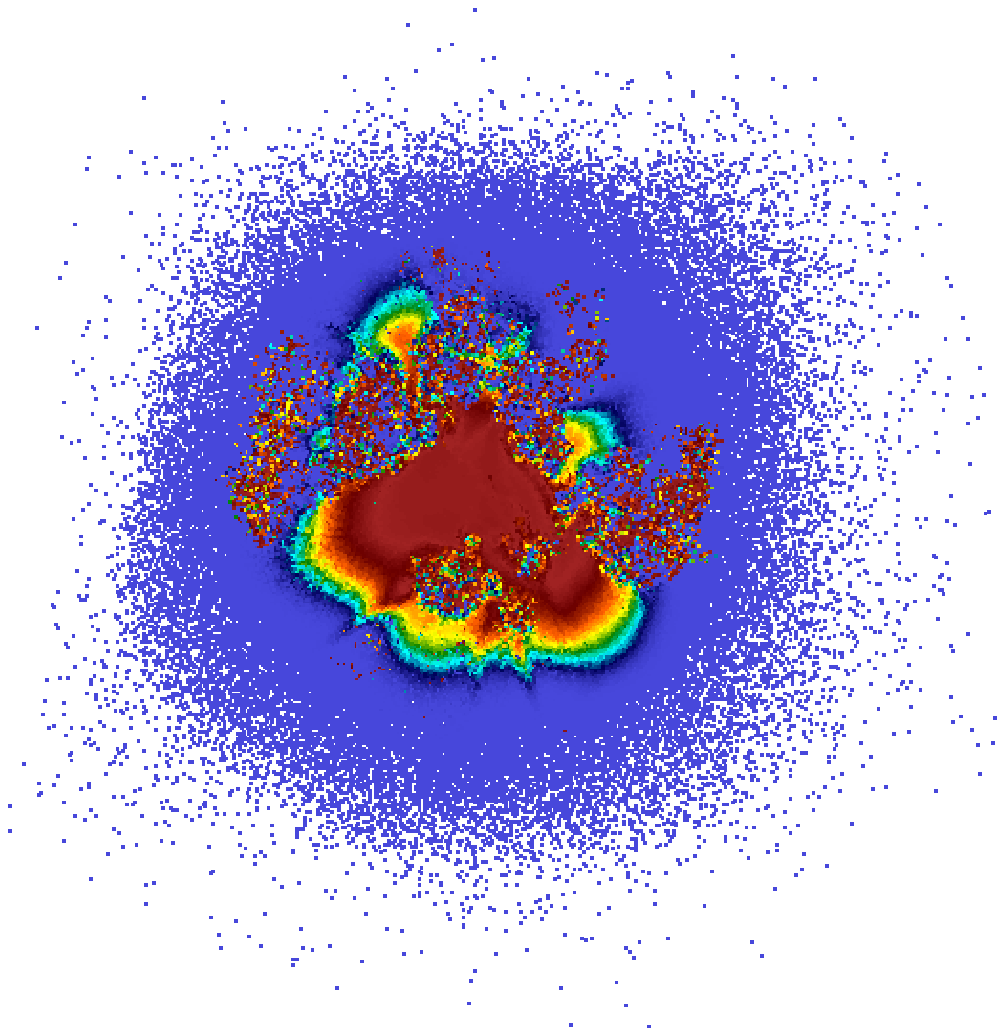}}
    \subfigure[X(\nuc{54}{Fe}+\nuc{56}{Fe}+\nuc{58}{Ni})]
    {\includegraphics[height=3.2cm]{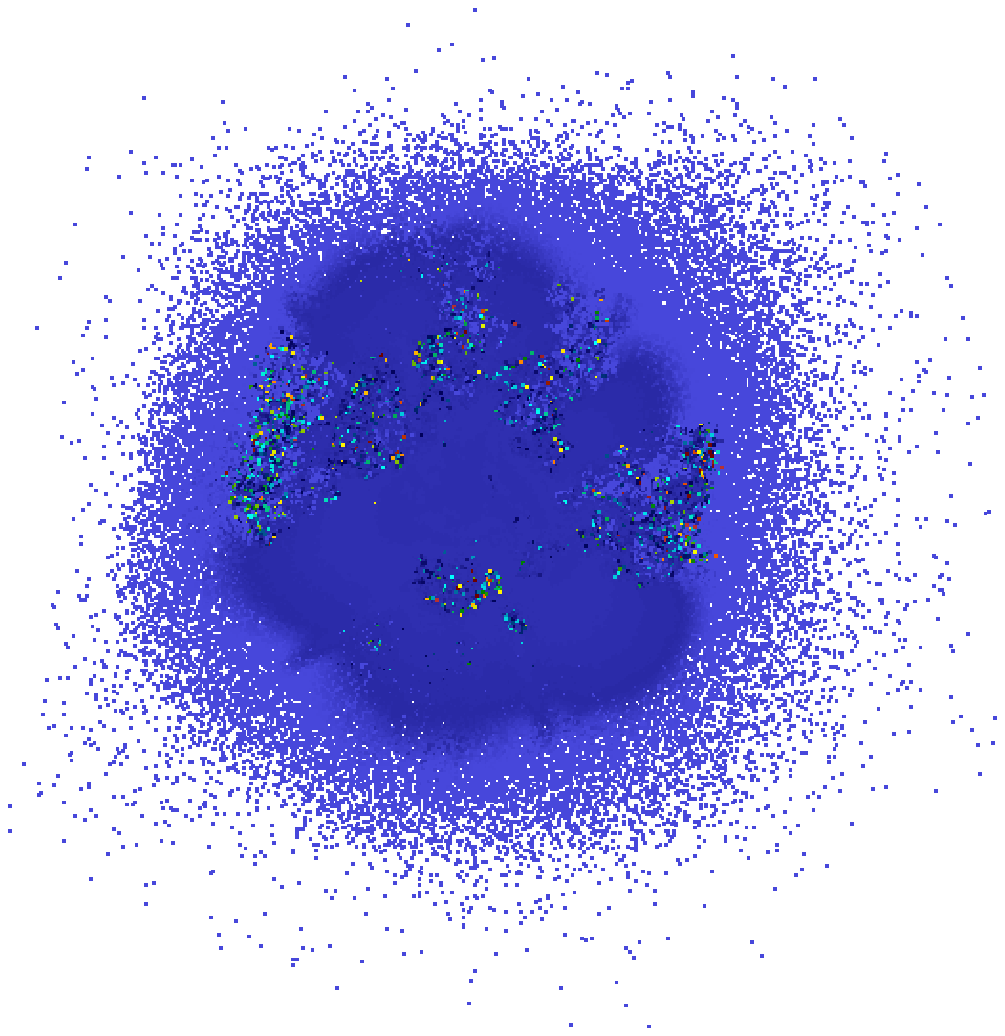}}
    \subfigure[X(\nuc{28}{Si})]
    {\includegraphics[height=3.2cm]{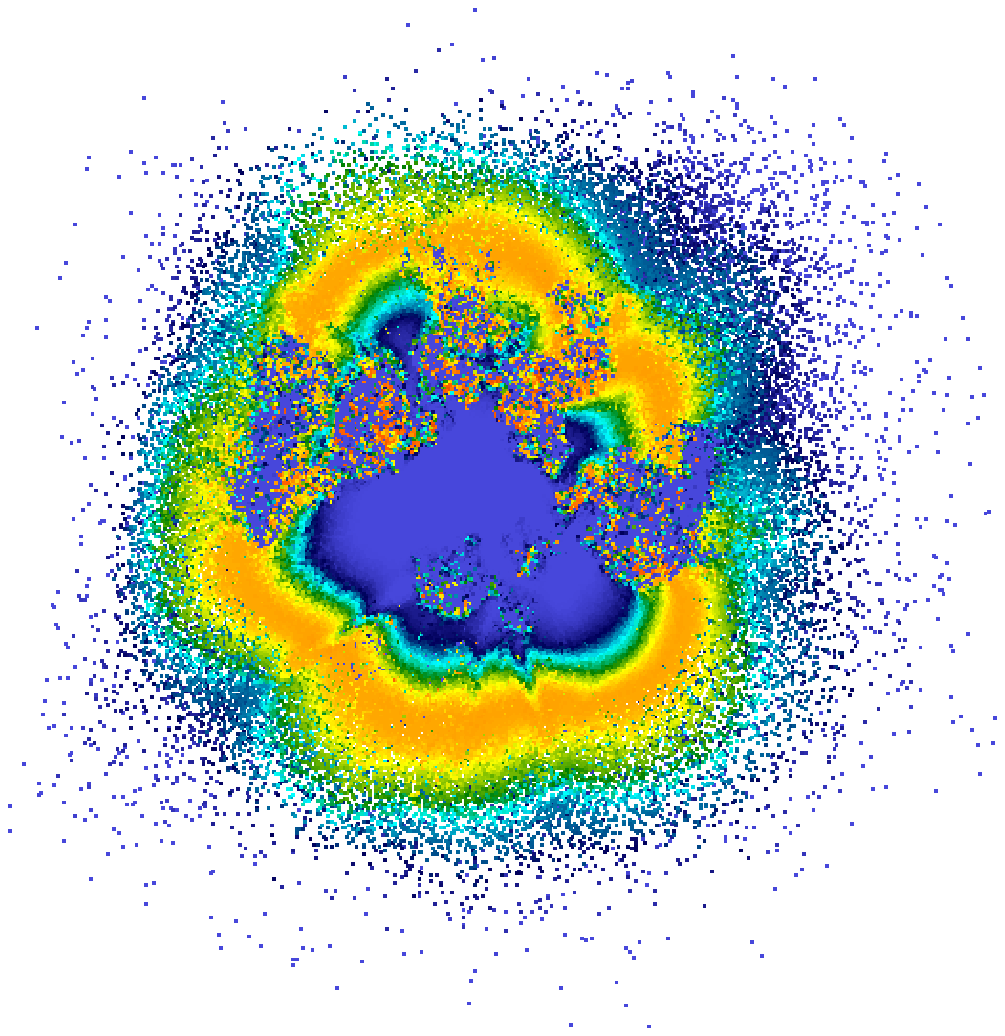}}
    \subfigure[X(\nuc{16}{O})]
    {\includegraphics[height=3.2cm]{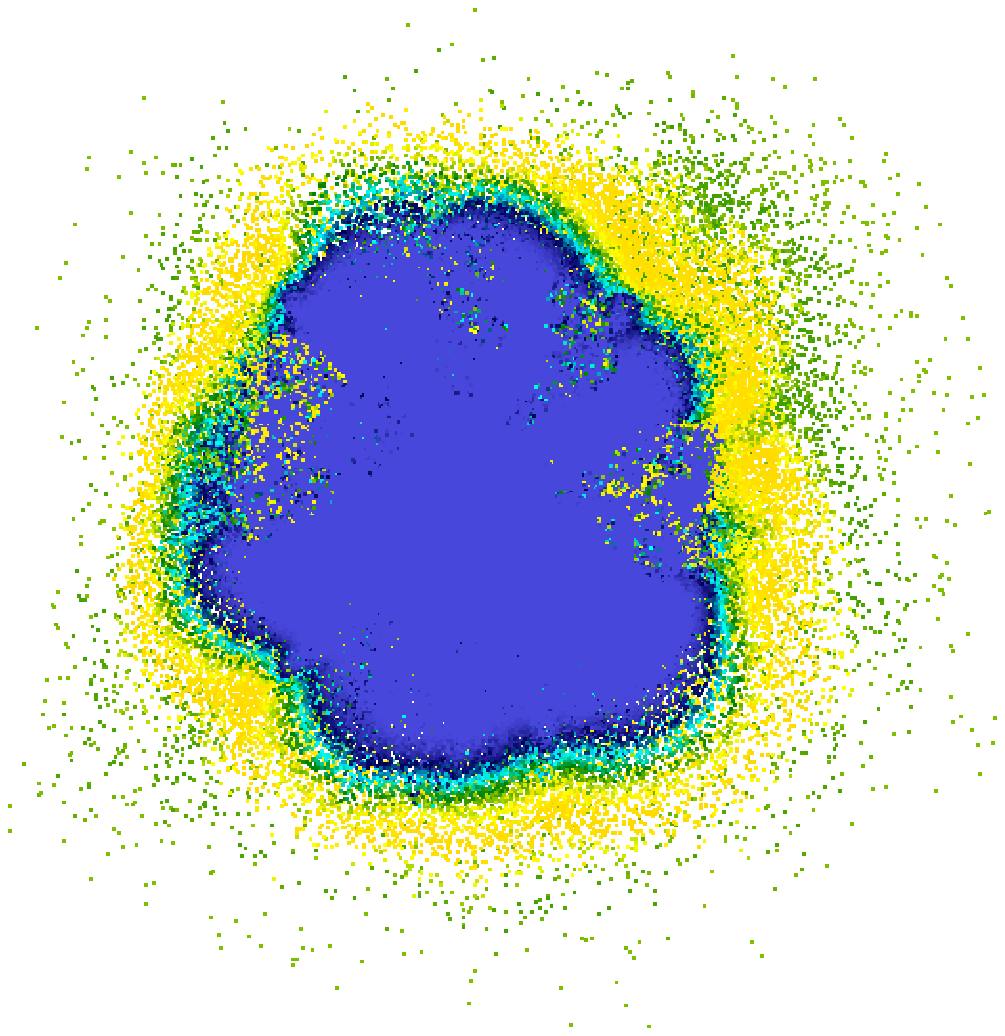}}
    \subfigure[X(\nuc{12}{C})]
    {\includegraphics[height=3.2cm]{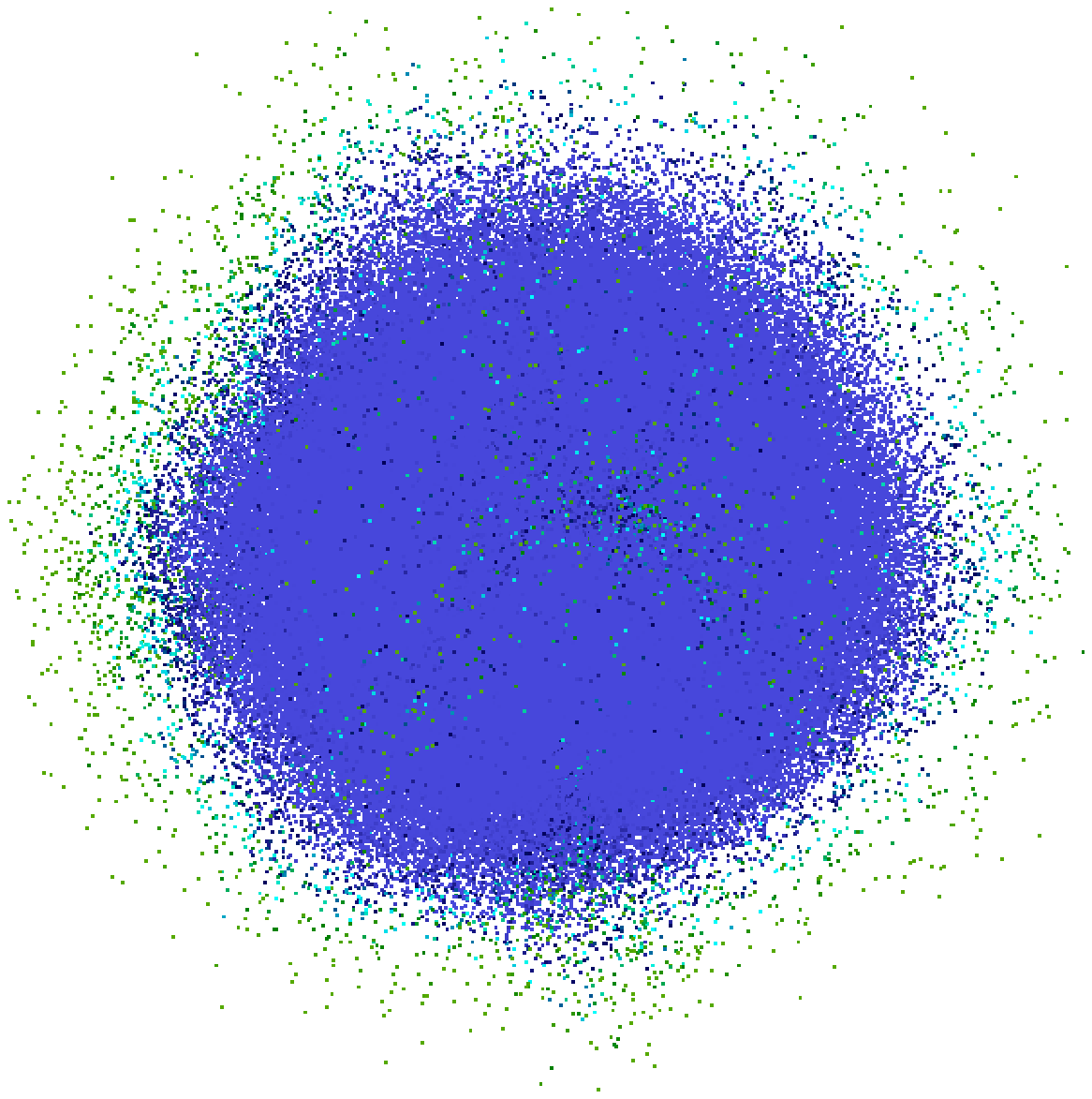}} \subfigure[N100;
    X(\nuc{56}{Ni})] {\includegraphics[height=3.2cm]{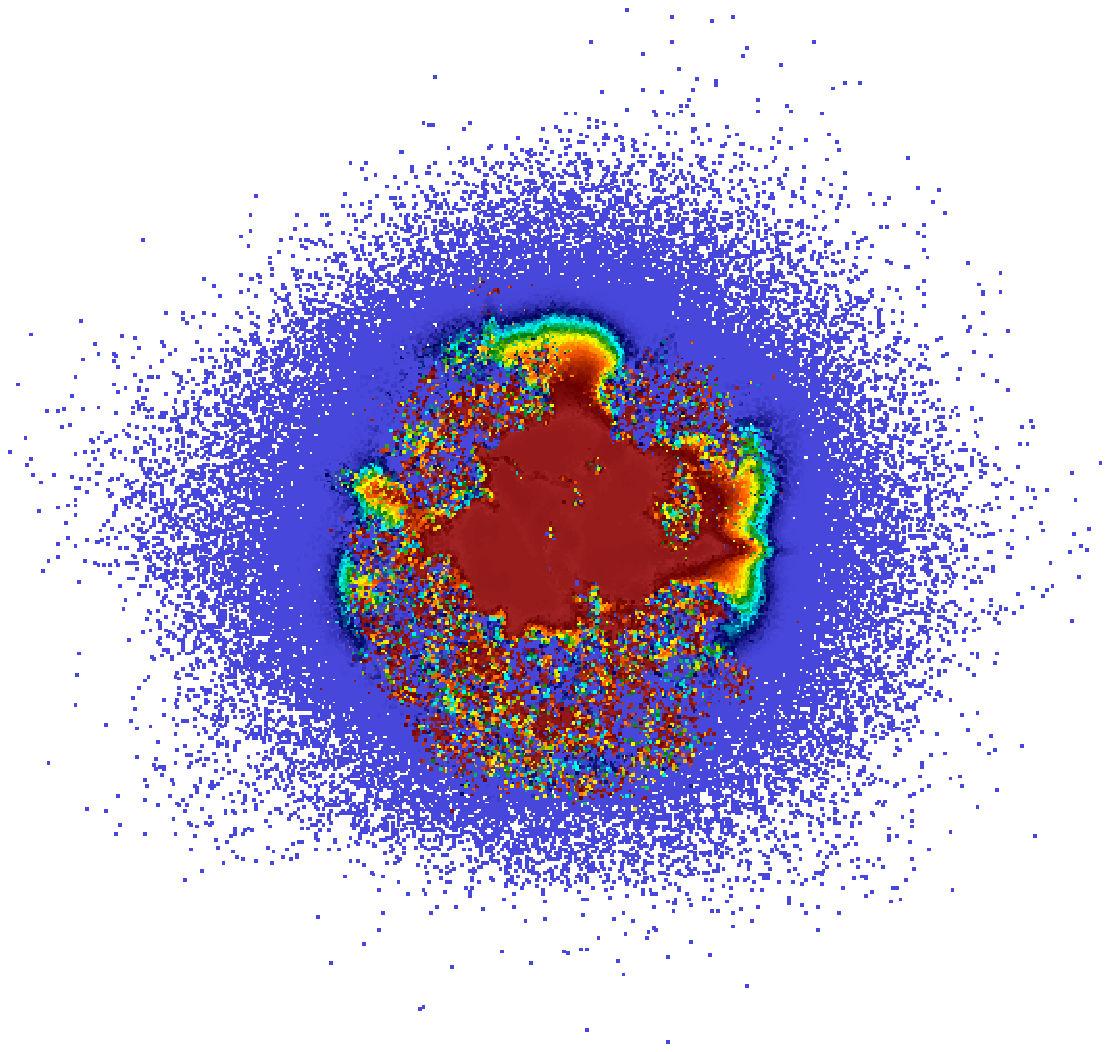}}
    \subfigure[X(\nuc{54}{Fe}+\nuc{56}{Fe}+\nuc{58}{Ni})]
    {\includegraphics[height=3.2cm]{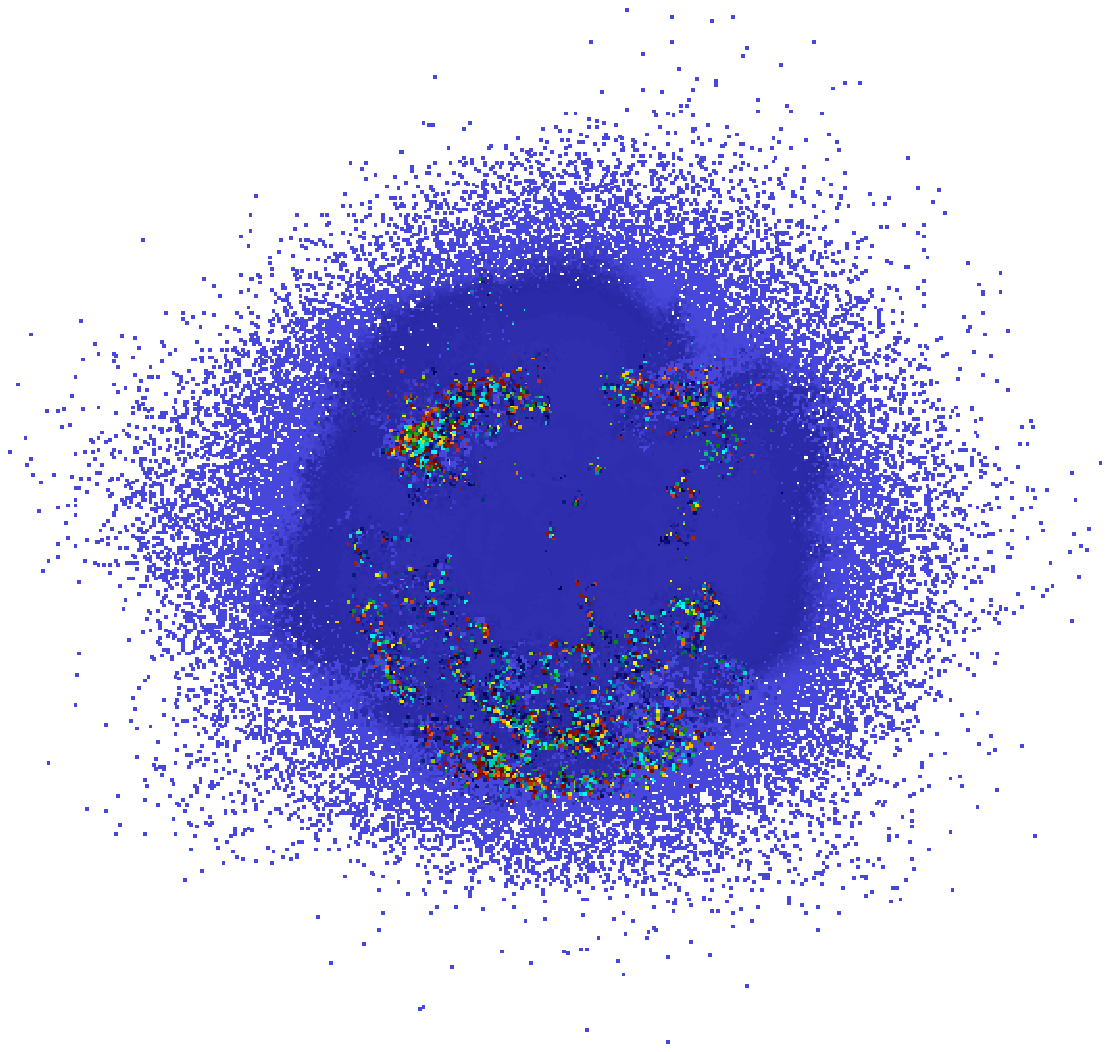}}
    \subfigure[X(\nuc{28}{Si})]
    {\includegraphics[height=3.2cm]{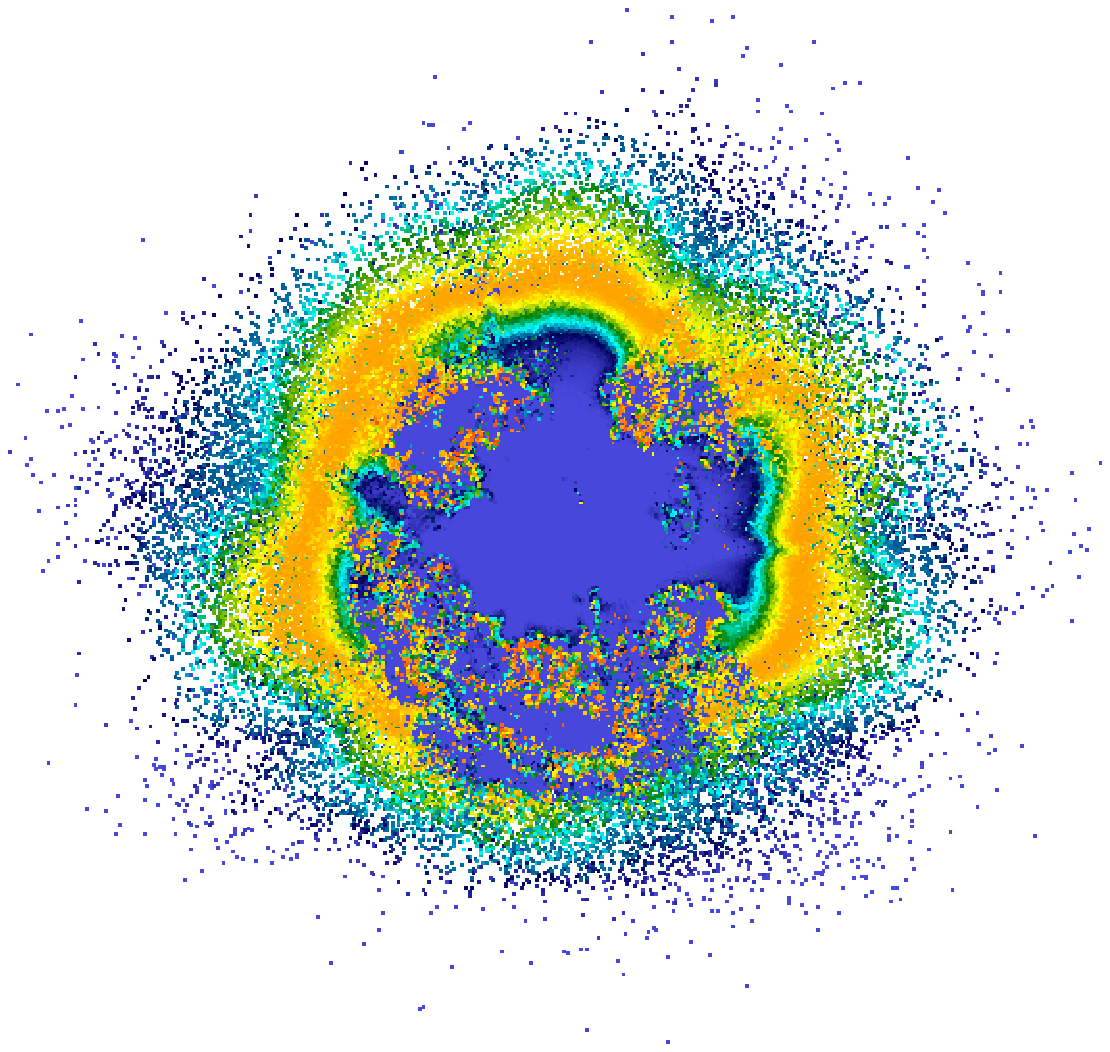}}
    \subfigure[X(\nuc{16}{O})]
    {\includegraphics[height=3.2cm]{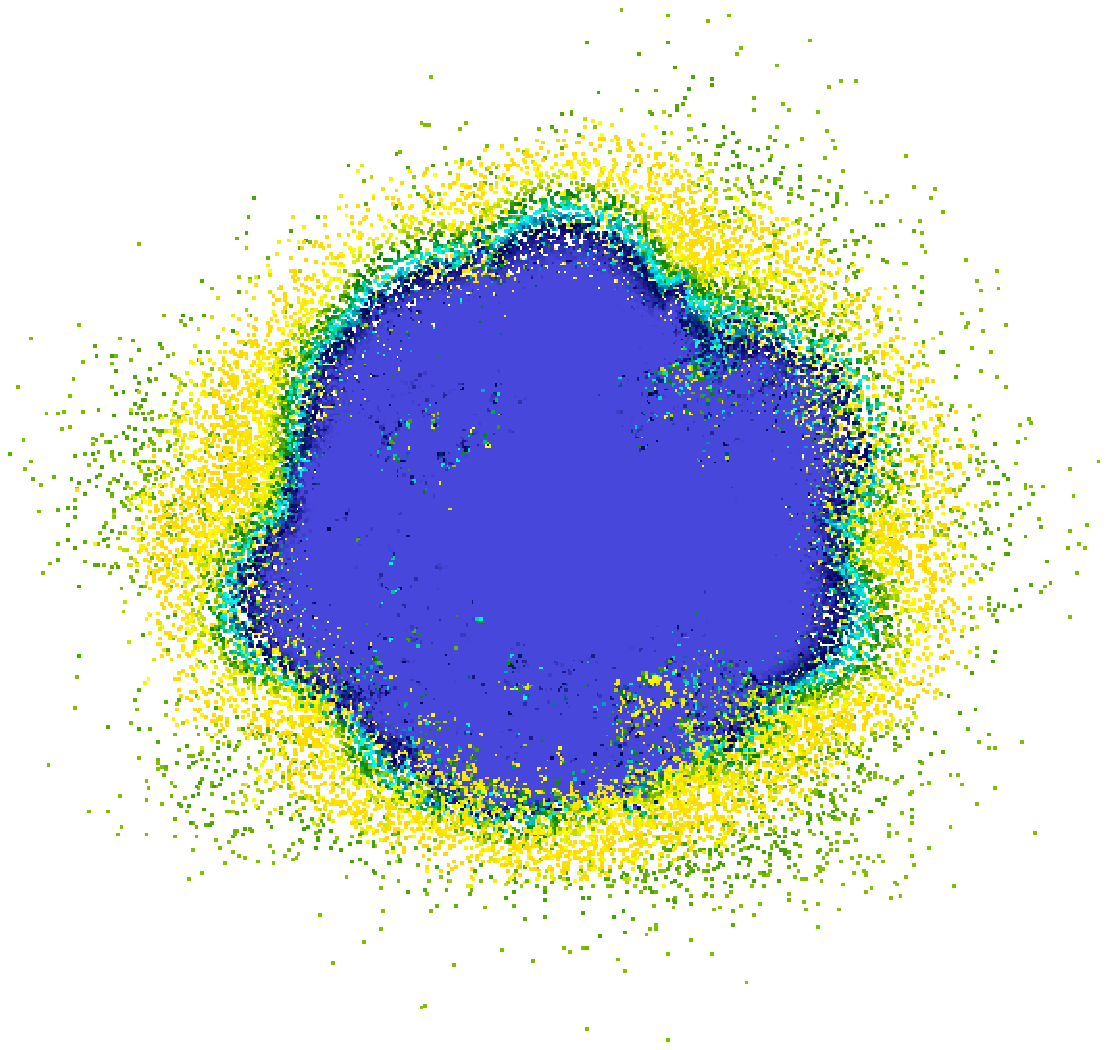}}
    \subfigure[X(\nuc{12}{C})]
    {\includegraphics[height=3.2cm]{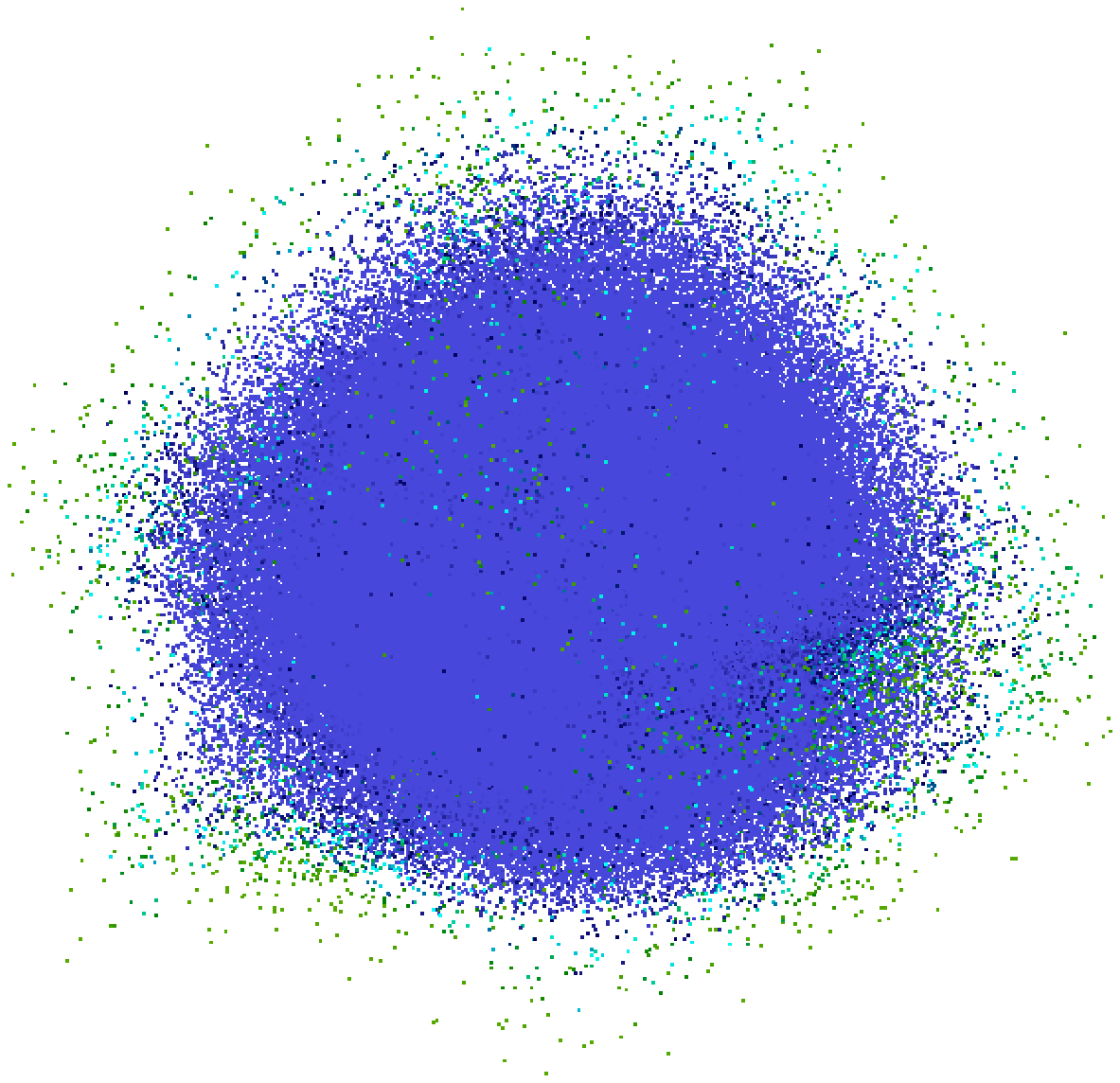}} \subfigure[N100H;
    X(\nuc{56}{Ni})] {\includegraphics[height=3.2cm]{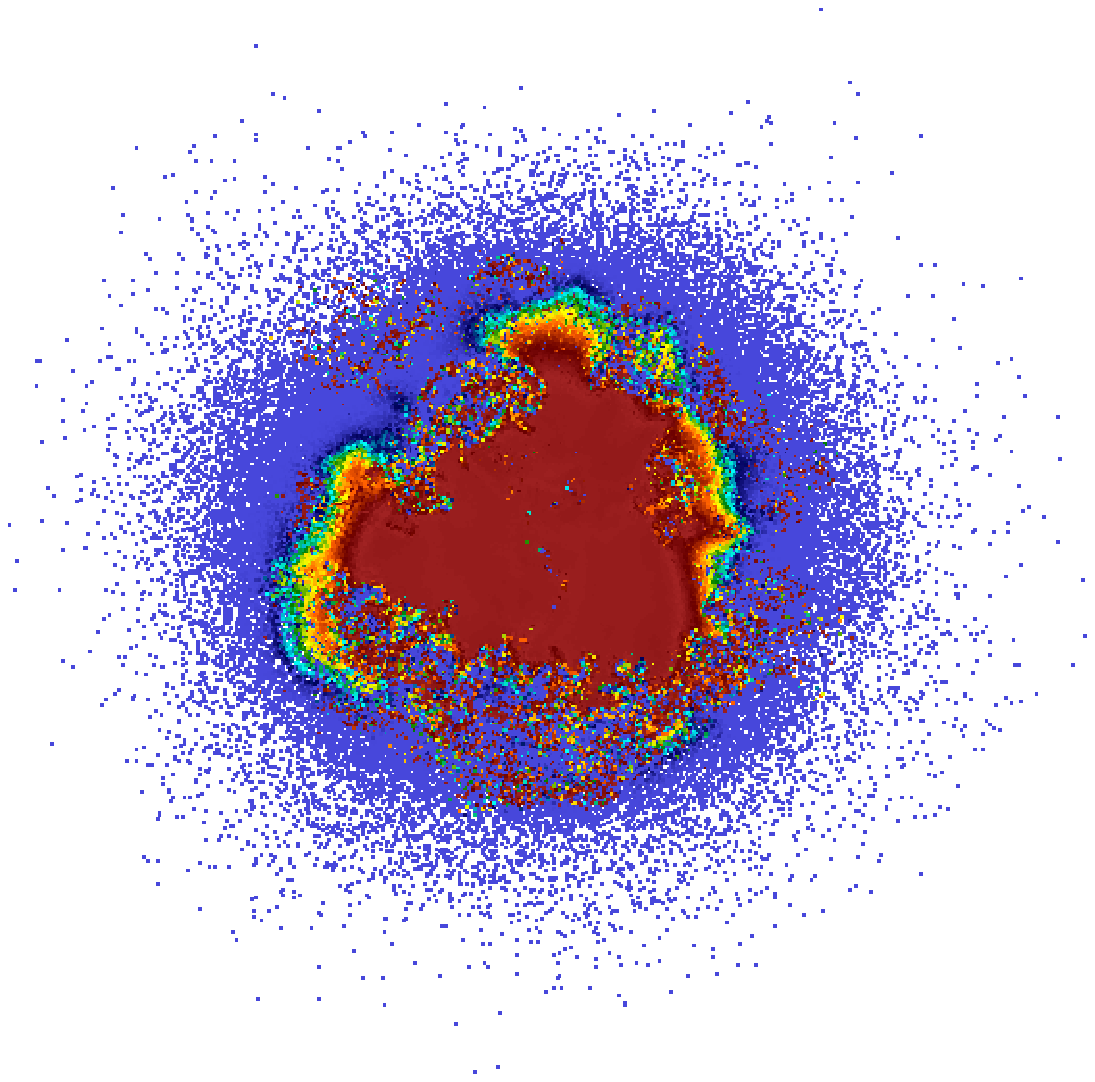}}
    \subfigure[X(\nuc{54}{Fe}+\nuc{56}{Fe}+\nuc{58}{Ni})]
    {\includegraphics[height=3.2cm]{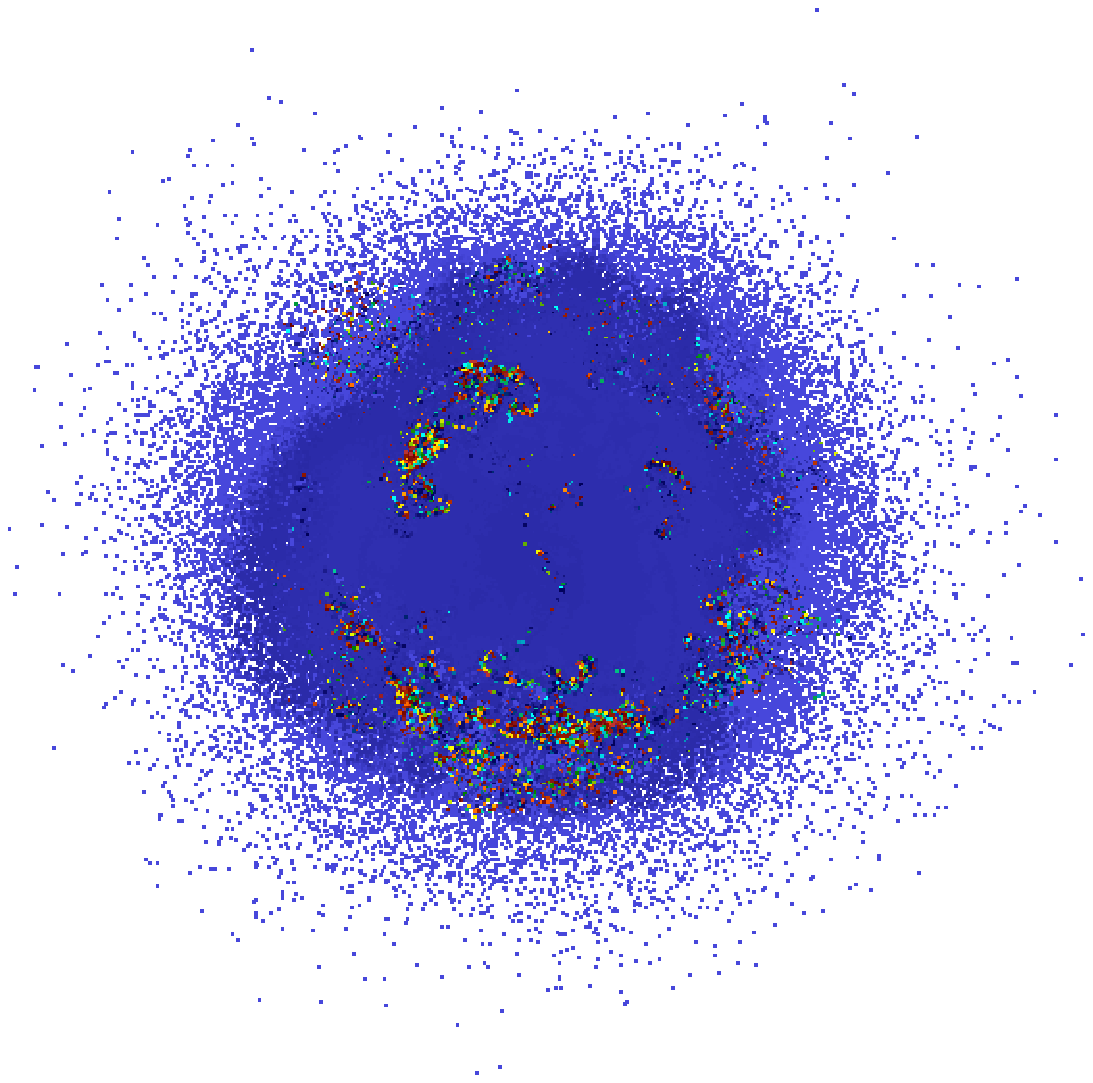}}
    \subfigure[X(\nuc{28}{Si})]
    {\includegraphics[height=3.2cm]{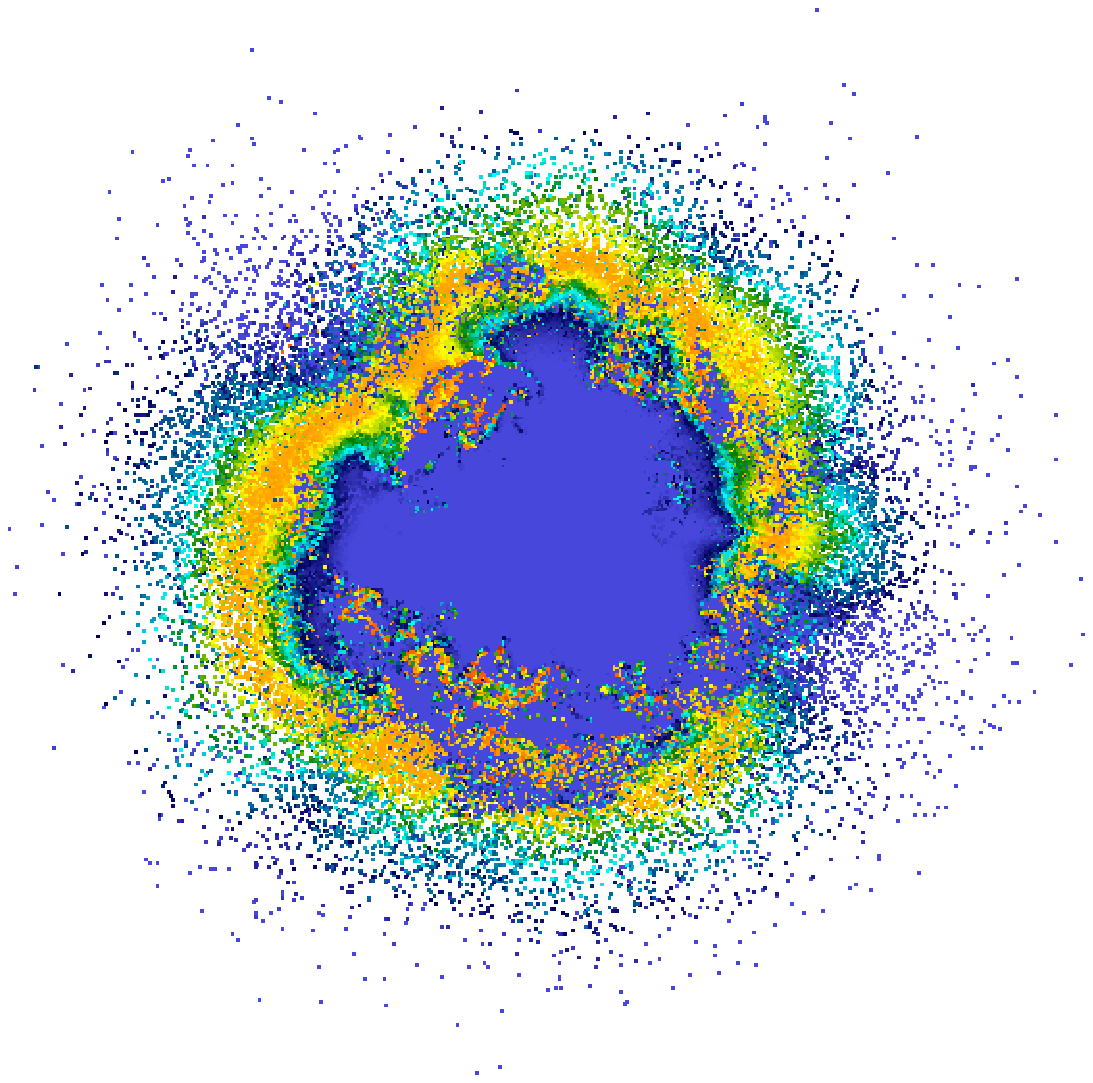}}
    \subfigure[X(\nuc{16}{O})]
    {\includegraphics[height=3.2cm]{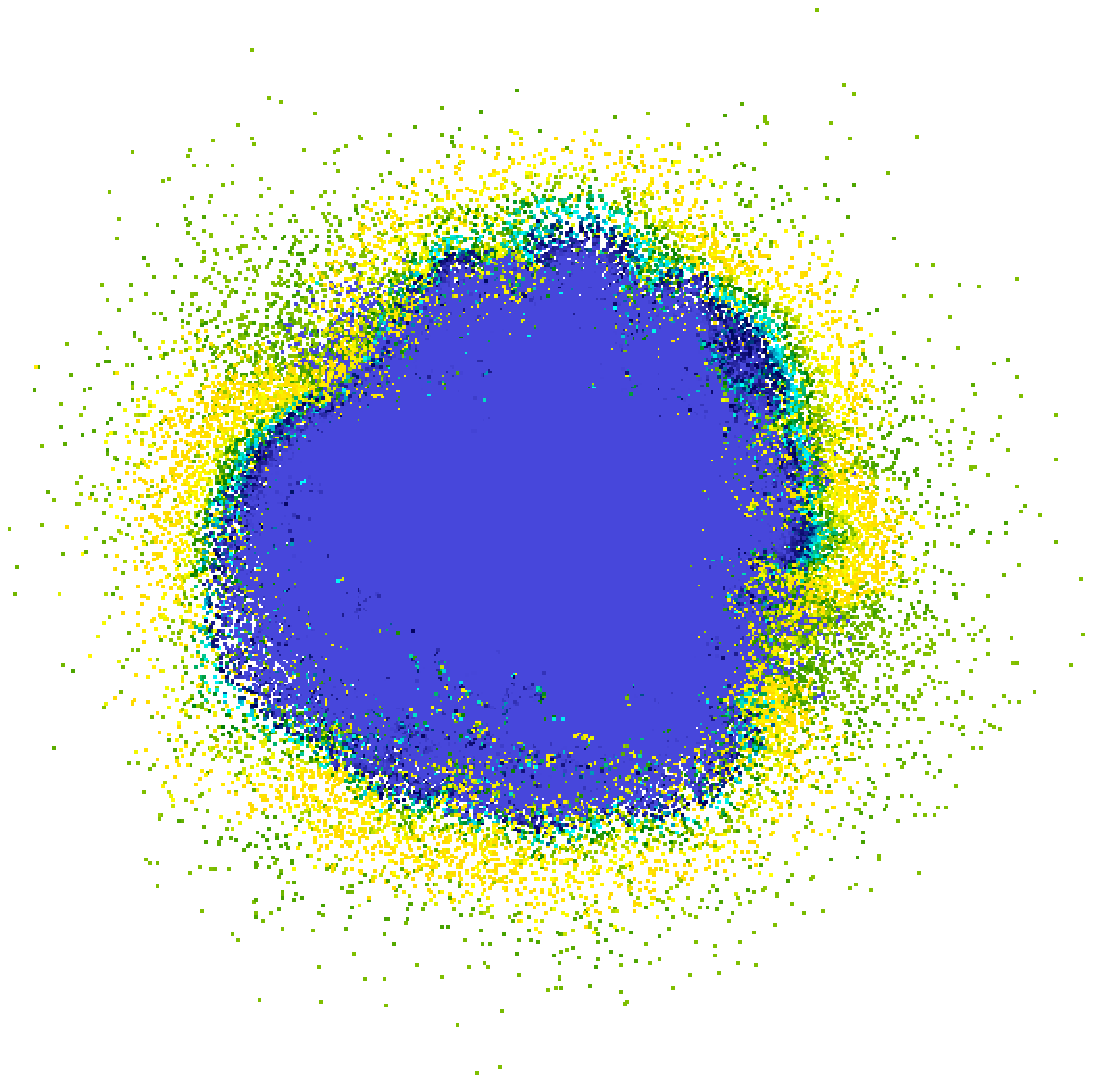}}
    \subfigure[X(\nuc{12}{C})]
    {\includegraphics[height=3.2cm]{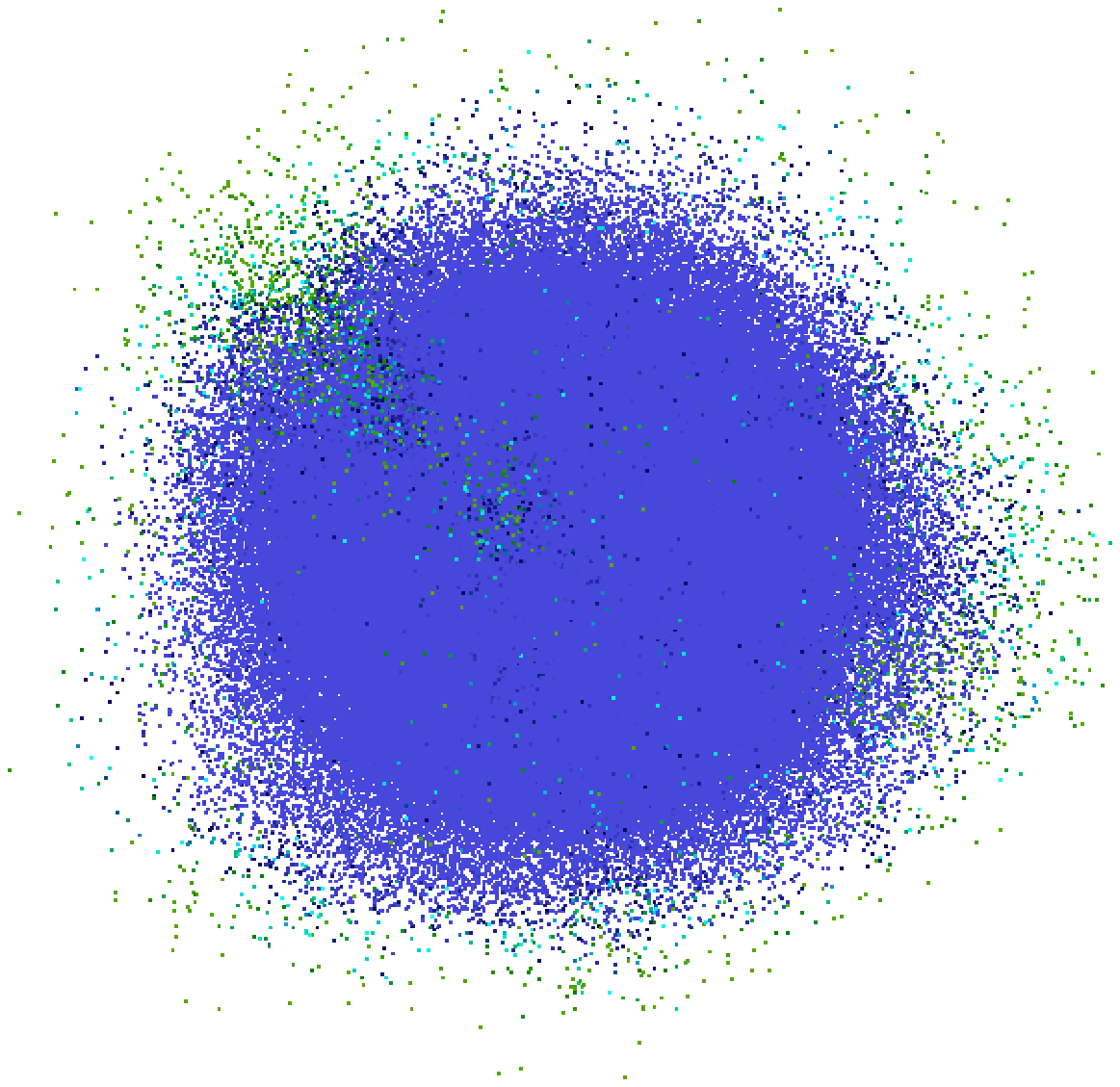}}
    {\includegraphics[width=0.7\columnwidth]{legend}}

    \caption{Tracer particle positions for the central density
      sequence of N100 models (from the low central density N100L
      model in the top row to the high central density model N100H in
      the bottom row). Just as in Fig.~\ref{fig:tracer} the tracer
      particles are colored by the mass fractions of \nuc{56}{Ni}
      (first column), \nuc{54}{Fe}+\nuc{56}{Fe}+\nuc{58}{Ni}
      (representing stable iron group nuclei, second column),
      \nuc{28}{Si} (third column), \nuc{16}{O} (forth column) and
      \nuc{12}{C} (fifth column) at $t=100\s$.  Again, for \nuc{12}{C}
      we show the whole cloud of tracer particles, whereas only one
      hemisphere is shown for all other species.}
    \label{fig:tracer2}
  \end{center}
\end{figure*}

Since we have essentially parametrized the strength of the
deflagration phase by ignition kernel number (see
Sec.~\ref{sec:inimod} and Fig.~\ref{fig:IGN}), the models with the
most ignition kernels burn the least amount of fuel in the detonation
and vice versa \citep{roepke2007a,mazzali2007a,kasen2009a}.
Furthermore, it is evident that models with fewer ignition kernels are
more asymmetric.  A prime example for this is N3, see left column of
Fig.~\ref{fig:lsets}.  The small number of ignition kernels
necessarily leads to a highly asymmetric seed configuration of the
RT-unstable deflagration plumes, which results in a highly asymmetric
distribution of the deflagration ashes in the star. For our N3 case,
only one hemisphere is burned by the deflagration. The fact that the
detonation can only be triggered at the interface of the deflagration
flame front and the fuel reinforces this asymmetry. The resulting
distribution of the nucleosynthesis products fully reflects the
asymmetric evolution of the deflagration and detonation (see
Fig.~\ref{fig:tracer}).  In contrast, consider the N100 model (right
column in Fig.~\ref{fig:lsets}). The rather symmetric ignition
configuration results in deflagration plumes rising in all
directions. As a consequence, DDTs also occur on all sides of the star
and the explosion asymmetries are very modest, which again reflects in
a more symmetric distribution of burning products (see
Fig.~\ref{fig:tracer2}).

\section{Nucleosynthesis}
\label{sec:nucleo}
 
For each of the fourteen models, we have determined the isotopic
composition of the ejecta using our tracer particle method
\citep{travaglio2004a,roepke2006b,seitenzahl2010a}.  In each initial
model we have distributed one million tracer particles of equal mass
in a way that the underlying density profile of the WD is reproduced
by the tracer particles.  Extrapolating the tracer resolution and
yield convergence study (done in 2D) of \citet{seitenzahl2010a} to 3D,
this number of tracer particles (100 per axis) is sufficient to
reliably predict the yields for the most abundant nuclides.  
{For all models, the initial chemical composition for the
  post-processing is taken to be 47.5 per
cent \nuc{12}{C}, 50 per cent \nuc{16}{O}, and 2.5 per cent
\nuc{22}{Ne} by mass.  The \nuc{22}{Ne} fraction roughly
corresponds to solar metallicity of the zero-age main-sequence
progenitor under the often made approximation that \nuc{14}{N} is
processed to \nuc{22}{Ne} during core helium burning via
$\nuc{14}{N}(\alpha,\gamma)\nuc{18}{F}(\beta^+)\nuc{18}{O}(\alpha,\gamma)\nuc{22}{Ne}$.
The \nuc{22}{Ne} content introduces an excess of neutrons and results
in an electron fraction of $\ye = 0.49886$.  This composition is
assumed to be homogeneous throughout the star.}
The yield distributions can be used as an input to radiative transfer
calculations
\citep[e.g.][]{kasen2006a,sim2007b,sim2010a,kromer2010a,pakmor2010a,roepke2012a}
to derive model light curves and spectra.  Furthermore, the yields can
be used as an input for galactic chemical evolution calculations or to
make predictions of the shapes of the late-time light curves
\citep[e.g.][]{seitenzahl2009d,seitenzahl2011b,roepke2012a}.

\subsection{Total integrated nucleosynthetic yields}
\label{sec:intyields}
\begin{figure}
  \includegraphics[width=\columnwidth]{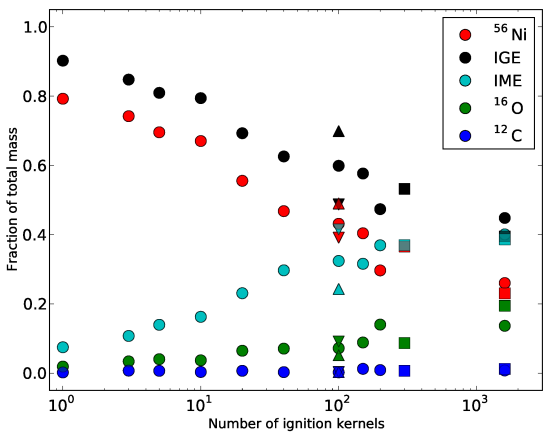}
  \caption{Fraction of the total mass at $t=100 \s$ in \nuc{56}{Ni},
    IGEs (including \nuc{56}{Ni}), IMEs, \nuc{16}{O}, and \nuc{12}{C}
    as a function of ignition kernel number.  We use upward-pointing
    triangles for N100H, downward-pointing triangles for N100L,
    squares for N300C and N1600C, and circles for all other models.}
  \label{fig:yields_vs_kernels}
\end{figure}

The yields of stable nuclides are presented in
Table~\ref{tab:syields}.  For this table we have completely decayed
all radioactive nuclides with half-lives less than 2 Gyr (such as
\nuc{40}{K} or \nuc{53}{Mn}) to stability; nuclides with half-lives
longer than 2 Gyr (i.e. \nuc{50}{V}) are tabulated with their
production yields at $t=100\s$.  The yields at $t=100\s$ of some
abundant and long-lived radioactive nuclides are tabulated in
Table~\ref{tab:ryields}. For convenience, we also show the yields of
\nuc{56}{Ni}, IGEs, IMEs, \nuc{16}{O}, and \nuc{12}{C} in
Fig.~\ref{fig:yields_vs_kernels}

We note a few obvious trends in the total integrated yields. As
expected, there is a clear trend that the total mass of \nuc{56}{Ni}
produced decreases with increasing ignition kernel number -- from
1.11\,\msun\ for N1 down to 0.32\,\msun\ for N1600C. This covers the
range of expected \nuc{56}{Ni} masses of normal SNe~Ia
\citep{stritzinger2006b}.  Similarly, the total mass produced as
intermediate mass elements (see e.g.  \nuc{28}{Si} or \nuc{32}{Si})
shows a clear trend of increasing yields with increasing ignition
kernel number.

This can be understood in the following way. For small numbers of
ignition kernels, the deflagration is relatively weak and burns only a
small fraction of the total mass of the star. The associated energy
release results in only moderately strong expansion. Consequently, at
the first occurrence of a DDT, the central density of the WD is still
quite high and most of the remaining fuel will be burned to IGE (most
of which is \nuc{56}{Ni}) by the detonation. In contrast, the larger
the number of ignition sparks, the stronger is the expansion during
the deflagration phase. The low central density at the onset of the
first DDT now results in much of the remaining fuel to be located
below a density of $1 \times 10^7 \gcc$, approximately the cutoff
where a detonation in equal mass carbon-oxygen material ceases to burn
completely to NSE and instead produces IMEs as a result of incomplete
burning.

\nuc{58}{Ni}, which is by far the most abundant stable nickel isotope
in all models, shows remarkably little variation.  {The
  models produce \nuc{58}{Ni} masses from 6.2 to $7.5 \times 10^{-2}\
  \msun$. The only exception is the low central density model N100L,
  which undergoes less in situ neutronization and only synthesizes
  $3.8 \times 10^{-2}\ \msun$ of \nuc{58}{Ni}.}

\begin{figure}
  \includegraphics[width=\columnwidth]{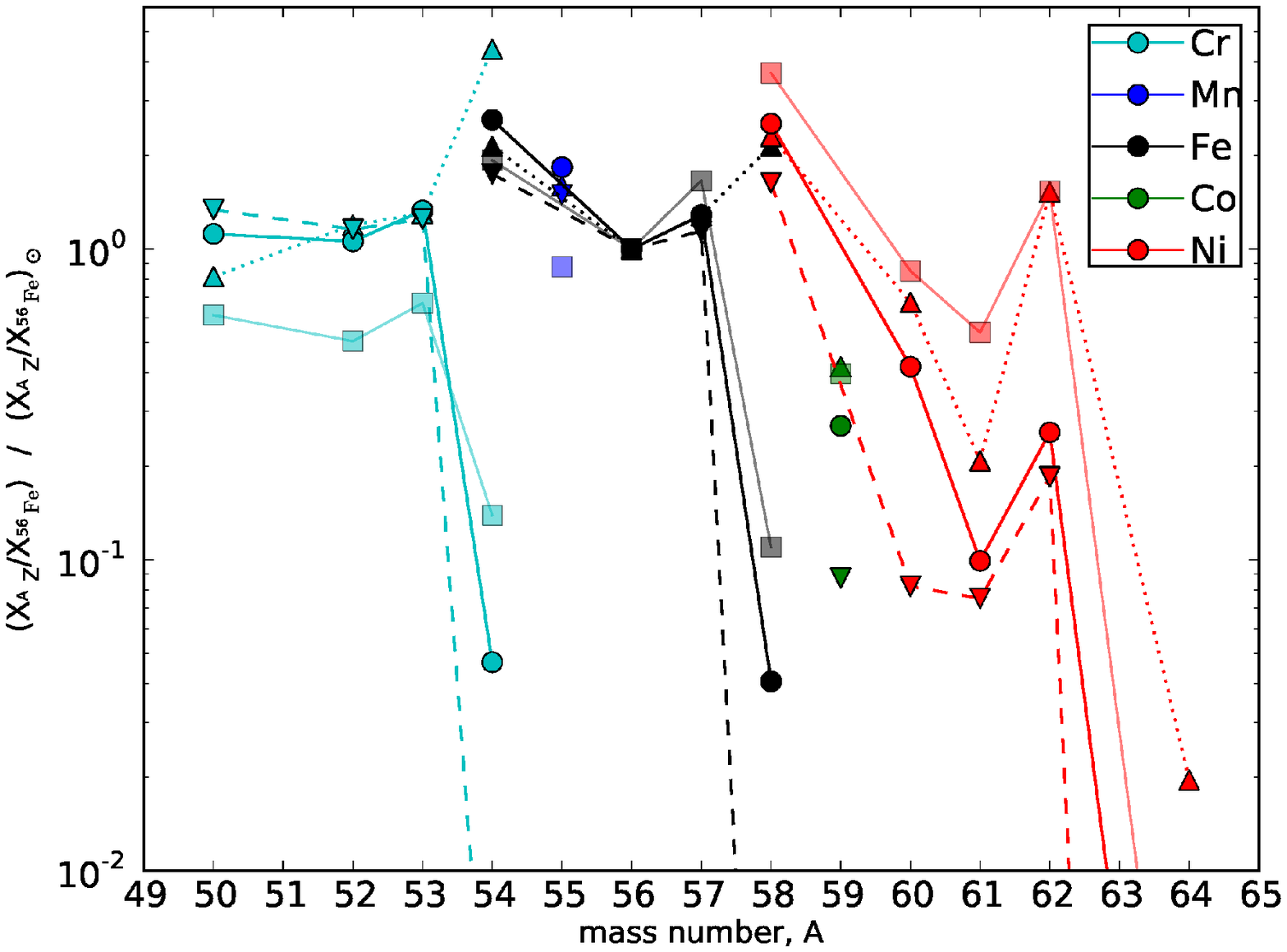}
  \caption{Production factors
    $(\nicefrac{X_\nuc{A}{Z}}{X_\nuc{56}{Fe}}) /
    (\nicefrac{X_\nuc{A}{Z}}{X_\nuc{56}{Fe}})_{\odot}$ of stable iron
    peak isotopes relative to \nuc{56}{Fe} normalized to the
    corresponding solar values \citep{lodders2003a} for models N100L
    (downward-pointing triangles/dashed lines), N100 (circles/solid
    lines), and N100H (upward-pointing triangles/dotted lines) and W7
    (transparent squares/thin solid lines; \citealt{maeda2010a}). The
    lower the central density, the smaller the overproduction of
    \nuc{54}{Fe} and \nuc{58}{Ni}.}
  \label{fig:solar_comp}
\end{figure}

While estimated \nuc{56}{Ni} masses of SNe~Ia cover quite a range,
most observed events cluster around $0.6\ \msun$
\citep[e.g.][]{stritzinger2006b}.  For this reason, we have chosen the
three N100 models for a comparison of their isotopic iron peak
production factors with the solar values,
$(\nicefrac{X_\nuc{A}{Z}}{X_\nuc{56}{Fe}}) /
(\nicefrac{X_\nuc{A}{Z}}{X_\nuc{56}{Fe}})_{\odot}$ (see
Fig.~\ref{fig:solar_comp}).  For the solar composition we use
\citet{lodders2003a}. Our model production factors exhibit
qualitatively the right behavior that is required for SN~Ia yields:
production factors of self-conjugate and slightly neutron rich
isotopes \nuc{50}{Cr}, \nuc{51}{Cr}, \nuc{52}{Cr}, \nuc{55}{Mn},
\nuc{54}{Fe}, and {\nuc{58}{Ni} only weakly depend on the
  particular choice of central density of the WD at the time of
  ignition.}
These isotopes are expected to be largely synthesized in nuclear
statistical equilibrium in SNe~Ia, which requires production factors
$\gtrsim 1$ as observed here. Large overproduction factors
($[\nicefrac{\nuc{A}{Z}}{\nuc{56}{Fe}}] \gtrsim 2$), are marginally in
conflict with the requirement that SNe~Ia produced roughly half of the
\nuc{56}{Fe} present in the Sun today \citep{clayton2003a}. We note,
however, that models with different central density at ignition yield
lower overproduction factors (Fig.~\ref{fig:solar_comp}) and that
lower progenitor metallicities also results in less neutron rich iron
group isotopes (see Section~\ref{sec:metallicity}).

The only isotope with an overproduction factor $> 3$ is \nuc{54}{Cr}
for the high central density model N100H. We interpret the large
overproduction factor we obtain there as an indication that delayed
detonation SNe~Ia that ignite at central densities $\gtrsim 5.5 \times
10^9 \gcc$ are rare and constitute at most a small fraction of all
SN~Ia events.  The most neutron-rich stable Fe-peak isotopes
\nuc{54}{Cr}, \nuc{58}{Fe}, and \nuc{64}{Ni} are shielded by
\nuc{54}{Fe}, \nuc{58}{Ni}, and \nuc{64}{Zn} from the $Z=N$ line and
thus require the most neutronization for direct production.
Consequently, these isotopes are the most sensitive to the central
density, with the highest production factors occurring for the N100H
model. The very pronounced underproduction of \nuc{58}{Fe} and
\nuc{64}{Ni} in all the models is not a problem, since the S-process
is the dominant source of nucleosynthesis for these isotopes.
{The solar abundances of the remaining isotopes
  \nuc{59}{Co}, \nuc{60}{Ni}, \nuc{61}{Ni}, and \nuc{62}{Ni} contain
  significant contributions from explosive Si-burning (core collapse
  SNe), alpha-rich freeze-out (including SNe~Ia), and the S-process.
  Our production factors for SNe~Ia on the order of a few to several
  tens of percent are therefore also very reasonable.}

For a detailed discussion of the nucleosynthesis origin in the Sun of
the isotopes discussed here see the book by \citet{clayton2003a},
which we have used as a reference for our comparison.

The subset of our three N100 models contains another interesting
trend. {We find that the \nuc{56}{Ni} mass in the N100
  model sequence increases with central density, i.e. the high central
  density model N100H produces the most and the low central density
  model N100L produces the least amount of \nuc{56}{Ni}.}  This is
exactly opposite to the behavior found by \citet{krueger2010a} when
they analyzed their statistical sample of 150 two-dimensional
delayed-detonation explosion simulations.  They considered a range of
central ignition densities (1 to $5 \times 10^9 \gcc$), a fixed
deflagration to detonation transition density of $10^{7.1} \gcc$, but
did not perform post-processing with a detailed nuclear reaction
network.

In our models, the fraction of NSE material \citep[for a recent
discussion of NSE in SN~Ia see][]{seitenzahl2009a} that is produced as
\nuc{56}{Ni} is highest for the low central density model and lowest
for the high central density model, which is what is expected.  The
total amount of iron group material synthesized is, however, strongly
increasing with central density. We have seen the same trend already
in \citet{seitenzahl2011a}, where we had estimated a roughly constant
\nuc{56}{Ni} yield as a function of central density. The \nuc{56}{Ni}
mass in that work was derived from the total mass in IGE and the
electron fraction \ye, essentially using the formula in
\citet{timmes2003a}.  Here we determine the nucleosynthesis in detail
with one million tracer particles and we find that the \nuc{56}{Ni}
mass increases with increasing central density.  We caution, however,
that our sample consisting of three models in a single ignition
configuration is not statistically significant.

\subsection{Dependence of yields on progenitor metallicity}
\label{sec:metallicity}
To assess the impact of varying the progenitor metallicity, we have
also post-processed the N100 model with one-half, one tenth, and one
hundredth of the canonical \nuc{22}{Ne} mass fraction of 0.025.  The
\nuc{12}{C} mass fractions were thus 0.4875, 0.4975 and 0.49975 and
X(\nuc{16}{O}) was kept constant at 0.5.  The models with reduced
\nuc{22}{Ne} are called N100{\textunderscore}Z0.5,
N100{\textunderscore}Z0.1, and N100{\textunderscore}Z0.01 respectively
and their yields are also presented in
Tables~\ref{tab:syields}~and~\ref{tab:ryields}.

\begin{figure}
  \includegraphics[width=\columnwidth]{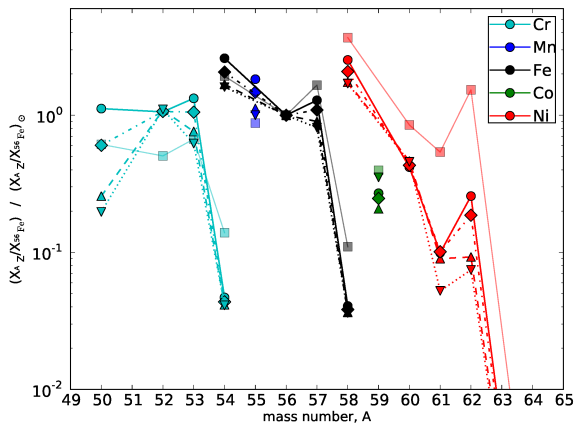}
  \caption{Production factors
    $(\nicefrac{X_\nuc{A}{Z}}{X_\nuc{56}{Fe}}) /
    (\nicefrac{X_\nuc{A}{Z}}{X_\nuc{56}{Fe}})_{\odot}$ of stable iron
    peak isotopes relative to \nuc{56}{Fe} normalized to the
    corresponding solar values \citep{lodders2003a} for models N100
    (circles/solid lines), N100{\textunderscore}Z0.5
    (diamonds/dot-dashed lines), N100{\textunderscore}Z0.1
    (upward-pointing triangles/dashed lines),
    N100{\textunderscore}Z0.01 (downward-pointing triangles/dotted
    lines) and W7 (transparent squares/thin solid lines;
    \citealt{maeda2010a}). The smaller the initial \nuc{22}{Ne} mass
    fraction, the smaller the overproduction of \nuc{54}{Fe},
    \nuc{55}{Mn}, and \nuc{58}{Ni}.}
  \label{fig:solar_comp_metal}
\end{figure}

As expected \citep{timmes2003a, travaglio2005a}, the \nuc{56}{Ni}
yields increases with decreasing initial \nuc{22}{Ne} due to the
decreasing electron fraction \ye, largely at the cost of stable iron
group isotopes such as \nuc{58}{Ni} or \nuc{54}{Fe}. As a consequence,
the overproduction factors of these isotopes are successively reduced
to $\lesssim 2$ even for the canonical central density case
(Fig.~\ref{fig:solar_comp_metal}).  Since the \nuc{56}{Fe} in the Sun
is largely due to supernovae that had progenitors with sub-solar
metallicity, we argue that the isotopic Fe-group yields of our
delayed-detonations are not inconsistent with solar isotopic ratios.

\subsection{Spatial and velocity distribution of nucleosynthetic
  yields}
\label{sec:distyields}
Total integrated yields constitute very important predictions and
diagnostics for an explosion model. How the yields are distributed in
mass and velocity space is, however, at least equally important,
especially when it comes to direct observables of the supernova such
as spectra or light curves.

It is not possible to show the spatial yield distributions of all
nuclei for all models. Therefore, we restrict ourselves to show only
\nuc{56}{Ni} (as the main source of radioactive heating),
\nuc{54}{Fe}+\nuc{56}{Fe}+\nuc{58}{Ni} (as most abundant stable IGE s
and important sources of opacity), \nuc{28}{Si} (as the generally most
abundant intermediate mass isotope) and the fuel isotopes \nuc{16}{O}
and \nuc{12}{C}, for a representative sample of the four models N3,
N40, N300C, and N1600 (see Fig.~\ref{fig:tracer}).  Globally, the
abundances are stratified as is expected for delayed-detonations: Near
the surface some unburned carbon sits on top of an oxygen-rich layer
that is composed of unburned oxygen fuel and products of low-density
carbon burning.  Further inwards IMEs such as \nuc{28}{Si} and
\nuc{32}{S} are the most abundant species, although a strict
onion-shell structure as is seen in one-dimensional models is not
present.  Instead, Rayleigh-Taylor plumes of material rich in stable
IGEs have penetrated into this layer.  Consequently, stable IGEs are
not found at the lowest velocities as predicted by one-dimensional
models \citep[e.g.][]{nomoto1984a,khokhlov1991b,hoeflich2004a}, but
rather at intermediate velocities (${\sim}3,000 - 10,000 \kms$).  This
agrees well with the results of \citet{maeda2010a}, who find typical
velocities of $5,000 - 10,000 \kms$ for the deflagration ash in their
two-dimensional O-DDT model.  This signature of the burning in the
deflagration is characteristic to multi-dimensional simulations that
are not ignited at the very center.  Unlike in the one-dimensional
case, the hot, less dense, buoyant ash can float towards the
surface. We note that \citet{stehle2005a} find substantial amounts of
stable iron out to velocities of about $9,000$ analyzing spectra of
SN~2002bo. In contrast to our model of comparable brightness (N100),
in their abundance tomography, which is based on one-dimensional
explosion models, the abundance of stable iron increases steadily
towards lower velocities and dominates in the very center.

In between the IGE-rich plumes, small pockets filled with
oxygen-rich material remain -- these were downdrafts in the deflagration that
were burned only in the latest phases of the detonation at low
densities.  The central regions, on the other hand, which were burned
by the detonation at high densities to NSE, form a homogeneous
\nuc{56}{Ni} clump. In the fainter models, this clump fragments
progressively.  This is a natural consequence of the fact that in
these models the majority of IGE material is produced by the
deflagration and not by the detonation.  In these models, the strong
initial expansion due to the deflagration results in such low
densities in the core that the ensuing detonation fails to process
this material to NSE (see Fig.~\ref{fig:tracer}(p)).  We also show the
same set of mass fractions for the three N100 models (see
Fig.~\ref{fig:tracer2}).  The global picture of chemical
stratification discussed above is rather insensitive to central
ignition density.

\begin{figure*}
  \includegraphics[width=2.1\columnwidth,trim=2pt 35pt 2pt
  35pt,clip]{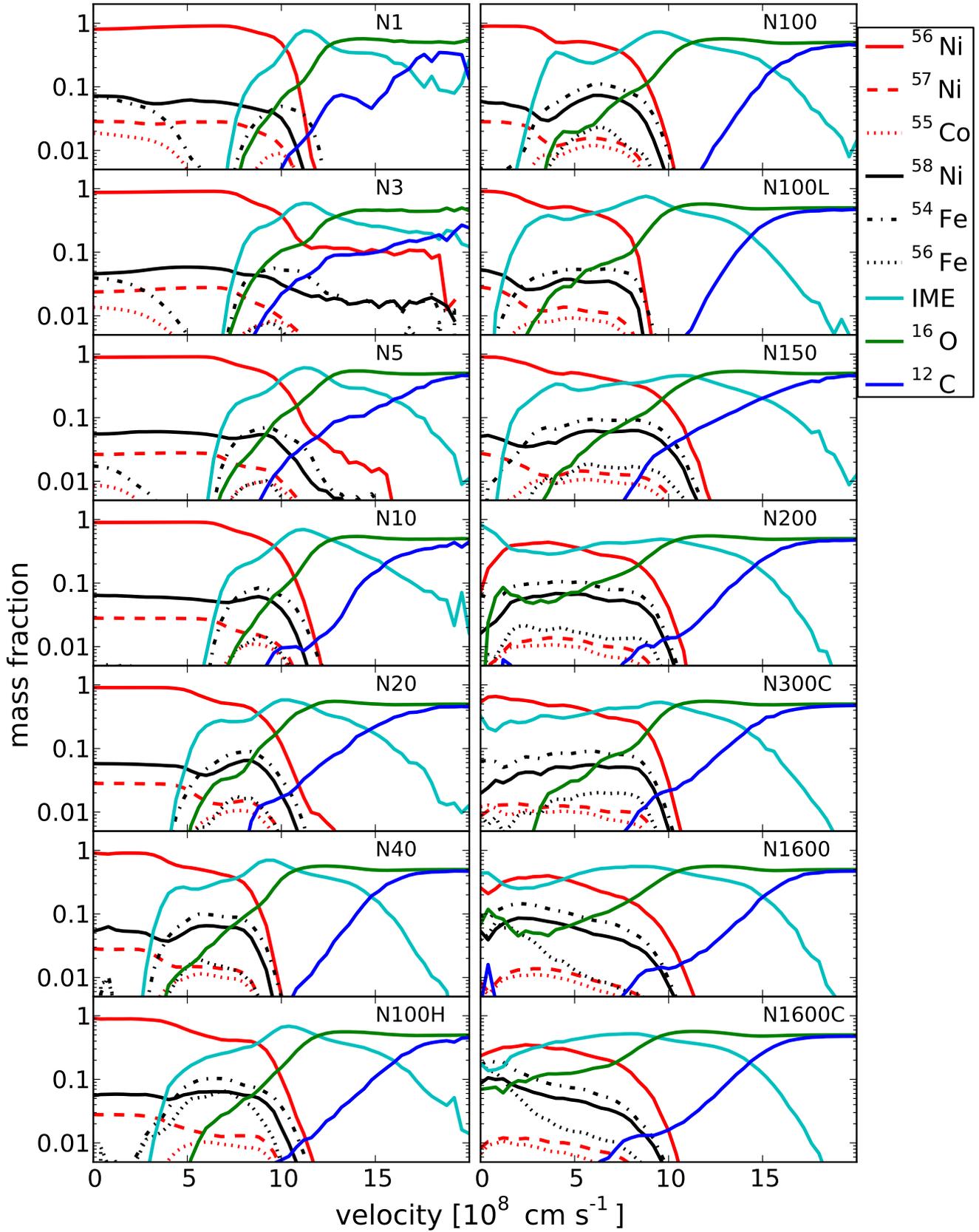}
  \caption{Shown are average mass fractions of some select isotopes
    and IME at time $t=100\s$ in velocity space.}
  \label{fig:fig1}
\end{figure*}

To better visualize the underlying main trends of the whole model
suite, and to facilitate a comparison of the three-dimensional models,
we reduce the information and show one-dimensional abundance profiles
in velocity space (see Fig.~\ref{fig:fig1}).  We emphasize that this
averaging or binning of the three dimensional data erases all
information about the inhomogeneities and (sometimes pronounced)
asymmetries brought about by e.g. the rising plumes of deflagration
ash or downdrafts of nuclear fuel.  This is also true for models with
a strong (turbulent) deflagration phase, which are rather symmetric
under rotations on large scales, but exhibit strong inhomogeneities in
the burning products on small scales (see the different morphology
plots in Sec.~\ref{sec:morph}).  For example, the presence of
e.g. \nuc{16}{O} and \nuc{56}{Ni} in these one-dimensional profiles at
the same velocity does not necessarily imply a co-spatial existence of
these nuclear species.

\citet{kozma2005a} showed that ejecta with oxygen at low velocities
(as we find it for our models with large $N_\mathrm{k}$) may lead to
strong [\ions{O}{i}] $\lambda\lambda$6300,6364 emission at late times,
which is not observed in SNe~Ia.  The mere presence of oxygen at low
velocities, however, is not the only relevant condition for
[\ions{O}{i}] emission.  Whether [\ions{O}{i}] features will arise,
depends strongly on the ionisation state and a possible microscopic
mixing of different species.  Compared to the deflagration model of
\citet{kozma2005a}, our delayed detonation models have lower density
ejecta, which could lead to a higher ionization.  The question of
microscopic mixing cannot be answered from present-day numerical
models since it takes place on scales which are not resolved.
However, if such a mixing is present, stronger transitions than
[\ions{O}{i}] will dominate the cooling.

As a general trend, \nuc{56}{Ni} is hardly present at velocities above
${\sim}12,000 \kms$.\footnote{One should be aware of the fact that
  already traces of IGEs on the equivalent level of solar abundance
  may affect the observables significantly.}  The exceptions are the
models N3 and N5, where parts of the asymmetrically rising
deflagration plumes have already risen to the stellar surface when the
first DDT occurs (see Fig.~\ref{fig:lsets}c).  In these models, the
deflagration ash is therefore not completely enclosed by burning
products of the detonation and there is some \nuc{56}{Ni} present at
very high velocity.  \nuc{57}{Ni} and \nuc{58}{Ni} are produced
co-spatially with \nuc{56}{Ni} and more or less follow its
distribution, albeit with a lower abundance.

The distribution of IMEs in velocity space shows a clear trend with
ignition kernel number.  The brightest models (small number of
ignition kernels) contain IMEs from the highest expansion velocities
down to ${\sim}6,000$--$7,000 \kms$.  For the fainter models (larger
number of ignition kernels) the inner boundary of the IMEs continues
to move inward and the IMEs at the highest velocities are more and
more replaced by unburned fuel \citep[see also][]{mazzali2007a}.

Notably, carbon and oxygen are extending down to low velocities. There
is always some carbon present down to or even below velocities of
$12,000 \kms$. Note that this is much lower than what has been found
for W7 \citep{nomoto1984a,iwamoto1999a,maeda2010a}.  Oxygen reaches
down even farther, always present below ${\sim}8,000 \kms$ and
sometimes even reaching to the very center. The presence of unburned
fuel in our models is a distinct multi-dimensional effect.  For
fainter models, the products of low-density detonation burning extend
down to lower velocities.  In extreme cases, pockets of unburned fuel
remain near the center that are not reached by the detonation at all.

The other abundant iron group isotopes \nuc{55}{Co}, \nuc{54}{Fe}, and
\nuc{56}{Fe} are mainly produced in the deflagration phase. As a
result, they are concentrated in an off center shell surrounding the
central \nuc{56}{Ni} bubble (see
Figs.~\ref{fig:tracer}--\ref{fig:fig1}).  Only models with a weak
deflagration phase (in particular N1, N3, N5) exhibit high enough
central densities during the detonation phase that some neutronization
and normal freeze-out from NSE also synthesizes significant amounts of
these isotopes in a second, centrally concentrated production site
(see Fig.~\ref{fig:fig1}).

{The idea suggests itself to compare the yield
  morphologies of our three-dimensional simulations to existing
  two-dimensional results from the literature. Such a comparison,
  however, is difficult to make quantitatively. For example, the
  results of \citet{kasen2006a} were not based on detailed yield
  information obtained by post-processing each model but
  mapped representative yields into the individual realizations.  A
  qualitative discussion, however, is possible: 
  Comparing Fig.~\ref{fig:tracer} to their figure 1, we can, however, identify a
  key commonality.  In both sets, the stable IGEs in bright models are 
  most abundant at intermediate velocities, surrounding a core rich in \nuc{56}{Ni}. 
  Conversely, the fainter models have the stable IGE much closer to the center.    
  To the best of our knowledge, the only published results of multi-dimensional
  delayed-detonations with detailed nucleosynthetic yields are the two
  models of \citet{maeda2010a}.  A meaningful comparison to their
  C-DDT model, which is based on a rather symmetrically and centrally
  ignited deflagration, is not possible.  In their C-DDT model, the
  DDT occurs at such a late time and low density that no more
  \nuc{56}{Ni} is produced by the detonation wave, resulting in a very
  low mass of final \nuc{56}{Ni} of ${\sim}0.25\msun$.  This is both
  outside the range of \nuc{56}{Ni} masses covered by our models as
  well as observed ``normal'' SNe~Ia.  On the other hand, their O-DDT
  model, which is based on an asymmetric ignition of the deflagration,
  produces ${\sim}0.54\msun$ of \nuc{56}{Ni}, which is rather similar
  to our N150 or N100 models.  In spite of the different symmetries in
  the ignition configuration, the angle averaged distribution of
  nuclides in velocity space (compare figure 11(c) of
  \citet{maeda2010a} and Fig.~\ref{fig:tracer}) are quite
  similar. In both O-DDT and our models, the lowest velocities are dominated by
  \nuc{56}{Ni} and \nuc{58}{Ni}, IMEs are present down to velocities
  of a few thousand $\kms$, \nuc{16}{O} down to just below $5,000 \kms$,
  and unburned \nuc{12}{C} down to around ${\sim}10,000 \kms$.  Stable
  IGE are only found at intermediate velocities
  between ${\sim}5,000 - 11,000 \kms$, and \nuc{56}{Ni} cuts off
  sharply around $11,000 \kms$.}
  
{In 2D axisymmtric simulations, the rising deflagration
  ``bubbles'' are in fact ``tori''.  This leads to differences in the 
  morphology, even if angle averages look alike.  3D models allow for 
  variation on smaller angular scales.  A rising torus is a more
  symmetric and also more extended object than a rising bubble. 
  Since it contains much more volume than a single bubble, a 2D 
  deflagration starting from the same ignition site releases more
  energy and is more symmetric than the corresponding case in 3D. 
  This may explain why the 2D O-DDT model, which was ignited in a rather
  asymmetric configuration in only 29 spots, ends up looking rather
  symmetric and not nearly as bright as e.g. our N20 or N40 model.}
  
\begin{table*}
\caption{Asymptotic nucleosynthetic yields (in solar masses) of stable nuclides.\label{tab:syields}}
\scriptsize
\begin{tabular}{cccccccccccccccccc}
\hline
                       &
           N1                  &
           N3                  &
           N5                  &
           N10                 &
           N20                 &
           N40                 &
           N100H               &
           N100                &
           N100L               &
           N150                &
           N200                &
           N300C               &
           N1600               &
           N1600C              &
           N100{\textunderscore}Z0.5            &
           N100{\textunderscore}Z0.1            &
           N100{\textunderscore}Z0.01           \\ \hline
\nuc{12}{C}  &2.61E-03 & 9.90E-03 & 9.05E-03 & 4.43E-03 & 9.20E-03 & 3.90E-03 & 3.87E-03 & 3.04E-03 & 3.85E-03 & 1.72E-02 & 1.21E-02 & 8.86E-03 & 1.06E-02 & 1.68E-02&3.10e-03&3.15e-03&3.16e-03\\
\nuc{13}{C}  &1.84E-08 & 6.15E-08 & 5.05E-08 & 2.57E-08 & 4.52E-08 & 2.18E-08 & 2.28E-08 & 1.74E-08 & 2.17E-08 & 1.00E-07 & 6.57E-08 & 4.95E-08 & 5.57E-08 & 8.44E-08&8.47e-09&1.91e-09&2.72e-10\\
\nuc{14}{N}  &2.92E-06 & 9.93E-06 & 8.46E-06 & 3.85E-06 & 8.34E-06 & 4.30E-06 & 4.25E-06 & 3.21E-06 & 3.98E-06 & 1.84E-05 & 1.33E-05 & 9.16E-06 & 1.04E-05 & 1.88E-05&1.80e-06&4.71e-07&7.22e-08\\
\nuc{15}{N}  &3.36E-09 & 1.22E-08 & 1.03E-08 & 4.47E-09 & 1.03E-08 & 5.16E-09 & 5.24E-09 & 3.67E-09 & 4.66E-09 & 2.29E-08 & 1.71E-08 & 1.14E-08 & 1.29E-08 & 2.41E-08&2.07e-09&2.98e-09&8.73e-08\\
\nuc{16}{O}  &2.63E-02 & 4.74E-02 & 5.63E-02 & 5.16E-02 & 9.04E-02 & 9.89E-02 & 7.30E-02 & 1.01E-01 & 1.24E-01 & 1.24E-01 & 1.96E-01 & 1.21E-01 & 1.91E-01 & 2.72E-01&9.87e-02&9.64e-02&9.47e-02\\
\hdashline[1pt/5pt]
\noalign{\smallskip}
\nuc{17}{O}  &3.96E-07 & 1.37E-06 & 1.16E-06 & 5.34E-07 & 1.12E-06 & 5.54E-07 & 5.61E-07 & 4.13E-07 & 5.14E-07 & 2.48E-06 & 1.74E-06 & 1.22E-06 & 1.36E-06 & 2.42E-06&2.84e-07&9.32e-08&5.43e-09\\
\nuc{18}{O}  &3.32E-09 & 1.33E-08 & 1.11E-08 & 4.54E-09 & 1.12E-08 & 5.29E-09 & 5.59E-09 & 3.53E-09 & 4.61E-09 & 2.52E-08 & 1.98E-08 & 1.25E-08 & 1.44E-08 & 2.73E-08&2.23e-09&1.21e-09&9.93e-10\\
\nuc{19}{F}  &3.73E-11 & 1.35E-10 & 1.17E-10 & 5.05E-11 & 1.22E-10 & 6.22E-11 & 6.32E-11 & 4.39E-11 & 5.68E-11 & 2.64E-10 & 2.13E-10 & 1.36E-10 & 1.58E-10 & 2.97E-10&2.20e-11&1.48e-11&4.79e-11\\
\nuc{20}{Ne} &1.47E-03 & 3.37E-03 & 3.75E-03 & 2.40E-03 & 5.41E-03 & 4.15E-03 & 3.66E-03 & 3.53E-03 & 4.33E-03 & 8.72E-03 & 1.15E-02 & 6.76E-03 & 9.40E-03 & 1.73E-02&3.60e-03&3.69e-03&3.74e-03\\
\nuc{21}{Ne} &3.08E-07 & 9.79E-07 & 8.81E-07 & 4.16E-07 & 9.63E-07 & 5.43E-07 & 5.20E-07 & 4.11E-07 & 5.17E-07 & 1.98E-06 & 1.68E-06 & 1.08E-06 & 1.29E-06 & 2.39E-06&1.97e-07&4.47e-08&6.93e-09\\
\hdashline[1pt/5pt]
\noalign{\smallskip}
\nuc{22}{Ne} &6.40E-05 & 3.26E-04 & 2.87E-04 & 1.31E-04 & 2.58E-04 & 5.77E-05 & 7.62E-05 & 4.07E-05 & 5.51E-05 & 4.83E-04 & 2.34E-04 & 2.11E-04 & 2.32E-04 & 2.97E-04&1.65e-05&2.30e-06&1.71e-07\\
\nuc{23}{Na} &2.20E-05 & 6.41E-05 & 6.09E-05 & 3.22E-05 & 7.30E-05 & 4.66E-05 & 4.25E-05 & 3.74E-05 & 4.68E-05 & 1.38E-04 & 1.38E-04 & 8.53E-05 & 1.09E-04 & 2.02E-04&2.63e-05&1.96e-05&1.72e-05\\
\nuc{24}{Mg} &3.93E-03 & 7.13E-03 & 8.53E-03 & 7.77E-03 & 1.46E-02 & 1.54E-02 & 1.15E-02 & 1.52E-02 & 1.83E-02 & 1.93E-02 & 3.32E-02 & 1.97E-02 & 3.08E-02 & 4.64E-02&2.02e-02&2.69e-02&2.90e-02\\
\nuc{25}{Mg} &3.35E-05 & 9.26E-05 & 8.92E-05 & 5.07E-05 & 1.11E-04 & 7.70E-05 & 6.86E-05 & 6.49E-05 & 8.02E-05 & 2.02E-04 & 2.14E-04 & 1.33E-04 & 1.75E-04 & 3.12E-04&3.09e-05&1.06e-05&8.99e-07\\
\nuc{26}{Mg} &5.15E-05 & 1.36E-04 & 1.34E-04 & 7.55E-05 & 1.71E-04 & 1.17E-04 & 1.04E-04 & 9.66E-05 & 1.19E-04 & 3.07E-04 & 3.27E-04 & 2.01E-04 & 2.63E-04 & 4.82E-04&4.44e-05&7.36e-06&1.04e-06\\
\hdashline[1pt/5pt]
\noalign{\smallskip}
\nuc{27}{Al} &1.98E-04 & 3.95E-04 & 4.56E-04 & 3.71E-04 & 7.32E-04 & 7.05E-04 & 5.47E-04 & 6.74E-04 & 8.32E-04 & 1.04E-03 & 1.64E-03 & 9.68E-04 & 1.48E-03 & 2.37E-03&5.88e-04&2.68e-04&8.71e-05\\
\nuc{28}{Si} &6.32E-02 & 8.99E-02 & 1.19E-01 & 1.38E-01 & 1.98E-01 & 2.59E-01 & 2.12E-01 & 2.84E-01 & 3.55E-01 & 2.71E-01 & 3.28E-01 & 3.19E-01 & 3.61E-01 & 3.44E-01&2.90e-01&2.94e-01&2.89e-01\\
\nuc{29}{Si} &2.69E-04 & 4.64E-04 & 5.68E-04 & 5.17E-04 & 9.49E-04 & 1.03E-03 & 7.73E-04 & 1.03E-03 & 1.25E-03 & 1.30E-03 & 2.08E-03 & 1.29E-03 & 1.97E-03 & 2.86E-03&7.28e-04&4.30e-04&1.35e-04\\
\nuc{30}{Si} &5.96E-04 & 1.00E-03 & 1.23E-03 & 1.18E-03 & 2.12E-03 & 2.35E-03 & 1.72E-03 & 2.36E-03 & 2.86E-03 & 2.77E-03 & 4.76E-03 & 2.87E-03 & 4.53E-03 & 6.55E-03&1.19e-03&1.44e-04&1.84e-05\\
\nuc{31}{P}  &1.40E-04 & 2.28E-04 & 2.87E-04 & 2.78E-04 & 4.87E-04 & 5.60E-04 & 4.20E-04 & 5.77E-04 & 7.09E-04 & 6.59E-04 & 1.08E-03 & 6.85E-04 & 1.05E-03 & 1.47E-03&3.58e-04&1.05e-04&3.54e-05\\
\hdashline[1pt/5pt]
\noalign{\smallskip}
\nuc{32}{S}  &2.62E-02 & 3.70E-02 & 4.79E-02 & 5.74E-02 & 7.74E-02 & 1.01E-01 & 8.55E-02 & 1.11E-01 & 1.38E-01 & 1.07E-01 & 1.10E-01 & 1.27E-01 & 1.22E-01 & 1.03E-01&1.12e-01&1.12e-01&1.15e-01\\
\nuc{33}{S}  &7.51E-05 & 1.06E-04 & 1.42E-04 & 1.53E-04 & 2.43E-04 & 3.14E-04 & 2.37E-04 & 3.39E-04 & 4.21E-04 & 3.27E-04 & 5.23E-04 & 3.65E-04 & 5.47E-04 & 6.79E-04&2.39e-04&1.04e-04&4.57e-05\\
\nuc{34}{S}  &8.26E-04 & 1.16E-03 & 1.57E-03 & 1.75E-03 & 2.84E-03 & 3.73E-03 & 2.74E-03 & 4.04E-03 & 5.02E-03 & 3.68E-03 & 6.22E-03 & 4.22E-03 & 6.56E-03 & 8.06E-03&1.86e-03&2.60e-04&7.14e-06\\
\nuc{36}{S}  &8.12E-08 & 1.35E-07 & 1.65E-07 & 1.41E-07 & 2.54E-07 & 2.57E-07 & 1.89E-07 & 2.47E-07 & 3.05E-07 & 3.59E-07 & 5.42E-07 & 3.23E-07 & 4.97E-07 & 7.68E-07&3.86e-08&1.64e-09&1.73e-11\\
\nuc{35}{Cl} &4.29E-05 & 6.20E-05 & 8.27E-05 & 8.37E-05 & 1.35E-04 & 1.67E-04 & 1.27E-04 & 1.78E-04 & 2.27E-04 & 1.89E-04 & 2.89E-04 & 2.00E-04 & 2.98E-04 & 3.84E-04&9.91e-05&2.64e-05&5.65e-06\\
\hdashline[1pt/5pt]
\noalign{\smallskip}
\nuc{37}{Cl} &7.32E-06 & 9.62E-06 & 1.34E-05 & 1.53E-05 & 2.26E-05 & 3.14E-05 & 2.44E-05 & 3.51E-05 & 4.49E-05 & 3.14E-05 & 4.66E-05 & 3.63E-05 & 5.22E-05 & 5.75E-05&2.27e-05&9.14e-06&3.52e-06\\
\nuc{36}{Ar} &4.52E-03 & 6.36E-03 & 8.03E-03 & 9.89E-03 & 1.28E-02 & 1.61E-02 & 1.43E-02 & 1.77E-02 & 2.17E-02 & 1.76E-02 & 1.50E-02 & 2.08E-02 & 1.64E-02 & 1.23E-02&1.85e-02&1.92e-02&2.04e-02\\
\nuc{38}{Ar} &3.77E-04 & 5.15E-04 & 7.18E-04 & 8.01E-04 & 1.25E-03 & 1.72E-03 & 1.29E-03 & 1.91E-03 & 2.48E-03 & 1.69E-03 & 2.69E-03 & 1.96E-03 & 2.98E-03 & 3.42E-03&8.31e-04&1.19e-04&5.40e-06\\
\nuc{40}{Ar} &1.68E-09 & 2.45E-09 & 3.13E-09 & 2.90E-09 & 4.79E-09 & 5.27E-09 & 3.81E-09 & 5.21E-09 & 6.60E-09 & 6.64E-09 & 1.04E-08 & 6.29E-09 & 9.87E-09 & 1.48E-08&4.66e-10&9.87e-12&5.11e-14\\
\nuc{39}{K}  &2.26E-05 & 2.98E-05 & 4.12E-05 & 4.65E-05 & 6.80E-05 & 9.52E-05 & 7.40E-05 & 1.07E-04 & 1.39E-04 & 9.53E-05 & 1.42E-04 & 1.10E-04 & 1.60E-04 & 1.75E-04&6.25e-05&1.89e-05&3.57e-06\\
\hdashline[1pt/5pt]
\noalign{\smallskip}
\nuc{41}{K}  &1.29E-06 & 1.65E-06 & 2.30E-06 & 2.66E-06 & 3.80E-06 & 5.38E-06 & 4.22E-06 & 6.08E-06 & 7.87E-06 & 5.36E-06 & 7.81E-06 & 6.21E-06 & 8.93E-06 & 9.65E-06&3.85e-06&1.42e-06&4.92e-07\\
\nuc{40}{Ca} &4.05E-03 & 5.74E-03 & 7.09E-03 & 8.82E-03 & 1.13E-02 & 1.35E-02 & 1.24E-02 & 1.47E-02 & 1.75E-02 & 1.50E-02 & 1.07E-02 & 1.78E-02 & 1.10E-02 & 7.50E-03&1.57e-02&1.66e-02&1.77e-02\\
\nuc{42}{Ca} &8.32E-06 & 1.11E-05 & 1.55E-05 & 1.74E-05 & 2.65E-05 & 3.69E-05 & 2.80E-05 & 4.15E-05 & 5.44E-05 & 3.63E-05 & 5.71E-05 & 4.22E-05 & 6.44E-05 & 7.30E-05&1.67e-05&2.08e-06&8.87e-08\\
\nuc{43}{Ca} &1.63E-08 & 1.93E-08 & 2.63E-08 & 2.92E-08 & 4.29E-08 & 5.62E-08 & 4.37E-08 & 6.12E-08 & 7.44E-08 & 5.57E-08 & 8.61E-08 & 6.03E-08 & 9.21E-08 & 1.14E-07&2.35e-08&6.22e-09&6.23e-09\\
\nuc{44}{Ca} &3.25E-06 & 4.40E-06 & 5.24E-06 & 6.49E-06 & 8.14E-06 & 9.14E-06 & 8.92E-06 & 1.00E-05 & 1.11E-05 & 1.04E-05 & 6.45E-06 & 1.19E-05 & 6.08E-06 & 4.06E-06&1.08e-05&1.18e-05&1.27e-05\\
\hdashline[1pt/5pt]
\noalign{\smallskip}
\nuc{46}{Ca} &1.58E-11 & 1.99E-11 & 2.66E-11 & 2.64E-11 & 4.01E-11 & 4.76E-11 & 7.13E-10 & 4.85E-11 & 6.06E-11 & 5.50E-11 & 8.60E-11 & 5.65E-11 & 8.43E-11 & 1.24E-10&2.24e-12&5.88e-13&5.82e-13\\
\nuc{48}{Ca} &1.97E-15 & 3.88E-15 & 7.61E-15 & 7.56E-15 & 1.86E-14 & 5.84E-15 & 9.97E-11 & 7.85E-15 & 3.04E-15 & 6.17E-15 & 5.21E-15 & 8.89E-14 & 4.52E-15 & 6.99E-15&5.04e-15&4.85e-15&4.82e-15\\
\nuc{45}{Sc} &5.12E-08 & 6.53E-08 & 8.56E-08 & 1.03E-07 & 1.36E-07 & 1.83E-07 & 1.56E-07 & 2.05E-07 & 2.51E-07 & 1.92E-07 & 2.20E-07 & 2.20E-07 & 2.48E-07 & 2.45E-07&1.43e-07&7.69e-08&3.94e-08\\
\nuc{46}{Ti} &4.33E-06 & 5.72E-06 & 8.01E-06 & 9.13E-06 & 1.33E-05 & 1.87E-05 & 1.46E-05 & 2.11E-05 & 2.75E-05 & 1.86E-05 & 2.72E-05 & 2.17E-05 & 3.11E-05 & 3.34E-05&9.34e-06&1.37e-06&2.19e-07\\
\nuc{47}{Ti} &1.80E-07 & 2.28E-07 & 3.01E-07 & 3.58E-07 & 4.84E-07 & 6.33E-07 & 5.44E-07 & 7.05E-07 & 8.51E-07 & 6.53E-07 & 7.66E-07 & 7.44E-07 & 8.46E-07 & 8.59E-07&3.37e-07&9.30e-08&9.90e-08\\
\hdashline[1pt/5pt]
\noalign{\smallskip}
\nuc{48}{Ti} &1.02E-04 & 1.41E-04 & 1.67E-04 & 2.07E-04 & 2.67E-04 & 2.85E-04 & 2.87E-04 & 3.14E-04 & 3.36E-04 & 3.31E-04 & 1.80E-04 & 3.78E-04 & 1.50E-04 & 9.30E-05&3.43e-04&3.73e-04&3.87e-04\\
\nuc{49}{Ti} &8.06E-06 & 1.07E-05 & 1.29E-05 & 1.61E-05 & 2.07E-05 & 2.35E-05 & 2.35E-05 & 2.59E-05 & 2.76E-05 & 2.69E-05 & 1.72E-05 & 3.05E-05 & 1.62E-05 & 1.03E-05&2.06e-05&1.27e-05&6.83e-06\\
\nuc{50}{Ti} &2.48E-08 & 5.52E-08 & 1.33E-07 & 1.44E-07 & 3.43E-07 & 1.62E-07 & 1.02E-04 & 2.60E-07 & 6.88E-10 & 1.61E-07 & 4.58E-08 & 1.25E-06 & 1.19E-09 & 1.72E-09&2.52e-07&2.47e-07&2.46e-07\\
\nuc{50}{V}  &8.95E-10 & 1.21E-09 & 1.94E-09 & 2.34E-09 & 3.70E-09 & 4.68E-09 & 3.20E-08 & 5.56E-09 & 3.84E-09 & 5.07E-09 & 6.16E-09 & 6.72E-09 & 6.11E-09 & 7.52E-09&3.01e-09&2.30e-09&2.28e-09\\
\nuc{51}{V}  &3.41E-05 & 4.14E-05 & 5.01E-05 & 6.18E-05 & 8.06E-05 & 9.51E-05 & 1.51E-04 & 1.05E-04 & 1.01E-04 & 1.09E-04 & 8.19E-05 & 1.20E-04 & 8.09E-05 & 6.01E-05&7.65e-05&3.99e-05&2.62e-05\\
\hdashline[1pt/5pt]
\noalign{\smallskip}
\nuc{50}{Cr} &1.01E-04 & 1.39E-04 & 1.84E-04 & 2.27E-04 & 3.08E-04 & 4.06E-04 & 3.95E-04 & 4.51E-04 & 4.63E-04 & 4.47E-04 & 4.48E-04 & 5.05E-04 & 5.08E-04 & 4.01E-04&2.53e-04&1.11e-04&8.59e-05\\
\nuc{52}{Cr} &3.55E-03 & 4.06E-03 & 4.56E-03 & 5.45E-03 & 7.23E-03 & 7.70E-03 & 1.16E-02 & 8.57E-03 & 8.02E-03 & 8.90E-03 & 5.62E-03 & 9.85E-03 & 4.78E-03 & 3.94E-03&8.97e-03&9.44e-03&9.64e-03\\
\nuc{53}{Cr} &5.25E-04 & 5.27E-04 & 5.91E-04 & 7.09E-04 & 9.48E-04 & 1.12E-03 & 1.46E-03 & 1.24E-03 & 9.98E-04 & 1.29E-03 & 1.06E-03 & 1.36E-03 & 1.06E-03 & 9.20E-04&1.02e-03&7.61e-04&6.27e-04\\
\nuc{54}{Cr} &9.85E-07 & 2.06E-06 & 4.54E-06 & 5.18E-06 & 1.08E-05 & 7.59E-06 & 1.25E-03 & 2.61E-07 & 1.11E-05 & 8.55E-06 & 4.94E-06 & 3.24E-05 & 3.17E-06 & 4.06E-06&1.07e-05&1.05e-05&1.05e-05\\
\nuc{55}{Mn} &8.74E-03 & 6.93E-03 & 6.80E-03 & 7.61E-03 & 9.70E-03 & 1.21E-02 & 1.39E-02 & 9.29E-03 & 1.33E-02 & 1.39E-02 & 1.31E-02 & 1.35E-02 & 1.31E-02 & 1.13E-02&1.11e-02&8.73e-03&7.84e-03\\
\hdashline[1pt/5pt]
\noalign{\smallskip}
\nuc{54}{Fe} &3.02E-02 & 3.07E-02 & 3.84E-02 & 4.77E-02 & 6.71E-02 & 9.00E-02 & 9.80E-02 & 9.94E-02 & 5.70E-02 & 1.04E-01 & 1.13E-01 & 1.03E-01 & 1.22E-01 & 1.00E-01&8.19e-02&6.87e-02&6.62e-02\\
\nuc{56}{Fe} &1.11E-00 & 1.04E-00 & 9.80E-01 & 9.46E-01 & 7.90E-01 & 6.70E-01 & 7.47E-01 & 6.22E-01 & 5.33E-01 & 5.84E-01 & 4.36E-01 & 5.33E-01 & 3.90E-01 & 3.54E-01&6.46e-01&6.66e-01&6.72e-01\\
\nuc{57}{Fe} &3.37E-02 & 3.05E-02 & 2.91E-02 & 2.82E-02 & 2.31E-02 & 2.03E-02 & 2.25E-02 & 1.88E-02 & 1.43E-02 & 1.79E-02 & 1.53E-02 & 1.56E-02 & 1.45E-02 & 1.26E-02&1.66e-02&1.41e-02&1.28e-02\\
\nuc{58}{Fe} &7.14E-06 & 1.41E-05 & 3.21E-05 & 3.63E-05 & 7.37E-05 & 5.33E-05 & 5.09E-03 & 8.02E-05 & 1.27E-07 & 6.05E-05 & 2.83E-05 & 2.13E-04 & 7.02E-06 & 8.70E-06&7.85e-05&7.72e-05&7.70e-05\\
\nuc{59}{Co} &4.18E-04 & 4.97E-04 & 5.62E-04 & 5.98E-04 & 6.12E-04 & 5.59E-04 & 9.96E-04 & 5.34E-04 & 1.48E-04 & 5.23E-04 & 5.28E-04 & 5.03E-04 & 6.23E-04 & 6.20E-04&5.09e-04&4.40e-04&7.53e-04\\
\hdashline[1pt/5pt]
\noalign{\smallskip}
\nuc{58}{Ni} &7.26E-02 & 6.78E-02 & 6.98E-02 & 7.13E-02 & 6.63E-02 & 6.89E-02 & 7.54E-02 & 6.90E-02 & 3.81E-02 & 7.01E-02 & 7.29E-02 & 6.26E-02 & 7.48E-02 & 6.16E-02&5.90e-02&5.09e-02&5.01e-02\\
\nuc{60}{Ni} &2.01E-03 & 2.63E-03 & 3.22E-03 & 3.58E-03 & 4.31E-03 & 4.40E-03 & 8.79E-03 & 4.54E-03 & 7.64E-04 & 4.56E-03 & 4.67E-03 & 4.57E-03 & 5.79E-03 & 6.75E-03&4.87e-03&5.39e-03&5.36e-03\\
\nuc{61}{Ni} &8.45E-05 & 9.36E-05 & 1.02E-04 & 1.01E-04 & 9.18E-05 & 6.57E-05 & 1.20E-04 & 4.76E-05 & 3.09E-05 & 3.51E-05 & 1.69E-05 & 2.56E-05 & 1.48E-05 & 1.61E-05&5.04e-05&4.62e-05&2.71e-05\\
\nuc{62}{Ni} &6.55E-04 & 7.31E-04 & 8.10E-04 & 8.06E-04 & 7.49E-04 & 5.27E-04 & 2.85E-03 & 4.00E-04 & 2.46E-04 & 2.94E-04 & 1.31E-04 & 3.01E-04 & 9.83E-05 & 1.05E-04&3.01e-04&1.54e-04&1.25e-04\\
\nuc{64}{Ni} &4.24E-09 & 8.11E-09 & 1.94E-08 & 2.15E-08 & 4.72E-08 & 2.40E-08 & 9.61E-06 & 3.83E-08 & 5.26E-14 & 2.44E-08 & 5.97E-09 & 1.63E-07 & 1.78E-11 & 2.55E-11&3.72e-08&3.65e-08&3.63e-08\\
\hdashline[1pt/5pt]
\noalign{\smallskip}
\nuc{63}{Cu} &4.44E-07 & 4.94E-07 & 5.59E-07 & 5.71E-07 & 5.62E-07 & 4.43E-07 & 2.32E-06 & 3.78E-07 & 1.72E-07 & 3.20E-07 & 2.10E-07 & 3.23E-07 & 1.91E-07 & 2.29E-07&3.17e-07&3.27e-06&1.29e-05\\
\nuc{65}{Cu} &1.09E-07 & 1.15E-07 & 1.32E-07 & 1.32E-07 & 1.31E-07 & 1.05E-07 & 3.23E-07 & 8.02E-08 & 6.04E-08 & 6.14E-08 & 2.58E-08 & 4.13E-08 & 2.41E-08 & 2.77E-08&9.98e-08&1.26e-07&1.45e-07\\
\nuc{64}{Zn} &1.72E-06 & 1.83E-06 & 2.09E-06 & 2.11E-06 & 2.09E-06 & 1.71E-06 & 1.94E-06 & 1.32E-06 & 9.19E-07 & 1.06E-06 & 5.45E-07 & 6.42E-07 & 5.66E-07 & 6.38E-07&2.22e-06&1.73e-05&2.64e-05\\
\nuc{66}{Zn} &2.55E-06 & 2.70E-06 & 3.11E-06 & 3.13E-06 & 3.04E-06 & 2.41E-06 & 2.74E-06 & 1.79E-06 & 1.44E-06 & 1.39E-06 & 5.97E-07 & 7.64E-07 & 5.74E-07 & 6.45E-07&1.43e-06&6.29e-07&4.37e-07\\
\nuc{67}{Zn} &1.48E-09 & 1.53E-09 & 1.77E-09 & 1.79E-09 & 1.75E-09 & 1.42E-09 & 2.26E-09 & 1.09E-09 & 8.79E-10 & 8.61E-10 & 3.91E-10 & 4.91E-10 & 3.76E-10 & 4.24E-10&7.08e-10&2.47e-10&1.73e-10\\
\hdashline[1pt/5pt]
\noalign{\smallskip}
\nuc{68}{Zn} &9.01E-10 & 9.38E-10 & 1.10E-09 & 1.12E-09 & 1.13E-09 & 8.87E-10 & 1.07E-08 & 7.24E-10 & 5.26E-10 & 5.68E-10 & 2.58E-10 & 5.91E-10 & 2.29E-10 & 2.53E-10&3.70e-10&1.71e-10&1.51e-10\\
\nuc{70}{Zn} &3.39E-16 & 6.43E-16 & 1.52E-15 & 1.63E-15 & 3.67E-15 & 1.36E-15 & 2.23E-12 & 2.23E-15 & 5.97E-24 & 1.25E-15 & 1.76E-16 & 1.48E-14 & 7.40E-22 & 1.25E-21&2.16e-15&2.10e-15&2.09e-15\\
\nuc{69}{Ga} &5.66E-14 & 8.72E-14 & 1.77E-13 & 2.00E-13 & 3.30E-13 & 2.72E-13 & 1.04E-11 & 3.93E-13 & 6.71E-15 & 3.27E-13 & 1.39E-13 & 8.44E-13 & 9.10E-15 & 1.15E-14&3.81e-13&3.73e-13&3.72e-13\\
\nuc{71}{Ga} &2.06E-15 & 2.25E-15 & 3.31E-15 & 3.49E-15 & 4.73E-15 & 3.43E-15 & 3.47E-13 & 4.03E-15 & 1.22E-15 & 3.15E-15 & 1.24E-15 & 9.47E-15 & 6.14E-16 & 7.02E-16&3.06e-15&2.69e-15&2.66e-15\\ \hline
\normalsize
\end{tabular}
\end{table*}
 \begin{table*}
\caption{Nucleosynthetic yields (in solar masses) of select radioactive nuclides at time
  $t=100\s$.\label{tab:ryields}}
\scriptsize
\begin{tabular}{cccccccccccccccccc}
\hline
                       &
           N1                  &
           N3                  &
           N5                  &
           N10                  &
           N20                  &
           N40                  &
           N100H                 &
           N100                  &
           N100L                 &
           N150                  &
           N200                  &
           N300C                 &
           N1600                 &
           N1600C                &
           N100{\textunderscore}Z0.5              &
           N100{\textunderscore}Z0.1              &
           N100{\textunderscore}Z0.01             \\ \hline

\nuc{14}{C}  &2.25E-06 & 7.64E-06 & 6.50E-06 & 2.96E-06 & 6.39E-06 & 3.30E-06 & 3.26E-06 & 2.47E-06 &3.06E-06 & 1.41E-05 & 1.02E-05 & 7.01E-06 & 7.96E-06 & 1.44E-05 & 1.20e-06&1.50e-07&6.57e-10\\
\nuc{22}{Na} &1.90E-09 & 4.64E-09 & 4.99E-09 & 3.03E-09 & 6.99E-09 & 5.11E-09 & 4.67E-09 & 4.27E-09 &5.25E-09 & 1.17E-08 & 1.46E-08 & 8.67E-09 & 1.18E-08 & 2.20E-08 & 4.43e-09&4.60e-09&7.84e-09\\
\nuc{26}{Al} &2.35E-07 & 5.83E-07 & 6.13E-07 & 3.86E-07 & 8.62E-07 & 6.57E-07 & 5.72E-07 & 5.68E-07 &7.16E-07 & 1.42E-06 & 1.82E-06 & 1.08E-06 & 1.51E-06 & 2.70E-06 & 3.73e-07&2.70e-07&9.30e-08\\
\nuc{32}{Si} &6.49E-09 & 1.61E-08 & 1.52E-08 & 8.25E-09 & 1.78E-08 & 1.14E-08 & 9.99E-09 & 9.47E-09 &1.10E-08 & 3.48E-08 & 3.04E-08 & 2.01E-08 & 2.42E-08 & 4.38E-08 & 1.58e-09&1.53e-11&4.58e-15\\
\nuc{32}{P}  &1.46E-07 & 2.55E-07 & 2.99E-07 & 2.65E-07 & 4.80E-07 & 5.01E-07 & 3.77E-07 & 4.96E-07 &5.88E-07 & 6.68E-07 & 1.03E-06 & 6.29E-07 & 9.56E-07 & 1.42E-06 & 1.89e-07&2.43e-08&2.02e-10\\
\hdashline[1pt/5pt]
\noalign{\smallskip}
\nuc{33}{P}  &1.00E-07 & 1.67E-07 & 2.01E-07 & 1.93E-07 & 3.38E-07 & 3.74E-07 & 2.74E-07 & 3.76E-07 &4.54E-07 & 4.44E-07 & 7.57E-07 & 4.56E-07 & 7.17E-07 & 1.05E-06 & 1.39e-07&1.33e-08&7.44e-11\\
\nuc{35}{S}  &1.78E-07 & 2.87E-07 & 3.59E-07 & 3.08E-07 & 5.52E-07 & 5.60E-07 & 4.16E-07 & 5.39E-07 &6.74E-07 & 7.86E-07 & 1.19E-06 & 7.13E-07 & 1.09E-06 & 1.71E-06 & 8.66e-08&3.58e-09&2.53e-11\\
\nuc{36}{Cl} &2.15E-07 & 3.24E-07 & 4.13E-07 & 3.98E-07 & 6.72E-07 & 7.66E-07 & 5.61E-07 & 7.77E-07 &9.58E-07 & 9.13E-07 & 1.48E-06 & 9.15E-07 & 1.43E-06 & 2.07E-06 & 2.81e-07&3.59e-08&8.47e-10\\
\nuc{37}{Ar} &7.11E-06 & 9.32E-06 & 1.30E-05 & 1.49E-05 & 2.19E-05 & 3.06E-05 & 2.38E-05 & 3.43E-05 &4.39E-05 & 3.06E-05 & 4.52E-05 & 3.54E-05 & 5.08E-05 & 5.55E-05 & 2.25e-05&9.11e-06&3.51e-06\\
\nuc{39}{Ar} &4.13E-09 & 5.53E-09 & 7.30E-09 & 6.96E-09 & 1.13E-08 & 1.28E-08 & 9.56E-09 & 1.29E-08 &1.57E-08 & 1.58E-08 & 2.39E-08 & 1.50E-08 & 2.28E-08 & 3.30E-08 & 2.46e-09&1.19e-10&1.47e-12\\
\hdashline[1pt/5pt]
\noalign{\smallskip}
\nuc{40}{K}  &1.76E-08 & 2.39E-08 & 3.18E-08 & 3.07E-08 & 4.89E-08 & 5.68E-08 & 4.21E-08 & 5.81E-08 &7.39E-08 & 6.86E-08 & 1.06E-07 & 6.72E-08 & 1.04E-07 & 1.48E-07 & 1.48e-08&1.30e-09&5.80e-11\\
\nuc{41}{Ca} &1.28E-06 & 1.65E-06 & 2.30E-06 & 2.65E-06 & 3.80E-06 & 5.37E-06 & 4.22E-06 & 6.07E-06 &7.86E-06 & 5.35E-06 & 7.79E-06 & 6.20E-06 & 8.92E-06 & 9.63E-06 & 3.85e-06&1.42e-06&4.92e-07\\
\nuc{44}{Ti} &3.24E-06 & 4.38E-06 & 5.22E-06 & 6.47E-06 & 8.11E-06 & 9.10E-06 & 8.88E-06 & 9.98E-06 &1.11E-05 & 1.03E-05 & 6.37E-06 & 1.19E-05 & 6.00E-06 & 3.96E-06 & 1.08e-05&1.18e-05&1.27e-05\\
\nuc{48}{V}  &1.88E-08 & 2.47E-08 & 3.53E-08 & 4.29E-08 & 6.23E-08 & 8.03E-08 & 7.01E-08 & 9.12E-08 &1.09E-07 & 8.69E-08 & 9.80E-08 & 9.77E-08 & 1.05E-07 & 1.10E-07 & 5.72e-08&3.84e-08&3.01e-08\\
\nuc{49}{V}  &6.80E-08 & 9.28E-08 & 1.31E-07 & 1.63E-07 & 2.41E-07 & 3.15E-07 & 3.21E-07 & 3.57E-07 &3.93E-07 & 3.51E-07 & 3.72E-07 & 4.00E-07 & 4.00E-07 & 4.01E-07 & 1.95e-07&9.93e-08&6.32e-08\\
\hdashline[1pt/5pt]
\noalign{\smallskip}
\nuc{48}{Cr} &1.02E-04 & 1.41E-04 & 1.66E-04 & 2.07E-04 & 2.67E-04 & 2.85E-04 & 2.86E-04 & 3.14E-04 &3.36E-04 & 3.30E-04 & 1.79E-04 & 3.77E-04 & 1.49E-04 & 9.24E-05 & 3.43e-04&3.73e-04&3.87e-04\\
\nuc{49}{Cr} &7.99E-06 & 1.06E-05 & 1.27E-05 & 1.59E-05 & 2.04E-05 & 2.32E-05 & 2.29E-05 & 2.56E-05 &2.72E-05 & 2.66E-05 & 1.68E-05 & 3.01E-05 & 1.58E-05 & 9.89E-06 & 2.05e-05&1.26e-05&6.76e-06\\
\nuc{51}{Cr} &1.39E-06 & 2.12E-06 & 3.13E-06 & 3.89E-06 & 5.95E-06 & 8.11E-06 & 1.13E-05 & 9.29E-06 &7.34E-06 & 9.13E-06 & 1.11E-05 & 1.03E-05 & 1.34E-05 & 1.42E-05 & 5.26e-06&3.53e-06&3.36e-06\\
\nuc{51}{Mn} &3.27E-05 & 3.91E-05 & 4.67E-05 & 5.76E-05 & 7.41E-05 & 8.66E-05 & 8.61E-05 & 9.52E-05 &9.39E-05 & 9.91E-05 & 7.06E-05 & 1.08E-04 & 6.74E-05 & 4.58E-05 & 7.06e-05&3.58e-05&2.23e-05\\
\nuc{52}{Mn} &1.44E-06 & 1.70E-06 & 2.10E-06 & 2.63E-06 & 3.80E-06 & 4.57E-06 & 4.82E-06 & 5.18E-06 &4.77E-06 & 5.45E-06 & 4.58E-06 & 5.87E-06 & 4.60E-06 & 3.95E-06 & 4.61e-06&4.29e-06&4.14e-06\\
\hdashline[1pt/5pt]
\noalign{\smallskip}
\nuc{53}{Mn} &3.06E-05 & 5.07E-05 & 7.30E-05 & 9.66E-05 & 1.53E-04 & 2.01E-04 & 3.78E-04 & 2.35E-04 &8.55E-05 & 2.48E-04 & 2.78E-04 & 2.67E-04 & 3.56E-04 & 3.95E-04 & 1.96e-04&1.77e-04&1.72e-04\\
\nuc{54}{Mn} &2.51E-07 & 5.52E-07 & 9.36E-07 & 1.22E-06 & 2.02E-06 & 2.44E-06 & 1.58E-05 & 3.03E-06 &2.60E-07 & 3.07E-06 & 3.04E-06 & 3.86E-06 & 3.15E-06 & 4.02E-06 & 2.84e-06&2.77e-06&2.77e-06\\
\nuc{52}{Fe} &3.49E-03 & 3.94E-03 & 4.36E-03 & 5.20E-03 & 6.80E-03 & 7.18E-03 & 7.29E-03 & 7.93E-03 &7.91E-03 & 8.27E-03 & 4.98E-03 & 9.01E-03 & 4.06E-03 & 3.03E-03 & 8.40e-03&8.90e-03&9.10e-03\\
\nuc{53}{Fe} &4.94E-04 & 4.76E-04 & 5.17E-04 & 6.11E-04 & 7.92E-04 & 9.12E-04 & 9.08E-04 & 1.00E-03 &9.12E-04 & 1.04E-03 & 7.76E-04 & 1.08E-03 & 6.99E-04 & 5.25E-04 & 8.17e-04&5.81e-04&4.51e-04\\
\nuc{55}{Fe} &1.33E-04 & 3.40E-04 & 5.33E-04 & 7.40E-04 & 1.20E-03 & 1.59E-03 & 2.94E-03 & 1.86E-03 &2.75E-04 & 2.01E-03 & 2.37E-03 & 2.12E-03 & 3.08E-03 & 3.29E-03 & 1.73e-03&1.66e-03&1.66e-03\\
\hdashline[1pt/5pt]
\noalign{\smallskip}
\nuc{59}{Fe} &2.93E-10 & 5.80E-10 & 1.34E-09 & 1.47E-09 & 3.16E-09 & 1.71E-09 & 5.10E-07 & 2.72E-09 &1.04E-15 & 1.81E-09 & 5.03E-10 & 1.07E-08 & 5.75E-13 & 9.55E-13 & 2.65e-09&2.60e-09&2.59e-09\\
\nuc{60}{Fe} &4.95E-11 & 1.03E-10 & 2.45E-10 & 2.67E-10 & 6.17E-10 & 2.60E-10 & 2.43E-07 & 4.20E-10 &3.64E-18 & 2.47E-10 & 4.91E-11 & 2.35E-09 & 3.02E-15 & 5.33E-15 & 4.07e-10&3.98e-10&3.96e-10\\
\nuc{55}{Co} &8.61E-03 & 6.58E-03 & 6.26E-03 & 6.86E-03 & 8.48E-03 & 1.05E-02 & 1.03E-02 & 1.14E-02 &9.01E-03 & 1.19E-02 & 1.08E-02 & 1.13E-02 & 1.00E-02 & 8.03E-03 & 9.33e-03&7.04e-03&6.17e-03\\
\nuc{56}{Co} &5.08E-05 & 4.84E-05 & 5.45E-05 & 6.44E-05 & 8.25E-05 & 1.08E-04 & 1.22E-04 & 1.18E-04 &5.43E-05 & 1.26E-04 & 1.39E-04 & 1.19E-04 & 1.48E-04 & 1.21E-04 & 1.07e-04&9.92e-05&9.66e-05\\
\nuc{57}{Co} &5.71E-05 & 1.58E-04 & 2.50E-04 & 3.52E-04 & 5.69E-04 & 7.44E-04 & 1.36E-03 & 8.70E-04 &6.54E-05 & 9.52E-04 & 1.11E-03 & 9.92E-04 & 1.42E-03 & 1.50E-03 & 8.43e-04&8.28e-04&8.25e-04\\
\hdashline[1pt/5pt]
\noalign{\smallskip}
\nuc{58}{Co} &3.11E-07 & 7.94E-07 & 1.30E-06 & 1.79E-06 & 2.88E-06 & 3.63E-06 & 1.07E-05 & 4.35E-06 &1.23E-07 & 4.72E-06 & 5.04E-06 & 5.14E-06 & 6.08E-06 & 7.29E-06 & 4.27e-06&4.22e-06&4.21e-06\\
\nuc{60}{Co} &2.01E-09 & 3.77E-09 & 8.36E-09 & 9.40E-09 & 1.78E-08 & 1.36E-08 & 8.78E-07 & 2.03E-08 &6.97E-13 & 1.59E-08 & 6.87E-09 & 4.97E-08 & 2.04E-10 & 3.00E-10 & 1.99e-08&1.96e-08&1.96e-08\\
\nuc{56}{Ni} &1.11E-00 & 1.04E-00 & 9.74E-01 & 9.39E-01 & 7.78E-01 & 6.55E-01 & 6.94E-01 & 6.04E-01 &5.32E-01 & 5.66E-01 & 4.15E-01 & 5.12E-01 & 3.64E-01 & 3.22E-01 & 6.29e-01&6.49e-01&6.55e-01\\
\nuc{57}{Ni} &3.36E-02 & 3.04E-02 & 2.89E-02 & 2.78E-02 & 2.25E-02 & 1.96E-02 & 2.06E-02 & 1.79E-02 &1.42E-02 & 1.69E-02 & 1.42E-02 & 1.45E-02 & 1.31E-02 & 1.11E-02 & 1.57e-02&1.33e-02&1.19e-02\\
\nuc{59}{Ni} &8.22E-05 & 1.32E-04 & 1.78E-04 & 2.22E-04 & 3.01E-04 & 3.57E-04 & 5.67E-04 & 3.93E-04 &5.78E-05 & 4.22E-04 & 4.80E-04 & 4.23E-04 & 5.82E-04 & 5.77E-04 & 3.80e-04&3.63e-04&4.14e-04\\
\hdashline[1pt/5pt]
\noalign{\smallskip}
\nuc{63}{Ni} &1.85E-09 & 3.43E-09 & 7.79E-09 & 8.72E-09 & 1.74E-08 & 1.13E-08 & 1.38E-06 & 1.76E-08 &1.58E-14 & 1.27E-08 & 4.26E-09 & 5.31E-08 & 1.85E-11 & 2.92E-11 & 1.72e-08&1.69e-08&1.68e-08\\
\nuc{62}{Zn} &6.48E-04 & 7.18E-04 & 7.80E-04 & 7.72E-04 & 6.86E-04 & 4.74E-04 & 5.25E-04 & 3.22E-04 &2.45E-04 & 2.31E-04 & 9.92E-05 & 1.34E-04 & 9.39E-05 & 9.86E-05 & 2.24e-04&7.89e-05&5.01e-05\\
\nuc{65}{Zn} &3.56E-10 & 4.88E-10 & 6.69E-10 & 7.18E-10 & 8.76E-10 & 8.30E-10 & 1.62E-09 & 7.35E-10 &4.09E-10 & 6.35E-10 & 4.11E-10 & 5.23E-10 & 3.95E-10 & 5.08E-10 & 8.46e-10&9.95e-10&1.11e-09\\
\nuc{65}{Ge} &5.98E-08 & 6.80E-08 & 8.19E-08 & 8.30E-08 & 8.57E-08 & 7.07E-08 & 7.22E-08 & 5.29E-08 &4.32E-08 & 3.99E-08 & 1.66E-08 & 2.07E-08 & 1.69E-08 & 2.03E-08 & 6.64e-08&8.45e-08&9.79e-08\\
\nuc{68}{Ge} &8.91E-10 & 9.19E-10 & 1.06E-09 & 1.07E-09 & 1.03E-09 & 8.30E-10 & 8.90E-10 & 6.33E-10 &5.26E-10 & 5.05E-10 & 2.40E-10 & 2.84E-10 & 2.29E-10 & 2.53E-10 & 2.81e-10&8.32e-11&6.38e-11\\ \hline \normalsize
\end{tabular}
\end{table*}

\section{Summary and Conclusions}
\label{sec:discussion}
We have performed fourteen three-dimensional hydrodynamical
simulations for delayed detonation SNe~Ia for a range of ignition
conditions. For each simulation we have determined the complete
nucleosynthetic yields by post-processing one million tracer particles
with a nuclear reaction network. This set of models constitutes the
first suite of three-dimensional explosion models that covers the
range of expected \nuc{56}{Ni} masses of spectroscopically normal
SNe~Ia.  {From our study we conclude that, fixing all other parameters
of the exploding WD but the ignition configuration, a delayed-detonation of
``normal'' SN~Ia brightness likely requires
rather symmetrical, central ignition to occur in nature.} 
Only such a setup sufficiently pre-expands the WD in the deflagration phase to reduce the
\nuc{56}{Ni} production in the subsequent detonation.  Otherwise only
the brightest SNe~Ia could be explained with delayed detonations of
Chandrasekhar-mass WDs as realized in our models by few (and thus
asymmetrically distributed) ignition sparks. {Whether or
  not the demand for symmetric ignition is in conflict with
  \citet{maeda2010b}, who require a typical off-set of ${\sim}3,500
  \kms$ in the deflagration ashes, is not clear. In spite of rather symmetric
  ignition in our ``normal'' models, the fact that one or a few
  RT-modes grow faster than the others still results in a distribution of 
  stable Fe-peak isotopes that is not uniform (see second column of Fig.~\ref{fig:tracer2}).}

Our suite of models is the first published set of
three-dimensional delayed detonation simulations with detailed
isotopic nucleosynthetic yields.  As such, the yields presented here
lend themselves to be used as an input for Galactic chemical evolution
calculations.  They also set the stage for predicting observables by
radiative transfer calculations.  Overall, we expect the brightness
range of normal SNe~Ia to be covered by our set of models.  It remains
to be seen whether the predicted spectra match the observations and
whether the set of models follows the width-luminosity relation
\citep{phillips1993a,phillips1999a} and other observational trends, as
a set of two-dimensional delayed-detonation models did
\citep{kasen2009a,blondin2011a}.  The Fe-group isotopes in our models
of normal SN~Ia brightness (\nuc{56}{Ni} masses around $0.6\ \msun$)
are synthesized in the required proportions (see
Figs.~\ref{fig:solar_comp} and \ref{fig:solar_comp_metal}), which
tells us that delayed-detonations cannot be ruled out as the dominant
SN~Ia explosion channel based on solar isotopic Fe-group ratios.  The
presence of IGEs (in particular stable isotopes) at high velocities
and oxygen and carbon in the inner ejecta in our models is expected to
leave testable imprints on the observables.  Moreover, our set of
three-dimensional full-star models captures asymmetries that
potentially could be constrained by spectrapolarimetry measurements
{\citep[see e.g. the review by][]{wang2008a}}.

\section*{Acknowledgements}
The simulations presented here were carried out in part on the JUGENE
supercomputer at the Forschungszentrum J{\"u}lich within the
Partnership for Advanced Computing in Europe (PRA042), the grant HMU13
and in part at the Computer Center of the Max Planck Society,
Garching, Germany.  This work was also supported by the Deutsche
Forschungsgemeinschaft via the Transregional Collaborative Research
Center TRR 33 ``The Dark Universe'', the Emmy Noether Program (RO
3676/1-1), the ARCHES prize of the German Ministry of Education and
Research (BMBF), the graduate school ``Theoretical Astrophysics and
Particle Physics'' at the University of W\"urzburg (GRK 1147) and the
Excellence Cluster EXC~153. FKR, MF and SAS acknowledge travel support
by the DAAD/Go8 German-Australian exchange program.

\bibliography{astrofritz} \bibliographystyle{mn2e}

\begin{thebibliography}{115}
\expandafter\ifx\csname natexlab\endcsname\relax\def\natexlab#1{#1}\fi

\bibitem[{{Arnett}(1969)}]{arnett1969a}
{Arnett} W.~D., 1969, \apss, 5, 180

\bibitem[{{Blinnikov} \& {Khokhlov}(1986)}]{blinnikov1986a}
{Blinnikov} S.~I., {Khokhlov} A.~M., 1986, Soviet Astronomy Letters, 12, 131

\bibitem[{{Blondin} {et~al}\mbox{.}(2011){Blondin}, {Kasen}, {R{\"o}pke},
  {Kirshner}, \& {Mandel}}]{blondin2011a}
{Blondin} S., {Kasen} D., {R{\"o}pke} F.~K., {Kirshner} R.~P., {Mandel} K.~S.,
  2011, \mnras, 417, 1280

\bibitem[{{Bloom} {et~al}\mbox{.}(2012){Bloom}, {Kasen}, {Shen}, {Nugent},
  {Butler}, {Graham}, {Howell}, {Kolb}, {Holmes}, {Haswell}, {Burwitz},
  {Rodriguez}, \& {Sullivan}}]{bloom2012a}
{Bloom} J.~S. {et~al.}, 2012, \apjl, 744, L17

\bibitem[{{Brachwitz} {et~al}\mbox{.}(2000){Brachwitz}, {Dean}, {Hix},
  {Iwamoto}, {Langanke}, {Mart{\'{\i}}nez-Pinedo}, {Nomoto}, {Strayer},
  {Thielemann}, \& {Umeda}}]{brachwitz2000a}
{Brachwitz} F. {et~al.}, 2000, \apj, 536, 934

\bibitem[{{Bravo} \& {Garc{\'{\i}}a-Senz}(2008)}]{bravo2008a}
{Bravo} E., {Garc{\'{\i}}a-Senz} D., 2008, \aap, 478, 843

\bibitem[{{Bravo} \& {Garc{\'{\i}}a-Senz}(2009)}]{bravo2009a}
{Bravo} E., {Garc{\'{\i}}a-Senz} D., 2009, \apj, 695, 1244

\bibitem[{{Bravo} {et~al}\mbox{.}(2009){Bravo}, {Garc{\'{\i}}a-Senz},
  {Cabez{\'o}n}, \& {Dom{\'{\i}}nguez}}]{bravo2009b}
{Bravo} E., {Garc{\'{\i}}a-Senz} D., {Cabez{\'o}n} R.~M., {Dom{\'{\i}}nguez}
  I., 2009, \apj, 695, 1257

\bibitem[{{Chan} \& {Lingenfelter}(1993)}]{chan1993a}
{Chan} K.-W., {Lingenfelter} R.~E., 1993, \apj, 405, 614

\bibitem[{{Chomiuk} {et~al}\mbox{.}(2012){Chomiuk}, {Soderberg}, {Moe},
  {Chevalier}, {Rupen}, {Badenes}, {Margutti}, {Fransson}, {Fong}, \&
  {Dittmann}}]{chomiuk2012a}
{Chomiuk} L. {et~al.}, 2012, \apj, 750, 164

\bibitem[{{Clayton}(2003)}]{clayton2003a}
{Clayton} D., 2003, Handbook of Isotopes in the Cosmos

\bibitem[{{Colella} \& {Woodward}(1984)}]{colella1984a}
{Colella} P., {Woodward} P.~R., 1984, Journal of Computational Physics, 54, 174

\bibitem[{{Dilday} {et~al}\mbox{.}(2012){Dilday}, {Howell}, {Cenko},
  {Silverman}, {Nugent}, {Sullivan}, {Ben-Ami}, {Bildsten}, {Bolte}, {Endl},
  {Filippenko}, {Gnat}, {Horesh}, {Hsiao}, {Kasliwal}, {Kirkman}, {Maguire},
  {Marcy}, {Moore}, {Pan}, {Parrent}, {Podsiadlowski}, {Quimby}, {Sternberg},
  {Suzuki}, {Tytler}, {Xu}, {Bloom}, {Gal-Yam}, {Hook}, {Kulkarni}, {Law},
  {Ofek}, {Polishook}, \& {Poznanski}}]{dilday2012a}
{Dilday} B. {et~al.}, 2012, ArXiv e-prints

\bibitem[{{Dursi} \& {Timmes}(2006)}]{dursi2006a}
{Dursi} L.~J., {Timmes} F.~X., 2006, \apj, 641, 1071

\bibitem[{{Fink} {et~al}\mbox{.}(2007){Fink}, {Hillebrandt}, \&
  {R{\"o}pke}}]{fink2007a}
{Fink} M., {Hillebrandt} W., {R{\"o}pke} F.~K., 2007, \aap, 476, 1133

\bibitem[{{Fink} {et~al}\mbox{.}(2010){Fink}, {R{\"o}pke}, {Hillebrandt},
  {Seitenzahl}, {Sim}, \& {Kromer}}]{fink2010a}
{Fink} M., {R{\"o}pke} F.~K., {Hillebrandt} W., {Seitenzahl} I.~R., {Sim}
  S.~A., {Kromer} M., 2010, \aap, 514, A53

\bibitem[{{Fryxell} {et~al}\mbox{.}(1989){Fryxell}, {M{\"u}ller}, \&
  {Arnett}}]{fryxell1989a}
{Fryxell} B.~A., {M{\"u}ller} E., {Arnett} W.~D., 1989, Hydro\-dynamics and
  nuclear burning. MPA Green Report 449, Max-Planck-Institut f\"ur Astrophysik,
  Garching

\bibitem[{{Gamezo} {et~al}\mbox{.}(2005){Gamezo}, {Khokhlov}, \&
  {Oran}}]{gamezo2005a}
{Gamezo} V.~N., {Khokhlov} A.~M., {Oran} E.~S., 2005, \apj, 623, 337

\bibitem[{{Gamezo} {et~al}\mbox{.}(1999){Gamezo}, {Wheeler}, {Khokhlov}, \&
  {Oran}}]{gamezo1999a}
{Gamezo} V.~N., {Wheeler} J.~C., {Khokhlov} A.~M., {Oran} E.~S., 1999, \apj,
  512, 827

\bibitem[{{Garcia-Senz} \& {Woosley}(1995)}]{garcia1995a}
{Garcia-Senz} D., {Woosley} S.~E., 1995, \apj, 454, 895

\bibitem[{{Gilfanov} \& {Bogd{\'a}n}(2010)}]{gilfanov2010a}
{Gilfanov} M., {Bogd{\'a}n} {\'A}., 2010, \nat, 463, 924

\bibitem[{{Han} \& {Podsiadlowski}(2004)}]{han2004a}
{Han} Z., {Podsiadlowski} P., 2004, \mnras, 350, 1301

\bibitem[{{H{\"o}flich} {et~al}\mbox{.}(2004){H{\"o}flich}, {Gerardy},
  {Nomoto}, {Motohara}, {Fesen}, {Maeda}, {Ohkubo}, \&
  {Tominaga}}]{hoeflich2004a}
{H{\"o}flich} P., {Gerardy} C.~L., {Nomoto} K., {Motohara} K., {Fesen} R.~A.,
  {Maeda} K., {Ohkubo} T., {Tominaga} N., 2004, \apj, 617, 1258

\bibitem[{{H{\"o}flich} \& {Khokhlov}(1996)}]{hoeflich1996a}
{H{\"o}flich} P., {Khokhlov} A., 1996, \apj, 457, 500

\bibitem[{{H{\"o}flich} \& {Stein}(2002)}]{hoeflich2002a}
{H{\"o}flich} P., {Stein} J., 2002, \apj, 568, 779

\bibitem[{{Iapichino} {et~al}\mbox{.}(2006){Iapichino}, {Br{\"u}ggen},
  {Hillebrandt}, \& {Niemeyer}}]{iapichino2006a}
{Iapichino} L., {Br{\"u}ggen} M., {Hillebrandt} W., {Niemeyer} J.~C., 2006,
  \aap, 450, 655

\bibitem[{{Iwamoto} {et~al}\mbox{.}(1999){Iwamoto}, {Brachwitz}, {Nomoto},
  {Kishimoto}, {Umeda}, {Hix}, \& {Thielemann}}]{iwamoto1999a}
{Iwamoto} K., {Brachwitz} F., {Nomoto} K., {Kishimoto} N., {Umeda} H., {Hix}
  W.~R., {Thielemann} F.-K., 1999, \apjs, 125, 439

\bibitem[{{Jordan} {et~al}\mbox{.}(2008){Jordan}, {Fisher}, {Townsley},
  {Calder}, {Graziani}, {Asida}, {Lamb}, \& {Truran}}]{jordan2008a}
{Jordan}, IV G.~C., {Fisher} R.~T., {Townsley} D.~M., {Calder} A.~C.,
  {Graziani} C., {Asida} S., {Lamb} D.~Q., {Truran} J.~W., 2008, \apj, 681,
  1448

\bibitem[{{Jordan} {et~al}\mbox{.}(2012){Jordan}, {Graziani}, {Fisher},
  {Townsley}, {Meakin}, {Weide}, {Reid}, {Norris}, {Hudson}, \&
  {Lamb}}]{jordan2012a}
{Jordan}, IV G.~C. {et~al.}, 2012, ArXiv e-prints

\bibitem[{{Kasen} {et~al}\mbox{.}(2009){Kasen}, {R{\"o}pke}, \&
  {Woosley}}]{kasen2009a}
{Kasen} D., {R{\"o}pke} F.~K., {Woosley} S.~E., 2009, \nat, 460, 869

\bibitem[{{Kasen} {et~al}\mbox{.}(2006){Kasen}, {Thomas}, \&
  {Nugent}}]{kasen2006a}
{Kasen} D., {Thomas} R.~C., {Nugent} P., 2006, \apj, 651, 366

\bibitem[{{Kerstein}(1988)}]{kerstein1988a}
{Kerstein} A.~R., 1988, Combust. Sci. Technol., 60, 441

\bibitem[{{Khokhlov}(1989)}]{khokhlov1989a}
{Khokhlov} A.~M., 1989, \mnras, 239, 785

\bibitem[{{Khokhlov}(1991{\natexlab{a}})}]{khokhlov1991a}
{Khokhlov} A.~M., 1991{\natexlab{a}}, \aap, 245, 114

\bibitem[{{Khokhlov}(1991{\natexlab{b}})}]{khokhlov1991c}
{Khokhlov} A.~M., 1991{\natexlab{b}}, \aap, 246, 383

\bibitem[{{Khokhlov}(1991{\natexlab{c}})}]{khokhlov1991b}
{Khokhlov} A.~M., 1991{\natexlab{c}}, \aap, 245, L25

\bibitem[{{Khokhlov}(1995)}]{khokhlov1995a}
{Khokhlov} A.~M., 1995, \apj, 449, 695

\bibitem[{{Khokhlov} {et~al}\mbox{.}(1997){Khokhlov}, {Oran}, \&
  {Wheeler}}]{khokhlov1997a}
{Khokhlov} A.~M., {Oran} E.~S., {Wheeler} J.~C., 1997, \apj, 478, 678

\bibitem[{{Kobayashi} \& {Nomoto}(2009)}]{kobayashi2009a}
{Kobayashi} C., {Nomoto} K., 2009, \apj, 707, 1466

\bibitem[{{Kozma} {et~al}\mbox{.}(2005){Kozma}, {Fransson}, {Hillebrandt},
  {Travaglio}, {Sollerman}, {Reinecke}, {R{\"o}pke}, \&
  {Spyromilio}}]{kozma2005a}
{Kozma} C., {Fransson} C., {Hillebrandt} W., {Travaglio} C., {Sollerman} J.,
  {Reinecke} M., {R{\"o}pke} F.~K., {Spyromilio} J., 2005, \aap, 437, 983

\bibitem[{{Kromer} {et~al}\mbox{.}(2010){Kromer}, {Sim}, {Fink}, {R{\"o}pke},
  {Seitenzahl}, \& {Hillebrandt}}]{kromer2010a}
{Kromer} M., {Sim} S.~A., {Fink} M., {R{\"o}pke} F.~K., {Seitenzahl} I.~R.,
  {Hillebrandt} W., 2010, \apj, 719, 1067

\bibitem[{{Krueger} {et~al}\mbox{.}(2010){Krueger}, {Jackson}, {Townsley},
  {Calder}, {Brown}, \& {Timmes}}]{krueger2010a}
{Krueger} B.~K., {Jackson} A.~P., {Townsley} D.~M., {Calder} A.~C., {Brown}
  E.~F., {Timmes} F.~X., 2010, \apjl, 719, L5

\bibitem[{{Kuhlen} {et~al}\mbox{.}(2006){Kuhlen}, {Woosley}, \&
  {Glatzmaier}}]{kuhlen2006a}
{Kuhlen} M., {Woosley} S.~E., {Glatzmaier} G.~A., 2006, \apj, 640, 407

\bibitem[{{Lisewski} {et~al}\mbox{.}(2000){Lisewski}, {Hillebrandt}, \&
  {Woosley}}]{lisewski2000b}
{Lisewski} A.~M., {Hillebrandt} W., {Woosley} S.~E., 2000, \apj, 538, 831

\bibitem[{{Livne}(1990)}]{livne1990a}
{Livne} E., 1990, \apjl, 354, L53

\bibitem[{{Livne} \& {Glasner}(1990)}]{livne1990b}
{Livne} E., {Glasner} A.~S., 1990, \apj, 361, 244

\bibitem[{{Lodders}(2003)}]{lodders2003a}
{Lodders} K., 2003, \apj, 591, 1220

\bibitem[{{Maeda} {et~al}\mbox{.}(2010{\natexlab{a}}){Maeda}, {Benetti},
  {Stritzinger}, {R{\"o}pke}, {Folatelli}, {Sollerman}, {Taubenberger},
  {Nomoto}, {Leloudas}, {Hamuy}, {Tanaka}, {Mazzali}, \&
  {Elias-Rosa}}]{maeda2010b}
{Maeda} K. {et~al.}, 2010{\natexlab{a}}, \nat, 466, 82

\bibitem[{{Maeda} {et~al}\mbox{.}(2010{\natexlab{b}}){Maeda}, {R{\"o}pke},
  {Fink}, {Hillebrandt}, {Travaglio}, \& {Thielemann}}]{maeda2010a}
{Maeda} K., {R{\"o}pke} F.~K., {Fink} M., {Hillebrandt} W., {Travaglio} C.,
  {Thielemann} F., 2010{\natexlab{b}}, \apj, 712, 624

\bibitem[{{Maier} \& {Niemeyer}(2006)}]{maier2006a}
{Maier} A., {Niemeyer} J.~C., 2006, \aap, 451, 207

\bibitem[{{Mazzali} {et~al}\mbox{.}(2007){Mazzali}, {R{\"o}pke}, {Benetti}, \&
  {Hillebrandt}}]{mazzali2007a}
{Mazzali} P.~A., {R{\"o}pke} F.~K., {Benetti} S., {Hillebrandt} W., 2007,
  Science, 315, 825

\bibitem[{{Meakin} {et~al}\mbox{.}(2009){Meakin}, {Seitenzahl}, {Townsley},
  {Jordan}, {Truran}, \& {Lamb}}]{meakin2009a}
{Meakin} C.~A., {Seitenzahl} I., {Townsley} D., {Jordan} G.~C., {Truran} J.,
  {Lamb} D., 2009, \apj, 693, 1188

\bibitem[{{Mennekens} {et~al}\mbox{.}(2010){Mennekens}, {Vanbeveren}, {De
  Greve}, \& {De Donder}}]{mennekens2010a}
{Mennekens} N., {Vanbeveren} D., {De Greve} J.~P., {De Donder} E., 2010, ArXiv
  e-prints

\bibitem[{{Nelemans} {et~al}\mbox{.}(2012){Nelemans}, {Toonen}, \&
  {Bours}}]{nelemans2012a}
{Nelemans} G., {Toonen} S., {Bours} M., 2012, ArXiv e-prints

\bibitem[{{Niemeyer} \& {Woosley}(1997)}]{niemeyer1997b}
{Niemeyer} J.~C., {Woosley} S.~E., 1997, \apj, 475, 740

\bibitem[{{Nomoto} {et~al}\mbox{.}(1984){Nomoto}, {Thielemann}, \&
  {Yokoi}}]{nomoto1984a}
{Nomoto} K., {Thielemann} F.-K., {Yokoi} K., 1984, \apj, 286, 644

\bibitem[{{Nonaka} {et~al}\mbox{.}(2012){Nonaka}, {Aspden}, {Zingale},
  {Almgren}, {Bell}, \& {Woosley}}]{nonaka2012a}
{Nonaka} A., {Aspden} A.~J., {Zingale} M., {Almgren} A.~S., {Bell} J.~B.,
  {Woosley} S.~E., 2012, \apj, 745, 73

\bibitem[{{Nugent} {et~al}\mbox{.}(1997){Nugent}, {Baron}, {Branch}, {Fisher},
  \& {Hauschildt}}]{nugent1997a}
{Nugent} P., {Baron} E., {Branch} D., {Fisher} A., {Hauschildt} P.~H., 1997,
  \apj, 485, 812

\bibitem[{{Osher} \& {Sethian}(1988)}]{osher1988a}
{Osher} S., {Sethian} J.~A., 1988, Journal of Computational Physics, 79, 12

\bibitem[{Pakmor {et~al}\mbox{.}(2011)Pakmor, Edelmann, Roepke, \&
  Hillebrandt}]{pakmor2011a}
Pakmor R., Edelmann P., Roepke F.~K., Hillebrandt W., 2011, Stellar gadget, in
  preparation

\bibitem[{{Pakmor} {et~al}\mbox{.}(2010){Pakmor}, {Kromer}, {R{\"o}pke}, {Sim},
  {Ruiter}, \& {Hillebrandt}}]{pakmor2010a}
{Pakmor} R., {Kromer} M., {R{\"o}pke} F.~K., {Sim} S.~A., {Ruiter} A.~J.,
  {Hillebrandt} W., 2010, \nat, 463, 61

\bibitem[{{Pakmor} {et~al}\mbox{.}(2012){Pakmor}, {Kromer}, {Taubenberger},
  {Sim}, {R{\"o}pke}, \& {Hillebrandt}}]{pakmor2012a}
{Pakmor} R., {Kromer} M., {Taubenberger} S., {Sim} S.~A., {R{\"o}pke} F.~K.,
  {Hillebrandt} W., 2012, \apjl, 747, L10

\bibitem[{{Patat} {et~al}\mbox{.}(2007){Patat}, {Chandra}, {Chevalier},
  {Justham}, {Podsiadlowski}, {Wolf}, {Gal-Yam}, {Pasquini}, {Crawford},
  {Mazzali}, {Pauldrach}, {Nomoto}, {Benetti}, {Cappellaro}, {Elias-Rosa},
  {Hillebrandt}, {Leonard}, {Pastorello}, {Renzini}, {Sabbadin}, {Simon}, \&
  {Turatto}}]{patat2007a}
{Patat} F. {et~al.}, 2007, Science, 317, 924

\bibitem[{{Perlmutter} {et~al}\mbox{.}(1999){Perlmutter}, {Aldering},
  {Goldhaber}, {Knop}, {Nugent}, {Castro}, {Deustua}, {Fabbro}, {Goobar},
  {Groom}, {Hook}, {Kim}, {Kim}, {Lee}, {Nunes}, {Pain}, {Pennypacker},
  {Quimby}, {Lidman}, {Ellis}, {Irwin}, {McMahon}, {Ruiz-Lapuente}, {Walton},
  {Schaefer}, {Boyle}, {Filippenko}, {Matheson}, {Fruchter}, {Panagia},
  {Newberg}, {Couch}, \& {The Supernova Cosmology Project}}]{perlmutter1999a}
{Perlmutter} S. {et~al.}, 1999, \apj, 517, 565

\bibitem[{{Phillips}(1993)}]{phillips1993a}
{Phillips} M.~M., 1993, \apjl, 413, L105

\bibitem[{{Phillips} {et~al}\mbox{.}(1999){Phillips}, {Lira}, {Suntzeff},
  {Schommer}, {Hamuy}, \& {Maza}}]{phillips1999a}
{Phillips} M.~M., {Lira} P., {Suntzeff} N.~B., {Schommer} R.~A., {Hamuy} M.,
  {Maza} J., 1999, \aj, 118, 1766

\bibitem[{{Prantzos} {et~al}\mbox{.}(2011){Prantzos}, {Boehm}, {Bykov},
  {Diehl}, {Ferri{\`e}re}, {Guessoum}, {Jean}, {Knoedlseder}, {Marcowith},
  {Moskalenko}, {Strong}, \& {Weidenspointner}}]{prantzos2011a}
{Prantzos} N. {et~al.}, 2011, Reviews of Modern Physics, 83, 1001

\bibitem[{{Prialnik} \& {Kovetz}(1995)}]{prialnik1995a}
{Prialnik} D., {Kovetz} A., 1995, \apj, 445, 789

\bibitem[{{Reinecke} {et~al}\mbox{.}(1999){Reinecke}, {Hillebrandt},
  {Niemeyer}, {Klein}, \& {Gr{\"o}bl}}]{reinecke1999a}
{Reinecke} M., {Hillebrandt} W., {Niemeyer} J.~C., {Klein} R., {Gr{\"o}bl} A.,
  1999, \aap, 347, 724

\bibitem[{{Riess} {et~al}\mbox{.}(1998){Riess}, {Filippenko}, {Challis},
  {Clocchiatti}, {Diercks}, {Garnavich}, {Gilliland}, {Hogan}, {Jha},
  {Kirshner}, {Leibundgut}, {Phillips}, {Reiss}, {Schmidt}, {Schommer},
  {Smith}, {Spyromilio}, {Stubbs}, {Suntzeff}, \& {Tonry}}]{riess1998a}
{Riess} A.~G. {et~al.}, 1998, \aj, 116, 1009

\bibitem[{{R{\"o}pke}(2005)}]{roepke2005c}
{R{\"o}pke} F.~K., 2005, \aap, 432, 969

\bibitem[{{R{\"o}pke}(2007)}]{roepke2007d}
{R{\"o}pke} F.~K., 2007, \apj, 668, 1103

\bibitem[{{R{\"o}pke} {et~al}\mbox{.}(2006{\natexlab{a}}){R{\"o}pke},
  {Gieseler}, {Reinecke}, {Travaglio}, \& {Hillebrandt}}]{roepke2006b}
{R{\"o}pke} F.~K., {Gieseler} M., {Reinecke} M., {Travaglio} C., {Hillebrandt}
  W., 2006{\natexlab{a}}, \aap, 453, 203

\bibitem[{{R{\"o}pke} {et~al}\mbox{.}(2006{\natexlab{b}}){R{\"o}pke},
  {Hillebrandt}, {Niemeyer}, \& {Woosley}}]{roepke2006a}
{R{\"o}pke} F.~K., {Hillebrandt} W., {Niemeyer} J.~C., {Woosley} S.~E.,
  2006{\natexlab{b}}, \aap, 448, 1

\bibitem[{{R{\"o}pke} {et~al}\mbox{.}(2007{\natexlab{a}}){R{\"o}pke},
  {Hillebrandt}, {Schmidt}, {Niemeyer}, {Blinnikov}, \&
  {Mazzali}}]{roepke2007c}
{R{\"o}pke} F.~K., {Hillebrandt} W., {Schmidt} W., {Niemeyer} J.~C.,
  {Blinnikov} S.~I., {Mazzali} P.~A., 2007{\natexlab{a}}, \apj, 668, 1132

\bibitem[{{R{\"o}pke} {et~al}\mbox{.}(2012){R{\"o}pke}, {Kromer}, {Seitenzahl},
  {Pakmor}, {Sim}, {Taubenberger}, {Ciaraldi-Schoolmann}, {Hillebrandt},
  {Aldering}, {Antilogus}, {Baltay}, {Benitez-Herrera}, {Bongard}, {Buton},
  {Canto}, {Cellier-Holzem}, {Childress}, {Chotard}, {Copin}, {Fakhouri},
  {Fink}, {Fouchez}, {Gangler}, {Guy}, {Hachinger}, {Hsiao}, {Chen},
  {Kerschhaggl}, {Kowalski}, {Nugent}, {Paech}, {Pain}, {Pecontal}, {Pereira},
  {Perlmutter}, {Rabinowitz}, {Rigault}, {Runge}, {Saunders}, {Smadja},
  {Suzuki}, {Tao}, {Thomas}, {Tilquin}, \& {Wu}}]{roepke2012a}
{R{\"o}pke} F.~K. {et~al.}, 2012, \apjl, 750, L19

\bibitem[{{R{\"o}pke} \& {Niemeyer}(2007)}]{roepke2007b}
{R{\"o}pke} F.~K., {Niemeyer} J.~C., 2007, \aap, 464, 683

\bibitem[{{R{\"o}pke} {et~al}\mbox{.}(2007{\natexlab{b}}){R{\"o}pke},
  {Woosley}, \& {Hillebrandt}}]{roepke2007a}
{R{\"o}pke} F.~K., {Woosley} S.~E., {Hillebrandt} W., 2007{\natexlab{b}}, \apj,
  660, 1344

\bibitem[{{Ruiter} {et~al}\mbox{.}(2009){Ruiter}, {Belczynski}, \&
  {Fryer}}]{ruiter2009a}
{Ruiter} A.~J., {Belczynski} K., {Fryer} C., 2009, \apj, 699, 2026

\bibitem[{{Ruiter} {et~al}\mbox{.}(2011){Ruiter}, {Belczynski}, {Sim},
  {Hillebrandt}, {Fryer}, {Fink}, \& {Kromer}}]{ruiter2011a}
{Ruiter} A.~J., {Belczynski} K., {Sim} S.~A., {Hillebrandt} W., {Fryer} C.~L.,
  {Fink} M., {Kromer} M., 2011, \mnras, 1282

\bibitem[{{Scannapieco} {et~al}\mbox{.}(2008){Scannapieco}, {Tissera}, {White},
  \& {Springel}}]{scannapieco2008a}
{Scannapieco} C., {Tissera} P.~B., {White} S.~D.~M., {Springel} V., 2008,
  \mnras, 389, 1137

\bibitem[{{Schmidt} {et~al}\mbox{.}(1998){Schmidt}, {Suntzeff}, {Phillips},
  {Schommer}, {Clocchiatti}, {Kirshner}, {Garnavich}, {Challis}, {Leibundgut},
  {Spyromilio}, {Riess}, {Filippenko}, {Hamuy}, {Smith}, {Hogan}, {Stubbs},
  {Diercks}, {Reiss}, {Gilliland}, {Tonry}, {Maza}, {Dressler}, {Walsh}, \&
  {Ciardullo}}]{schmidt1998a}
{Schmidt} B.~P. {et~al.}, 1998, \apj, 507, 46

\bibitem[{{Schmidt} {et~al}\mbox{.}(2006{\natexlab{a}}){Schmidt}, {Niemeyer},
  \& {Hillebrandt}}]{schmidt2006b}
{Schmidt} W., {Niemeyer} J.~C., {Hillebrandt} W., 2006{\natexlab{a}}, \aap,
  450, 265

\bibitem[{{Schmidt} {et~al}\mbox{.}(2006{\natexlab{b}}){Schmidt}, {Niemeyer},
  {Hillebrandt}, \& {R{\"o}pke}}]{schmidt2006c}
{Schmidt} W., {Niemeyer} J.~C., {Hillebrandt} W., {R{\"o}pke} F.~K.,
  2006{\natexlab{b}}, \aap, 450, 283

\bibitem[{Seitenzahl(2011)}]{seitenzahl2011b}
Seitenzahl I., 2011, Progress in Particle and Nuclear Physics, 66, 329 ,
  particle and Nuclear Astrophysics, International Workshop on Nuclear Physics,
  32nd Course

\bibitem[{{Seitenzahl} {et~al}\mbox{.}(2011){Seitenzahl},
  {Ciaraldi-Schoolmann}, \& {R{\"o}pke}}]{seitenzahl2011a}
{Seitenzahl} I.~R., {Ciaraldi-Schoolmann} F., {R{\"o}pke} F.~K., 2011, \mnras,
  414, 2709

\bibitem[{{Seitenzahl} {et~al}\mbox{.}(2009{\natexlab{a}}){Seitenzahl},
  {Meakin}, {Lamb}, \& {Truran}}]{seitenzahl2009c}
{Seitenzahl} I.~R., {Meakin} C.~A., {Lamb} D.~Q., {Truran} J.~W.,
  2009{\natexlab{a}}, \apj, 700, 642

\bibitem[{{Seitenzahl} {et~al}\mbox{.}(2009{\natexlab{b}}){Seitenzahl},
  {Meakin}, {Townsley}, {Lamb}, \& {Truran}}]{seitenzahl2009b}
{Seitenzahl} I.~R., {Meakin} C.~A., {Townsley} D.~M., {Lamb} D.~Q., {Truran}
  J.~W., 2009{\natexlab{b}}, \apj, 696, 515

\bibitem[{{Seitenzahl} {et~al}\mbox{.}(2010){Seitenzahl}, {R{\"o}pke}, {Fink},
  \& {Pakmor}}]{seitenzahl2010a}
{Seitenzahl} I.~R., {R{\"o}pke} F.~K., {Fink} M., {Pakmor} R., 2010, \mnras,
  407, 2297

\bibitem[{{Seitenzahl} {et~al}\mbox{.}(2009{\natexlab{c}}){Seitenzahl},
  {Taubenberger}, \& {Sim}}]{seitenzahl2009d}
{Seitenzahl} I.~R., {Taubenberger} S., {Sim} S.~A., 2009{\natexlab{c}}, \mnras,
  400, 531

\bibitem[{{Seitenzahl} {et~al}\mbox{.}(2009{\natexlab{d}}){Seitenzahl},
  {Townsley}, {Peng}, \& {Truran}}]{seitenzahl2009a}
{Seitenzahl} I.~R., {Townsley} D.~M., {Peng} F., {Truran} J.~W.,
  2009{\natexlab{d}}, Atomic Data and Nuclear Data Tables, 95, 96

\bibitem[{{Sim}(2007)}]{sim2007b}
{Sim} S.~A., 2007, \mnras, 375, 154

\bibitem[{{Sim} {et~al}\mbox{.}(2010){Sim}, {R{\"o}pke}, {Hillebrandt},
  {Kromer}, {Pakmor}, {Fink}, {Ruiter}, \& {Seitenzahl}}]{sim2010a}
{Sim} S.~A., {R{\"o}pke} F.~K., {Hillebrandt} W., {Kromer} M., {Pakmor} R.,
  {Fink} M., {Ruiter} A.~J., {Seitenzahl} I.~R., 2010, \apjl, 714, L52

\bibitem[{{Smiljanovski} {et~al}\mbox{.}(1997){Smiljanovski}, {Moser}, \&
  {Klein}}]{smiljanovski1997a}
{Smiljanovski} V., {Moser} V., {Klein} R., 1997, Combustion Theory Modelling,
  1, 183

\bibitem[{{Sreenivasan}(1991)}]{sreenivasan1991a}
{Sreenivasan} K.~R., 1991, Annual Review of Fluid Mechanics, 23, 539

\bibitem[{{Stehle} {et~al}\mbox{.}(2005){Stehle}, {Mazzali}, {Benetti}, \&
  {Hillebrandt}}]{stehle2005a}
{Stehle} M., {Mazzali} P.~A., {Benetti} S., {Hillebrandt} W., 2005, \mnras,
  360, 1231

\bibitem[{{Sternberg} {et~al}\mbox{.}(2011){Sternberg}, {Gal-Yam}, {Simon},
  {Leonard}, {Quimby}, {Phillips}, {Morrell}, {Thompson}, {Ivans}, {Marshall},
  {Filippenko}, {Marcy}, {Bloom}, {Patat}, {Foley}, {Yong}, {Penprase},
  {Beeler}, {Allende Prieto}, \& {Stringfellow}}]{sternberg2011a}
{Sternberg} A. {et~al.}, 2011, Science, 333, 856

\bibitem[{{Stritzinger} {et~al}\mbox{.}(2006){Stritzinger}, {Mazzali},
  {Sollerman}, \& {Benetti}}]{stritzinger2006b}
{Stritzinger} M., {Mazzali} P.~A., {Sollerman} J., {Benetti} S., 2006, \aap,
  460, 793

\bibitem[{{Timmes} {et~al}\mbox{.}(2003){Timmes}, {Brown}, \&
  {Truran}}]{timmes2003a}
{Timmes} F.~X., {Brown} E.~F., {Truran} J.~W., 2003, \apjl, 590, L83

\bibitem[{{Timmes} \& {Woosley}(1992)}]{timmes1992a}
{Timmes} F.~X., {Woosley} S.~E., 1992, \apj, 396, 649

\bibitem[{{Timmes} {et~al}\mbox{.}(1995){Timmes}, {Woosley}, \&
  {Weaver}}]{timmes1995a}
{Timmes} F.~X., {Woosley} S.~E., {Weaver} T.~A., 1995, \apjs, 98, 617

\bibitem[{{Townsley} {et~al}\mbox{.}(2007){Townsley}, {Calder}, {Asida},
  {Seitenzahl}, {Peng}, {Vladimirova}, {Lamb}, \& {Truran}}]{townsley2007a}
{Townsley} D.~M., {Calder} A.~C., {Asida} S.~M., {Seitenzahl} I.~R., {Peng} F.,
  {Vladimirova} N., {Lamb} D.~Q., {Truran} J.~W., 2007, \apj, 668, 1118

\bibitem[{{Travaglio} {et~al}\mbox{.}(2005){Travaglio}, {Hillebrandt}, \&
  {Reinecke}}]{travaglio2005a}
{Travaglio} C., {Hillebrandt} W., {Reinecke} M., 2005, \aap, 443, 1007

\bibitem[{{Travaglio} {et~al}\mbox{.}(2004){Travaglio}, {Hillebrandt},
  {Reinecke}, \& {Thielemann}}]{travaglio2004a}
{Travaglio} C., {Hillebrandt} W., {Reinecke} M., {Thielemann} F.-K., 2004,
  \aap, 425, 1029

\bibitem[{{Travaglio} {et~al}\mbox{.}(2011){Travaglio}, {R{\"o}pke}, {Gallino},
  \& {Hillebrandt}}]{travaglio2011a}
{Travaglio} C., {R{\"o}pke} F.~K., {Gallino} R., {Hillebrandt} W., 2011, \apj,
  739, 93

\bibitem[{{van der Sluys} {et~al}\mbox{.}(2010){van der Sluys}, {Politano}, \&
  {Taam}}]{vandersluys2010a}
{van der Sluys} M., {Politano} M., {Taam} R.~E., 2010, 1314, 13

\bibitem[{{Wang} {et~al}\mbox{.}(2010){Wang}, {Li}, \& {Han}}]{wang2010a}
{Wang} B., {Li} X.-D., {Han} Z.-W., 2010, \mnras, 401, 2729

\bibitem[{{Wang} \& {Wheeler}(2008)}]{wang2008a}
{Wang} L., {Wheeler} J.~C., 2008, \araa, 46, 433

\bibitem[{{Woosley}(2007)}]{woosley2007a}
{Woosley} S.~E., 2007, \apj, 668, 1109

\bibitem[{{Woosley} {et~al}\mbox{.}(2011){Woosley}, {Kerstein}, \&
  {Aspden}}]{woosley2011a}
{Woosley} S.~E., {Kerstein} A.~R., {Aspden} A.~J., 2011, \apj, 734, 37

\bibitem[{{Woosley} {et~al}\mbox{.}(2009){Woosley}, {Kerstein}, {Sankaran},
  {Aspden}, \& {R{\"o}pke}}]{woosley2009a}
{Woosley} S.~E., {Kerstein} A.~R., {Sankaran} V., {Aspden} A.~J., {R{\"o}pke}
  F.~K., 2009, \apj, 704, 255

\bibitem[{{Woosley} \& {Weaver}(1994)}]{woosley1994a}
{Woosley} S.~E., {Weaver} T.~A., 1994, in Les Houches Session {LIV}:
  Supernovae, {Bludman} S.~A., {Mochkovitch} R., {Zinn-Justin} J., eds.,
  North-Holland, Amsterdam, pp. 63--154

\bibitem[{{Woosley} {et~al}\mbox{.}(2004){Woosley}, {Wunsch}, \&
  {Kuhlen}}]{woosley2004a}
{Woosley} S.~E., {Wunsch} S., {Kuhlen} M., 2004, \apj, 607, 921

\bibitem[{{Zel'dovich} {et~al}\mbox{.}(1970){Zel'dovich}, {Librovich},
  {Makhviladze}, \& {Sivashinskii}}]{zeldovich1970a}
{Zel'dovich} Y.~B., {Librovich} V.~B., {Makhviladze} G.~M., {Sivashinskii}
  G.~I., 1970, Journal of Applied Mechanics and Technical Physics, 11, 264

\bibitem[{{Zingale} {et~al}\mbox{.}(2009){Zingale}, {Almgren}, {Bell},
  {Nonaka}, \& {Woosley}}]{zingale2009a}
{Zingale} M., {Almgren} A.~S., {Bell} J.~B., {Nonaka} A., {Woosley} S.~E.,
  2009, \apj, 704, 196

\end{thebibliography}

\end{document}